\documentclass[lineno]{jfm}

\usepackage{graphicx}
\usepackage{newtxtext}
\usepackage{newtxmath}
\usepackage{natbib}
\usepackage{hyperref}
\usepackage{tikz}
\usetikzlibrary{calc}
\usetikzlibrary{shapes}

\hypersetup{
    colorlinks = true,
    urlcolor   = blue,
    citecolor  = black,
}

\newcommand{\RomanNumeralCaps}[1]
\nolinenumbers

\title{\nolinenumbers
Vortex ring–cylinder interactions: regimes, reconnection, and the role of topology}

\author{
\nolinenumbers
Andres Herrera-G\'omez\aff{1}, Juan \'Angel Tendero-Ventanas \aff{1}
 \and Rodolfo Ostilla-M\'onico\aff{1}
  \corresp{\email{rodolfo.ostilla@uca.es}}}

\affiliation{
\nolinenumbers
\aff{1}Dpto.~Ing. Mec\'anica y Dise\~no Industrial, Escuela Superior de Ingenier\'ia, Universidad de C\'adiz, Av.~de la Universidad de C\'adiz 10, 11519 Puerto Real, Espa\~na}

\begin{document}
\nolinenumbers
\maketitle
\nolinenumbers

\begin{abstract}
\nolinenumbers
We investigate the interaction between vortex rings and cylindrical obstacles using direct numerical simulations across a wide range of geometric and dynamical parameters. The flow is characterized in terms of the diameter ratio between ring and object $T_D = d/D$, the Reynolds number based on circulation $Re_\Gamma$, and the slenderness ratio $\Lambda$. By systematically varying $T_D$ and $Re_\Gamma$, we identify three distinct interaction regimes: the \emph{wire}, \emph{cutting}, and \emph{wall} regimes. In the \emph{wire} regime ($T_D \lesssim 0.05$), the primary vortex ring survives the interaction with limited deformation and carries a weak lobe of secondary vorticity generated by the object's boundary layer. As $T_D$ increases, the interaction transitions to the \emph{cutting} regime, where the ring is split into two secondary structures formed through the reconnection between boundary-layer and ring vorticity. For sufficiently large obstacles ($T_D \gtrsim 0.8$), the \emph{wall} regime emerges, in which boundary-layer vorticity dominates and the primary ring is deflected and stretched along the obstacle surface. The transition between regimes depends primarily on $T_D$, while increasing $Re_\Gamma$ enhances vortical dynamics producing additional small-scale and tertiary structures. Finally, by modifying the topology of the obstacle, we demonstrate that reconnection and recovery of the primary ring depend critically on the topology of the secondary vorticity. These results provide a unified framework for interpreting vortex–body interactions, bridging the gap between vortex ring, tube, and wall collision dynamics.
\end{abstract}


\nolinenumbers
\section{Introduction}
\label{sec:intro}

Vortex rings are canonical flow structures in fluid dynamics, providing a reproducible and well-defined system for investigating fundamental phenomena such as vorticity dynamics, stability, and energy transfer \citep{shariff1992vortex, lim1995vortex}. From a fundamental perspective, their study offers valuable insights into the behavior of more complex turbulent flows which can be rationalized from the dynamics of simpler structures. From an applied perspective, vortex rings arise in a variety of natural and technological settings, such as animal locomotion, rotorcraft wakes, volcanic eruptions, and cardiovascular flows. Advances in understanding their behavior have often led to improvements in engineering design and predictive modeling.

Among the many flow configurations involving vortex rings, their interaction with solid boundaries has received particular attention due to its simplicity and practical applications. Collisions with flat walls are relatively well understood: secondary vorticity is generated at the wall, which can organize into new ring structures or smaller-scale features, depending on the Reynolds number, and subsequently interact with the primary ring affecting its trajectory \citep{walker1987impact,orlandi1993vortex,swearingen1995dynamics,cheng2010numerical,mishra2021instability}. However, when the boundary is complex, or is a finite object, the interaction becomes complicated: the vortex may split into several components and new structures which are ejected from the collision area appear \citep{adhikari2009impact, hrynuk2012flow,hu2018vortex, new2020collision}. 

The interaction between a vortex ring and a cylindrical object is a particularly interesting case as it spans two limiting regimes. Thick cylinders serve as a model for studying curvature effects: a vortex interacting with such object will be affected by curvature in one direction, while ``feeling'' a flat wall in the other. On the other hand, slender wires offer a geometric setup for examining the topological cutting of vortex lines by extremely thin objects. 

As an example of the former, \cite{new2017head} investigated vortex rings of outer diameter $D$ colliding with cylindrical surfaces of diameter $d$, spanning $d/D\in[1,4]$. By using a combination of Laser-Induced Fluorescence (LIF) and Particle Image Velocimetry (PIV), they show that surface curvature significantly alters the interaction relative to a flat walls: small-scale vortex ringlets are generated and ejected away from the cylinder surface, something absent in collisions with flat no-slip walls. 

Furthermore, \cite{homa1988interaction} studied the formation of secondary vorticity in ring impact on very thin objects, showing that intense secondary vorticity forms rapidly on these structures. However, this work focused on the role of asymmetry during impacts. Later, \cite{naitoh1995vortex} studied the passage of vortex rings over wires with $d/D\in[0.00126,0.15]$ using smoke visualization. In this regime, the ring was found to survive the interaction: \cite{naitoh1995vortex} observed the formation of secondary vortices induced by boundary layer separation along the wire, and noted that the primary vortex ring deformed and adopted a trajectory more reminiscent of an elliptical ring following the interaction, while also observing a transition between what they denote a to be laminar and turbulent regimes. However, due to the qualitative nature of smoke visualization, that study offered limited insight into finer-scale phenomena such as vortex reconnection and amplification. 

Building on these findings, \cite{adhikari2009msc} provided a more detailed study and description of the ``cut-and-reconnect'' mechanism for thin wires. However, a significant dependence of the flow phenomenology on the stroke ratio used to generate the vortices was found, with the authors reporting that for several values of $d/D$, ring reconnection and reformation was present or not depending on it. We also note that studies of vortex rings impacting pourous surfaces also see similar phenomena of primary vortex survival \citep{adhikari2009impact, hrynuk2012flow}.

Numerical studies of this problem are challenging due to the disparate length scales involved, as the wires become increasingly thin and vortex sheets are increasingly concentrated and require fine resolution to resolve. \cite{verzicco1995numerical} studied the interaction between a vortex dipole and a cylinder in two-dimensions, while three-dimensional simulations have largely been limited to Large Eddy Simulations (LES). For instance, \cite{ren2015three} simulated collisions between rings and cylinders of equal diameter and observed tertiary ringlets and subsequent reconnection. \cite{new2021large} extended their experiments with LES to explore reconnection between secondary structures and ringlets, emphasizing the role of vortex stretching in cases with small cylinders. 

Despite these efforts, a systematic understanding of how the interaction transitions from the ``wire'' limit (where the ring is largely preserved) to the ``curved wall'' limit (where the ring breaks apart and ejects ringlets) remains lacking. It is unclear how ringlets relate to the complex collection of structures generated during ring-wall interaction. The precise parameters govern this transition are also unknown: the influence of Reynolds number and ring slenderness (i.e., the ratio of inner to outer radius) has not been systematically explored, and there is no consensus on the appropriate non-dimensional thickness parameter to characterize the interaction \citep{naitoh1995vortex, new2017head}.

Vortex ring-cylinder interactions also bear resemblance to body–vortex interaction (BVI), in which a vortex tube encounters a moving object. In this case, the secondary vorticity forming on the object plays a fundamental role in the flow dynamics. However, BVI typically results in deformation waves or detached vorticity propagating along the vortex tube, rather than vortical structures being formed and ejected \citep{soriano2024direct}. This raises further questions about the existence of fundamental differences between rings and tubes interacting with objects due to the different topology of the vortex lines.

In this manuscript, we aim to address these open questions through a systematic investigation of vortex ring–cylinder interactions using direct numerical simulations (DNS). We explore a wide range of diameter ratios: from slender wires to cylinders larger than the ring’s outer radius, and vary ring slenderness to identify the key parameters that control interaction outcomes. We also examine how changes in object topology influence the resulting flow structures by simulating the interaction between a ring and half a cylinder, or star-shaped objects with $N_p$ points composed of cylinders and analyzing the outcome of the collision. 

The manuscript is organized as follows: $\S$\ref{sec:num} details the numerical methods used in the manuscript. In $\S$\ref{sec:res1} we present an overview of the three regimes seen in our simulation: ``wire'', ``cutting'', and ``curved wall''. The ``wire'' regime is analyzed in depth in $\S$\ref{sec:res2}, while $\S$\ref{sec:res3} analyzes the ``cutting'' regime with a focus on how the topology of the object affects the outcome. The effect of topology with a focus on the comparison between vortex rings and vortex tubes is presented in $\S$\ref{sec:topology}. Finally, $\S$\ref{sec:cc} presents a summary of the findings as well as an outlook for future work.

\section{Methods}
\label{sec:num}

\subsection{System description}

\begin{figure}
\centering
\begin{tikzpicture}[scale=1.4]
\usetikzlibrary{arrows.meta}
\draw (1.5,1.5) circle(0.5cm);
\draw (1,1.5) -- (1,3.5);
\draw (2,1.5) -- (2,3.5);
\draw (1.5,3.5) circle(0.5cm);
\draw[fill] (1.5,3.5) circle [radius=0.02];

\draw [stealth-] (1.2,1.5) arc(180:360:0.3cm); 
\node at (1.5,1.5) {$\Gamma$};

\draw [latex-latex, thick] (1.018,3.371) -- (1.983,3.629);
\node [above left] at (1.5,3.5) {$2\sigma$};

\draw [dashed] (0.5,2.5) -- (2.5,2.5);
\draw [latex-latex] (0.5,2.5) -- (0.5,3.5);
\node [left] at (0.5,3) {$R$};
\draw [dotted] (1.5,3.5)--(0.5,3.5);

\filldraw[color=black, fill=gray!25] (4,2.5) circle (0.25);

\draw [dotted] (3.75,2.5) -- (3.75,3.5);
\draw [dotted] (4.25,2.5) -- (4.25,3.5);
\draw [latex-latex] (3.75,3.4) -- (4.25,3.4);
\node [above] at (4,3.4) {$d$};

\draw [-latex,thick] (2,2.5) -- (2.4,2.5);
\node [above] at (2.2,2.5) {$\hat{V}_a$};

\draw [-latex,thin] (4,1.5) -- (4.4,1.5);
\draw [-latex,thin] (4,1.5) -- (4,1.9);
\draw [-latex,thin] (4,1.5) -- (3.75,1.25);

\node [left] at (4,1.7) {$\hat{x}$};
\node [below] at (4.2,1.5) {$\hat{z}$};
\node [below] at (3.9,1.35) {$\hat{y}$};

\end{tikzpicture}
\hspace{1cm}
\includegraphics[width=.4\textwidth]{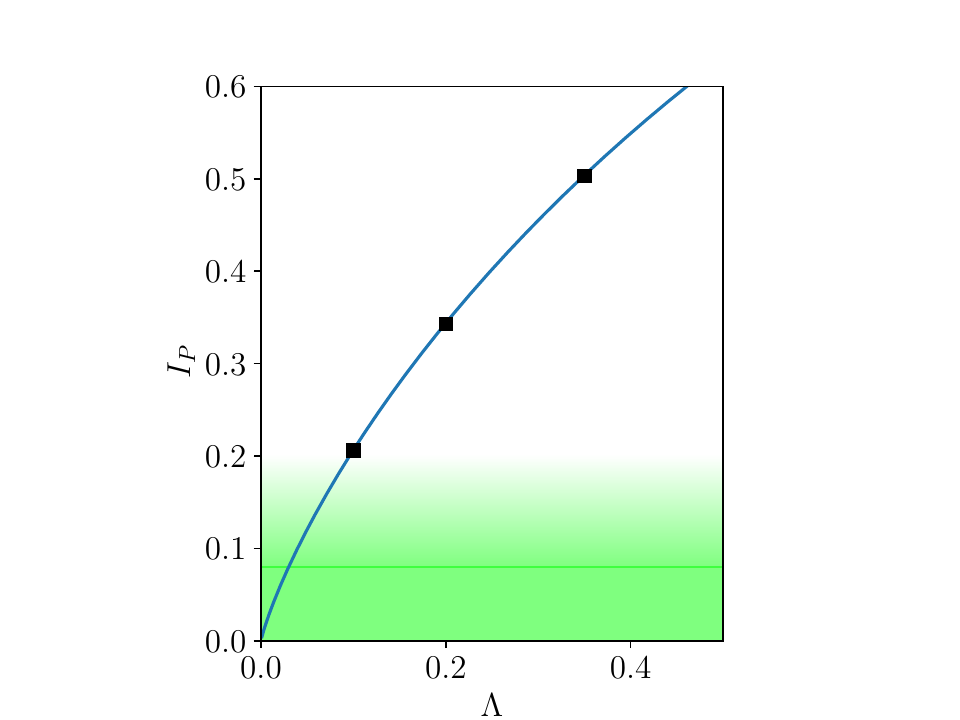}
\caption{Left: Schematic of a vortex ring approaching the cylinder-like object. The figure represents a cut of the system by a plane containing the ring's axis of symmetry, shown as a dashed line, and the cylinder's axis. Right: variation of impact parameter $I_P$ with ring slenderness $\sigma$ as in equation \ref{eq:ipsigma}. The shaded green region indicates the strong vortex regime.} 
\label{fig:schemaring}
\end{figure}

We consider the interaction in an incompressible fluid between a vortex ring of inner radius $\sigma$ and outer radius $R$ (outer diameter $D=2R$) and cylindrical objects of diameter $d$, as shown in the schematic of Figure \ref{fig:schemaring}.

The equations which model this problem are the incompressible Navier-Stokes equation: 

\begin{equation}
     \displaystyle\frac{\partial\hat{\textbf{u}}}{\partial \hat{t}} + \hat{\textbf{u}}\cdot\nabla\hat{\textbf{u}} = -\rho^{-1}\nabla \hat{p} +\nu\nabla^2 \hat{\textbf{u}}
 \label{eq:ns}
\end{equation}

\noindent and the continuity equation:

\begin{equation}
 \nabla\cdot \hat{\textbf{u}}=0
 \label{eq:compress}
\end{equation}

\noindent where $\hat{\textbf{u}}(\hat{\textbf{x}},\hat{t})$ is the dimensional velocity, $\hat{\textbf{x}}$ the dimensional position vector, $\hat{t}$ is dimensional time, $\hat{p}$ the fluid pressure, $\rho$ is the fluid density and $\nu$ is the fluid kinematic viscosity. 

The vortex ring is defined through an approximate Lamb-Ossen (Gaussian) vorticity profile with a circulation $\Gamma$ as in \cite{mishra2021instability}. The equations are non-dimensionalized using $R$ and $\Gamma$ to define length and time scales, resulting in (unhatted) non-dimensional time $t=\hat{t}\Gamma/R^2$, position $\textbf{x}=\hat{\textbf{x}}/R$ and velocity $\textbf{u}=\hat{\textbf{u}}R/\Gamma$. 

The computational domain is taken as a cylinder of radius $5R$ and height $14R$ bounded by free-slip walls. The vortex is released a distance of $2R$ from the top wall ($z=\hat{z}/R=0$), and travels a distance of $L_z=4R$ before encountering the cylindrical object, whose axis is located at $x=\hat{x}/R=0$, $z=\hat{z}/R=4$, i.e.~parallel to the $y$-axis. This leaves a total of $9R$ lengths in the axial direction for the flow to develop after the interaction. In addition, the $x$-$z$ plane will remain as perpendicular to the cylinder, while $y$-$z$ will be parallel to the cylinder. 

Three non-dimensional parameters characterize the problem:
\begin{equation}
 Re_\Gamma=\frac{\Gamma}{\nu}, \qquad \Lambda=\frac{\sigma}{R}, \qquad T_D=\frac{d}{D}=\frac{d}{2R}.
\end{equation}
\noindent The first two commonly appear in vortex ring systems: the circulation Reynolds number $Re_\Gamma$ and the ring slenderness ratio $\Lambda$. The third non-dimensional parameter compares the size of the object and the ring. We chose the characteristic size of the ring to be $D$, the outer diameter, which results in a the thickness ratio $T_D$. The outer diameter $D$ is chosen over the outer radius $R$ to follow \cite{new2017head}, but this is not significant, as choosing $R$ would result in a different parameter which would just differ by a factor of two. The relevant choice is choosing an outer scale over an inner one in the ring. We could also define a non-dimensional thickness ratio based on the inner vortex radius $\sigma$, 
\begin{equation}
    T_\sigma=\frac{d}{\sigma}=\frac{2T_D}{\Lambda}.
\end{equation} 
\noindent This definition is analogous to the thickness ratio used in the interaction between a vortex tube and an object, where $\sigma$ is the only length choice available. And indeed, a similar quantity ($\Pi_3\approx d/(2\sigma)=T_\sigma/2$) was the dimensionless parameter used by \cite{naitoh1995vortex} to classify the regimes of vortex ring-wire interaction. However, as we will show below, the distinction between flow regimes is better delineated by the value of $T_D$ and not $T_\sigma$, so $T_D$ will be the preferred parameter throughout the manuscript.

\subsection{The fixed relationship between slenderness and impact parameter}

In the interaction between a vortex tube and a body, the (dimensional) impact velocity $\hat{V}$ is an independent parameter, regardless of whether the vortex moves towards the body or the body towards the vortex, as the tube does not self-advect. When non-dimensionalized, we obtain the impact parameter:

\begin{equation}
I_P=\frac{2\pi\sigma \hat{V}}{\Gamma}.
\end{equation} 

\noindent This represents the ratio between the impact velocity and a characteristic velocity of the vortex, and is the prime non-dimensional number generally used to classify the interaction: depending on its value,  distinct flow phenomenology can be observed, which is usually classed into ``strong'' vortex and ``weak'' vortex regimes \citep{marshall1997instantaneous}. 

As already stated, the choice of $\hat{V}$ is generally free in BVI. However, when considering vortex rings, these structures self-propel at a velocity $\hat{V}_a$ which can be approximated as \citep{fukumoto2010global}:

\begin{equation}
    \hat{V}_a=\frac{\Gamma}{4\pi R}\left ( \log \frac{8}{\Lambda} -\frac{1}{4} \right ).
    \label{eq:va}
\end{equation}

\noindent If the object is kept fixed, which is usually the case in experiments, $\hat{V}_a=\hat{V}$, and this results in a fixed relationship between the impact parameter $I_P$ and the slenderness ratio $\Lambda$: 

\begin{equation}
    I_P = \frac{\Lambda}{2} \left ( \log \frac{8}{\Lambda} -\frac{1}{4} \right ) .
    \label{eq:ipsigma}
\end{equation}

\noindent This equation is plotted in the right panel of Figure \ref{fig:schemaring}. In general, smaller values of $\Lambda$ (i.e.~more slender rings) will propel faster, but will have even larger rotation velocities, which overall result in smaller values of $I_P$. Furthermore, unless $\Lambda$ is made very small, the range of values $I_P$ can take are rather limited and generally correspond to values of $I_P$ in the weak vortex regime of BVI as shown in the right panel of Figure \ref{fig:schemaring} \citep{marshall1997instantaneous, soriano2024direct}. Achieving small values of $I_P$ such that the interaction would correspond to a strong vortex, would require $\Lambda\lesssim 0.05$ and correspond to extremely thin rings which are both uncommon in experiments as they are hard to produce, and impractical to simulate as the resolution requirements of the simulation increase dramatically.

\subsection{Simulation details}

Equations \ref{eq:ns}-\ref{eq:compress} are discretized in space using an energy-conserving second-order centered finite difference approach, and are advanced in time using a fractional time-stepping method. The simulations are performed using the open-source code AFiD \citep{van2015pencil}, which has previously been used for head-on vortex ring collision and vortex impact against a wall among other problems \citep{mckeown2020turbulence,mckeown2023energy,mishra2021instability}. To avoid singularity problems at the axis, the radial velocity is substituted for the primitive variable $q_r=rv_r$ \citep{verzicco1996finite}. The cylindrical objects are implemented through the immersed boundary method detailed in \cite{fadlun2000combined}. 

The first set of simulations, discussed in $\S\S$\ref{sec:res1}-\ref{sec:res3}, will explore the interaction between a ring and a cylinder (wire), bridging the gap between the setups of \cite{naitoh1995vortex} and \cite{new2017head}. We will simulate two different Reynolds, $Re_\Gamma=1000$ and $Re_\Gamma=2000$, three different slenderness ratios $\Lambda=0.1$, $0.2$ and $0.35$, and vary $T_D$ in the range $T_D\in[0.025,2]$, which results in values of $T_\sigma\in[0.25,20]$. A summary of the cases studied is available in table \ref{tab:ring}. For both values of $Re_\Gamma$, the resolution is chosen as $N_r\times N_\theta \times N_z= 384\times384\times384$, with homogeneously distributed points in all directions. In Appendix \ref{appA} we provide a comparison of results obtained for different meshes for a case at at $Re_\Gamma=2000$ to justify that the grid resolution is sufficient. 

\begin{table} 
    \centering
    \begin{tabular}{|c|c|c|c|c|}
    \hline
    $\Lambda$ & $T_D$ & $T_\sigma$ & $I_P$ & $Re_\Gamma$ \\ \hline
    0.2 & 0.025 & 0.25 & 0.34 & 1000, 2000 \\
    0.2 & 0.05  & 0.5  & 0.34 & only 1000 \\
    0.2 & 0.1  & 1   & 0.34 & 1000, 2000 \\    
    0.2 & 0.2  & 2   & 0.34 & only 1000  \\        
    0.2 & 0.4  & 4   & 0.34 & 1000, 2000  \\
    0.2 & 0.7  & 7   & 0.34 & only 1000  \\
    0.2 & 1    & 10  & 0.34 & 1000, 2000  \\
    0.2 & 2    & 20  & 0.34 & 1000, 2000 \\
     \hline
    0.1  & 0.05 & 1   & 0.21 & 1000, 2000   \\
    0.1  & 0.4  & 8   & 0.21 & 1000, 2000   \\
    0.35 & 0.05 & 0.29 & 0.50 & 1000, 2000  \\
    0.35 & 0.4  & 2.29 & 0.50 & 1000, 2000  \\
    \hline
    \end{tabular}
    \caption{Control parameters for the simulations of tube-wire interaction in this manuscript. The ring's impact parameter is estimated from equation \ref{eq:ipsigma}.} 
    \label{tab:ring}
\end{table}

The second set of simulations, discussed in $\S$\ref{sec:topology}, explores the varying the topology of the impacted object on the collision. For these simulations, we fix $Re_\Gamma=1000$ and the slenderness ratio $\Lambda=0.2$ and set $T_D=0.1$ and vary the geometry of the wire-like objects. The effective $I_P$ is $0.34$, again in the weak vortex regime. The resolution is kept as $N_r\times N_\theta \times N_z= 384\times384\times384$ with points distributed homogeneously in the three dimensions. 

\section{An overview of the flow regimes through flow visualization}  \label{sec:res1}
\subsection{Flow regimes for $\Lambda=0.2$}

\begin{figure}
    \centering
    \includegraphics[width=0.9\linewidth]{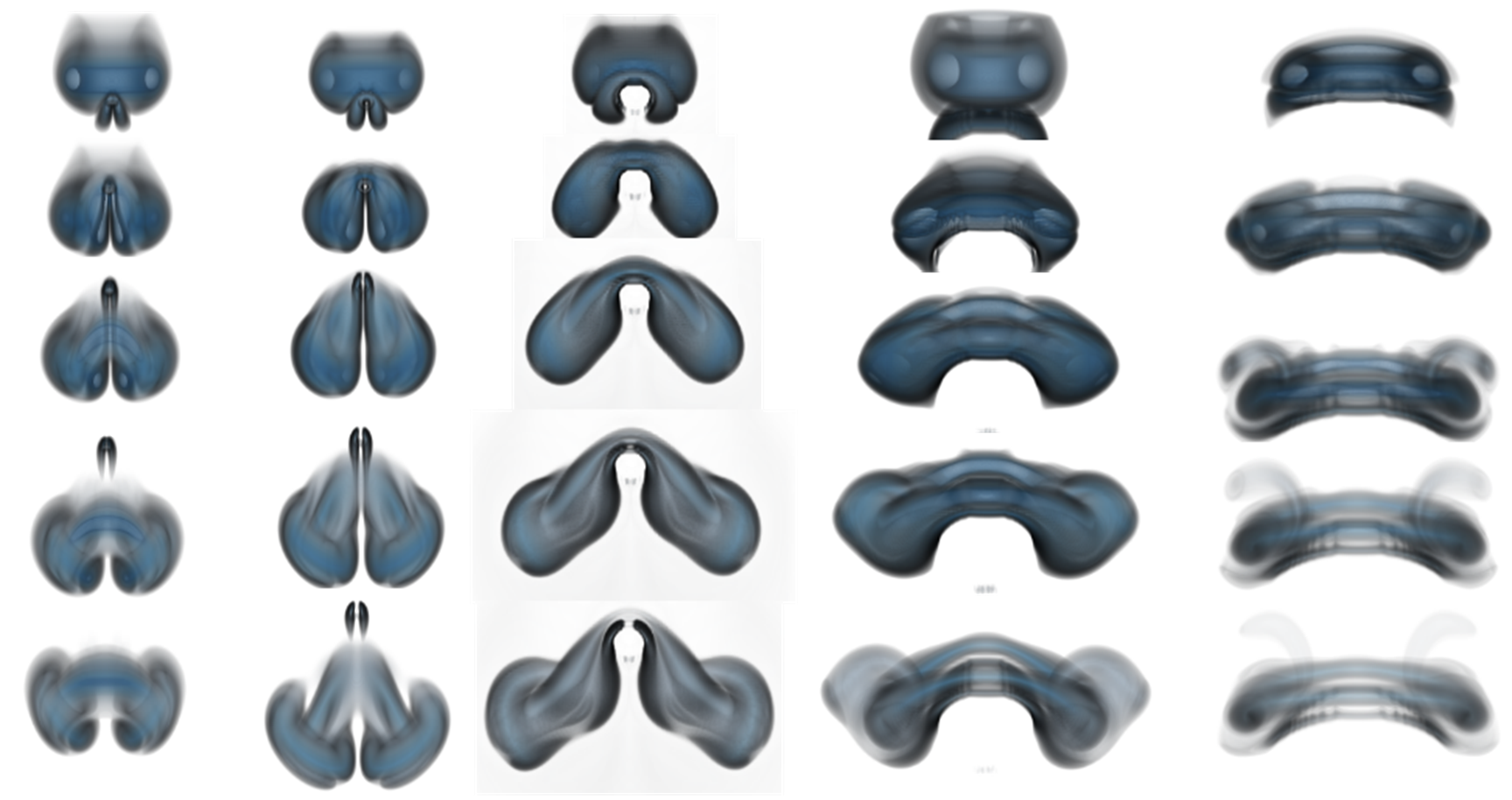}
    \caption{Volumetric visualization of the instantaneous vorticity modulus of the impact of a vortex ring with $\Lambda=0.2$ and $Re_\Gamma=1000$ upon a cylinder of varying diameter: (left to right: $T_D=0.025$, $0.1$, $0.4$, $1$ and $2$). Time ranges from $t=25$ to $65$ in intervals of $10$. }
    \label{fig:cr0p2-rcyls-re1000}
\end{figure}

We first present the different regimes in the flow as $T_D$ is varied. We start by showing the temporal evolution of the vorticity modulus for $Re_\Gamma=1000$ simulations with $T_D=0.025$, $0.1$, $0.4$, $1$ and $2$ in Figure \ref{fig:cr0p2-rcyls-re1000}. These are chosen as they exemplify the effect of the cylinder thickness on how the flow evolves, and which type of structures are formed. 

First, for $T_D=0.025$ we can see how a small boundary layer is formed around the wire. The ring is slightly slowed down by the object, but manages to transverse the wire, coming out with a deformed profile: the ring is no longer axisymmetric but instead  has two ``lobes'' of vorticity in the direction perpendicular to the wire. The secondary vorticity originating from the wire is dragged along for a short period, but does not interact strongly with the ring. We denote this as the ``wire'' regime explored by \cite{naitoh1995vortex}: a regime defined by the fact that the primary vortex survives the interaction albeit with some deformation and is no longer axisymmetric. The phenomenology of this regime will be explored in detail in $\S$\ref{sec:res1}. This regime is analogous to the ``cut-and-reconnect'' regime explored by \cite{adhikari2009msc}, however, as we observe several types of cutting and reconnection, we prefer to denote it as the ``wire'' regime. We are able to observe ring reconnection at higher values of $T_D$ than in \cite{adhikari2009msc} who only explore the phenomena up to $T_D\approx 0.03$. Indeed, we note that similar flow phenomena are seen for $T_D=0.05$, and they are not shown in the interest of space. We also note that as our vortices are generated numerically and not using a piston, the dependence of the behaviour on the piston stroke-ratio $L_p/D_p$, where $L_p$ is the piston stroke length, and $D_p$ the piston diameter does not appear in our phenomenology.

Once $T_D$ is increased to $0.1$, a new regime appears, which we denote as the ``cutting'' regime. The result of the interaction is to split the primary vortex into two vortices, which are launched diagonally outwards from the first interaction, the angle being larger with increasing $T_D$. For $T_D=0.2$ (not shown) and for $T_D=0.4$, vorticity is wrapped around the cylinder and stretched into sheets which are dissipated. In this regime, the primary vortex does not survive and is instead split into two.

Finally, the two right-most columns show the ``curved wall'' regime as analyzed by \cite{new2017head}. As the cylinder becomes larger relative to the outer radius of the ring, the vortex is no longer able to wrap itself around the object. The interaction increasingly resembles that of a ring against a wall. In the direction perpendicular to the cylinder's axis, the secondary vorticity from the object is lifted up, and curls around the primary ring. However this lifted up vorticity is unable to curl into a secondary vortex as seen in the experiments of \cite{new2017head} at larger Reynolds numbers. Instead, for both values of $T_D$ the lifted vorticity is dissipated due to viscosity, and at latter times the primary ring also dissipates. We also note that while some difference remains between the direction parallel to the cylinder axis, and that perpendicular to it, in the limit of $T_D\to\infty$ we expect them to become equal as the flow will simply become that of a ring impacting a wall head-on.

\begin{figure}
    \centering
    \includegraphics[width=0.9\linewidth]{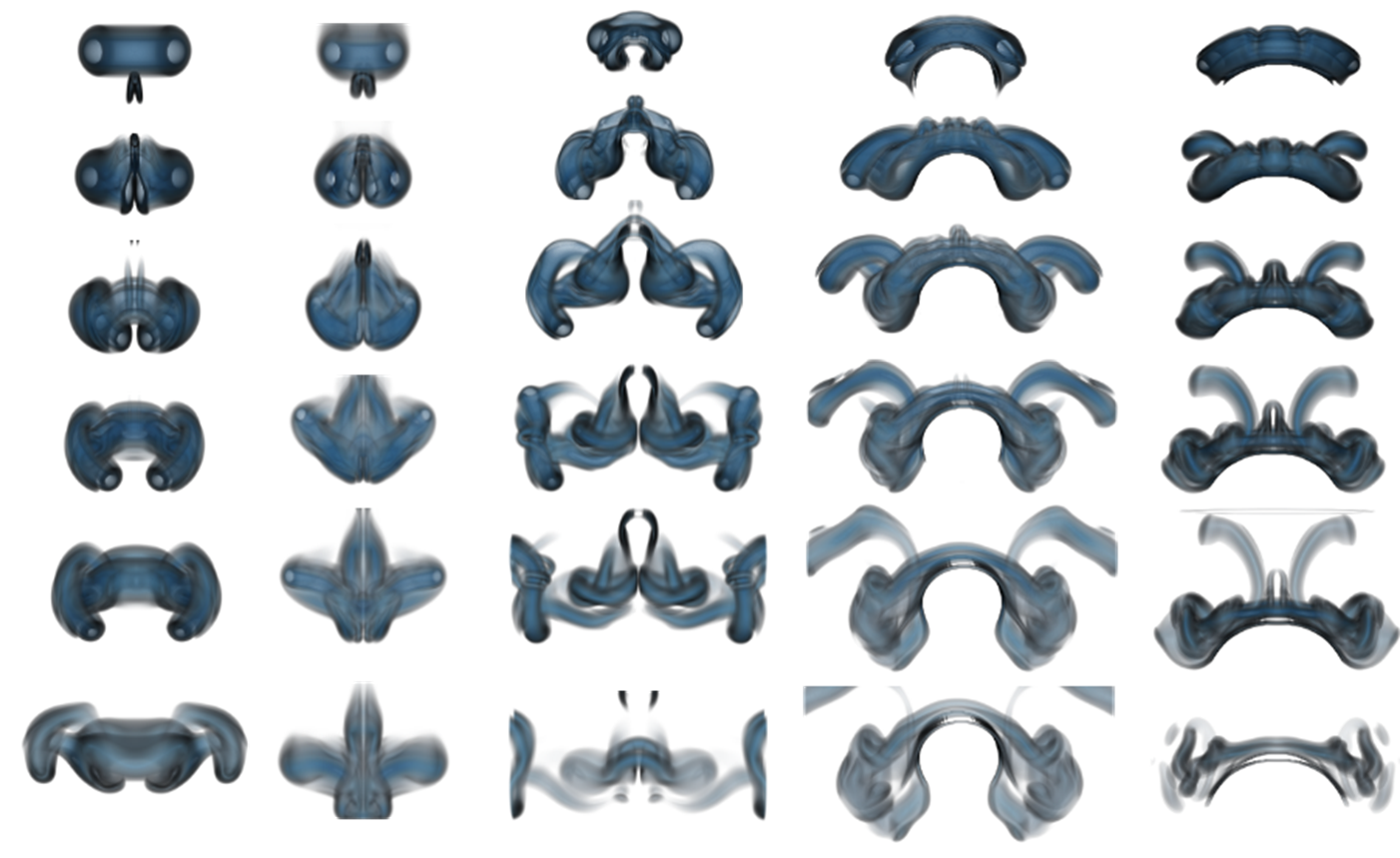}
    \caption{Volumetric visualization of the instantaneous vorticity modulus of the impact of a vortex ring with $\Lambda=0.2$ and $Re_\Gamma=2000$ upon a cylinder of varying diameter: (left to right: $T_D=0.025$, $0.1$, $0.4$, $1$ and $2$). Time ranges from $t=25$ to $65$ in intervals of $10$, with the last snapshot at $t=85$. }
    \label{fig:cr0p2-rcyls-re2000}
\end{figure}

We now increase the value of $Re_\Gamma$:  Figure \ref{fig:cr0p2-rcyls-re2000} shows the temporal evolution of the vorticity modulus for the same cases as figure \ref{fig:cr0p2-rcyls-re1000}, but for $Re_\Gamma=2000$. As the Reynolds number increases, the flow phenomenology is modified. While the broad contours between the three regimes above are more or less maintained, due to the decreased viscosity, the structures resulting from the first instants of time after the interaction can develop new physics. This happens because the vorticity layers are thinner, and can roll into new vortices, and in some cases, the secondary rings can interact and reconnect, to recreate a ring which resembles the primary ring, with some azimuthal variance. We note that vortex sheets rolling into vortex tubes was a phenomena observed and described in the head-on colission of two vortex rings \citep{mckeown2018cascade,mckeown2020turbulence}.

For instance, when $T_D=0.1$, the two rings which are laterally ejected from the wire grow enough to touch, and reconnect, while for $T_D=0.4$, two rings are ejected laterally, and two rings are ejected in a forwards direction which then reconnect to form a deformed ring which continues traveling in the same direction the primary ring was traveling before the interaction. Finally, for the large $T_D$ cases, the secondary vorticity which dissipated at $Re_\Gamma=1000$ is now able to lift-up and eject from the collision plane. In addition, we see that for $T_D=2$, the secondary vorticity causes the primary ring to slow down, and even to slightly lift off from the cylinder at later stages, giving the appearance of the ``bounce'' often seen for the interaction of rings with flat walls \citep{mishra2021instability}. 

Overall, we can state that for $Re_\Gamma=2000$, the flow phenomenology is quite complex: if the object is thin, the primary ring survives but is deformed after interaction. If the object is thicker, the presence of a curved wall causes secondary vortex rings to be ejected from the collision plane at an angle which increases with increasing $T_D$. Finally, for the largest $T_D$, the interaction increasingly resembles that between a cylinder and a wall, where the primary vortex is lifted up and away from the object. However, the transition between regimes is smooth. In intermediate cases, it is possible for two phenomena to occur simultaneously as seen for $T_D=0.4$: a ring-like structure keeps on propagating in the original direction (characteristic of the ``wire'' regime), while secondary rings are ejected laterally from the collision plane (characteristic of the ``cutting'' regime).

Having delineated the regimes, we can now analyze the flow phenomena associated to the ``wire'' and `cutting'' regimes in more detail. In $\S$\ref{sec:res2} we will focus on how and when the primary ring survives, while in $\S$\ref{sec:res3} we will analyze how secondary structures are ejected from the collision region. The ``curved wall'' regime and its associated phenomenology is not included in the present manuscript as detailed analysis are already available in \cite{new2017head, new2021large} and our simulations do not significantly diverge from the results presented there. But before doing so, we comment on the effect of $\Lambda$ and $T_\sigma$ on the results.

\subsection{The effect of $\Lambda$ and $T_\sigma$}

Up to this point, we have fixed $\Lambda=0.2$, and varied the cylinder diameter. In this subsection, we vary $\Lambda$. This not only serves to examine the role of ring slenderness, but it also serves to decouple $T_D$ and $T_\sigma$ as they both have varied by the same factor as the ring diameter was changed. By modifying $\Lambda$, we can decouple these two parameters to assess which of the two non-dimensional parameters characterizes the interaction more effectively.  

\begin{figure}
    \centering
    \includegraphics[width=0.8\linewidth]{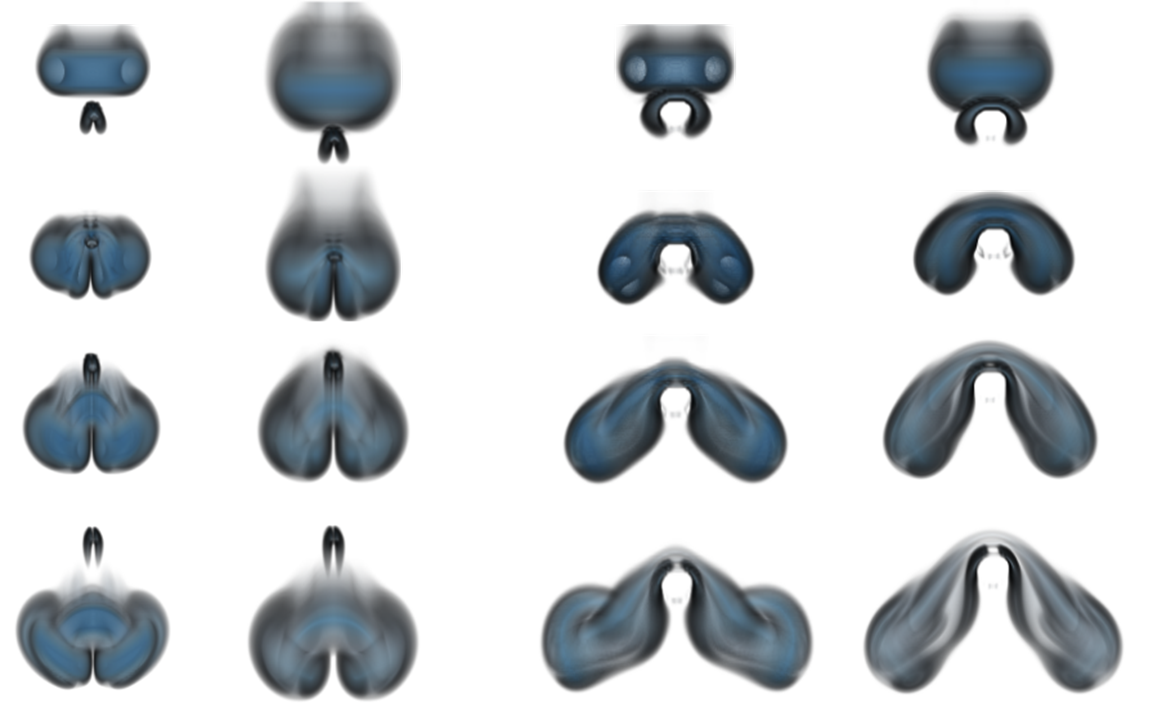}
    \caption{Volumetric visualization of the instantaneous vorticity modulus of a vortex ring impacting a wire at $Re_\Gamma=1000$. Left to right: $T_D=0.05$, $\Lambda=0.1$ ($T_\sigma=1$); $T_D=0.05$, $\Lambda=0.35$ ($T_\sigma=0.28$); $T_D=0.4$, $\Lambda=0.1$ ($T_\sigma=8$); and $T_D=0.4$, $\Lambda=0.35$ ($T_\sigma=2.3$). For $\Lambda=0.1$, time ranges from $t=20$ to $50$ in intervals of $10$, while for $\Lambda=0.35$ time ranges from $t=25$ to $70$ in intervals of $15$.}
    \label{fig:crs-rcyls-re1000}
\end{figure}

Figure \ref{fig:crs-rcyls-re1000} shows the temporal evolution of the vorticity modulus at $Re_\Gamma=1000$ for $\Lambda=0.1$ and $\Lambda=0.35$, with $T_D=0.025$ and $T_D=0.4$, corresponding to the ``wire'' and ``splitting'' regimes. Comparing the first and second columns, we note that $\Lambda$ differs by a factor of 3.5, so $T_\sigma$ also varies by that factor, yet the dynamics remain essentially the same. As $T_D=0.025$ for both cases, this supports the fact that $T_D$ is the controlling parameter. A similar comparison between the third and fourth column, where again $T_\sigma$ changes from $8$ to $2.3$ while $T_D$ is fixed, further confirms that large-scale behaviour is dictated by $T_D$, while changes in $\Lambda$ or $T_\sigma$ do not alter the interaction regime.

\begin{figure}
    \centering
    \includegraphics[width=0.8\linewidth]{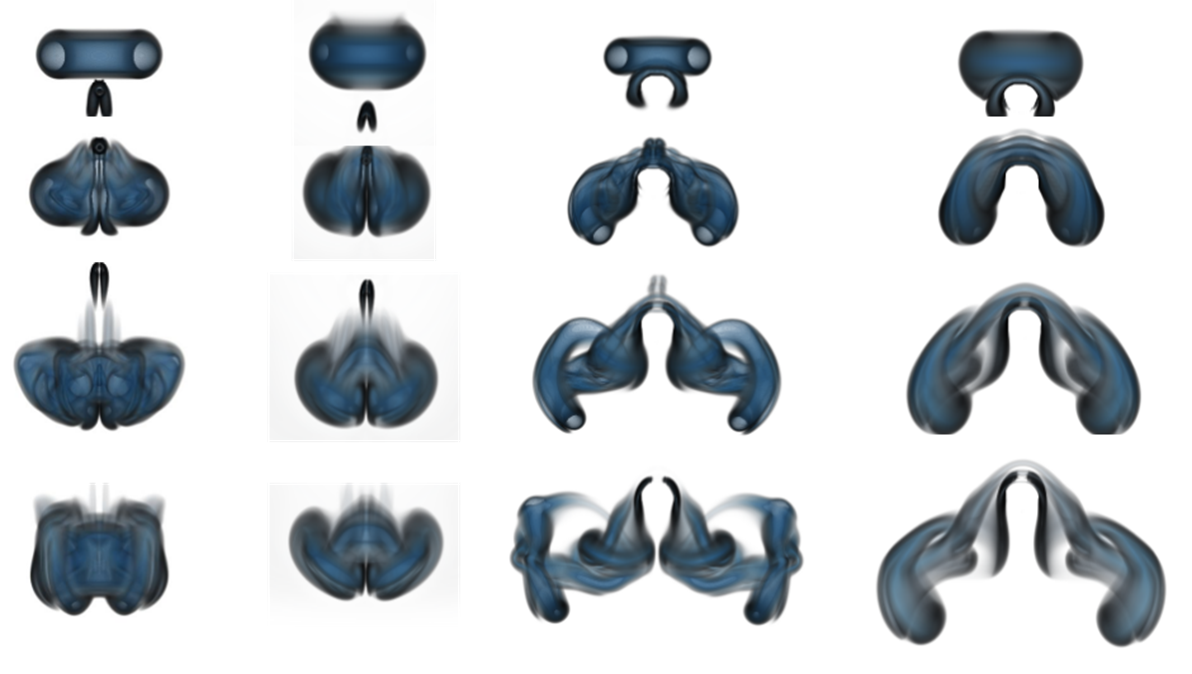}
    \caption{Volumetric visualization of the instantaneous vorticity modulus of a vortex ring impacting a wire at $Re_\Gamma=2000$. Cases and times are the same as Figure \ref{fig:crs-rcyls-re1000}.}
    \label{fig:crs-rcyls-re2000}
\end{figure}

As the Reynolds number increases, the same trend persists: higher $Re_\Gamma$ enables some vortex sheets to roll up into tubes or rings, influencing the dynamics, but the dominant parameter remains $T_D$ and not $T_\sigma$. This is consistent with the results of  \cite{mishra2021instability} for the impact of a ring on a flat wall, where vortex slenderness had little influence on the dynamics except that ``fat'' (high $\Lambda$) rings evolved more slowly. In practice, this also means that direct comparisons of rings with different $\Lambda$ at the same $Re_\Gamma$ can be misleading: a larger $\Lambda$ requires a higher $Re_\Gamma$ to trigger sheet roll-up, as the velocities, and hence the shears are slightly smaller. For example, in Figure \ref{fig:crs-rcyls-re2000}, the $\Lambda=0.35$ case behaves much like its $Re_\Gamma=1000$ counterpart, while the $\Lambda=0.1$: case shows the same transition observed earlier for $\Lambda=0.2$ in figures \ref{fig:cr0p2-rcyls-re1000} and \ref{fig:cr0p2-rcyls-re2000}. This difference reflects a small shift in the threshold Reynolds number for roll-up due to the lower shear rates, but is not a fundamentally new phenomena. Therefore, aside from noting these nuances, we can fix $\Lambda=0.2$, and use $T_D$ alone to classify the flow regimes. 

\section{The ``wire'' regime: Survival of the Primary Ring}
\label{sec:res2}

\subsection{Results for $Re_\Gamma=1000$}

\begin{figure}
    \centering
    \includegraphics[width=0.80\linewidth]{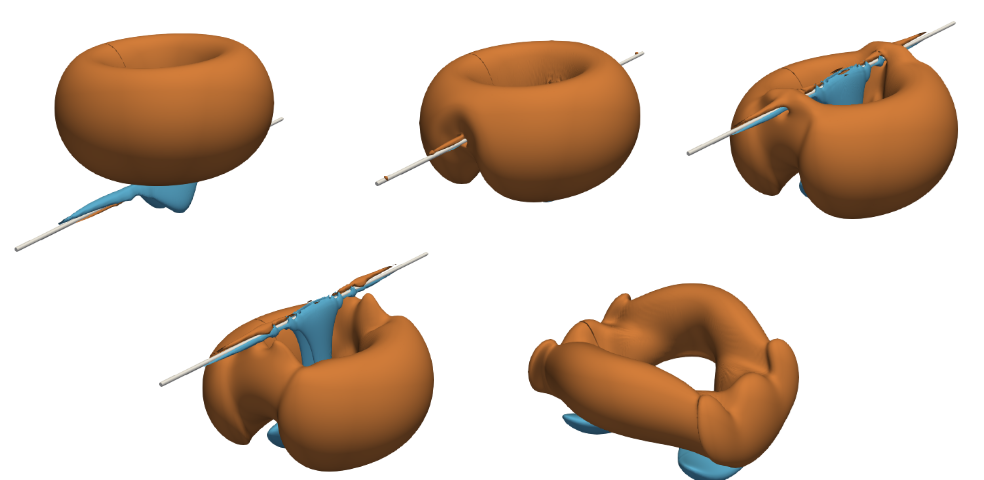}
    \caption{Evolution of the azimuthal vorticity for $T_D=0.025$, $\Lambda=0.2$, $Re_\Gamma=1000$. Isocontours at $\omega_\theta=0.2$ (orangish-red) and $\omega_\theta=-0.15$ (blue).  From top left to bottom right: $t=25$, $32.5$, $37.5$, $42.5$ and $60$. }
    \label{fig:cont-cr0p2-rcyl0p025-wth-re1000}
\end{figure}

We begin by examining the case with the smallest $T_D=0.025$. Figure~\ref{fig:cont-cr0p2-rcyl0p025-wth-re1000} shows two isocontours of the instantaneous azimuthal vorticity. Orangish-red is used to color the positive azimuthal vorticity, linked to the primary vortex, while blue is used to color negative azimuthal vorticity, linked to the boundary layer formation and later detached secondary vortices. In these panels we can clearly observe how the ring is able to transverse cleanly the wire: primary vorticity can wrap itself around the wire, and at moments there is vorticity on both sides of the wire. The primary vortex ring is also able to drag some of the oppositely-signed secondary vorticity with it, and this remains in the shape of two lobes in the $x$-$z$ plane.  

\begin{figure}
    \centering
    \includegraphics[trim={3cm 0 4cm 0},clip,height=0.22\linewidth]{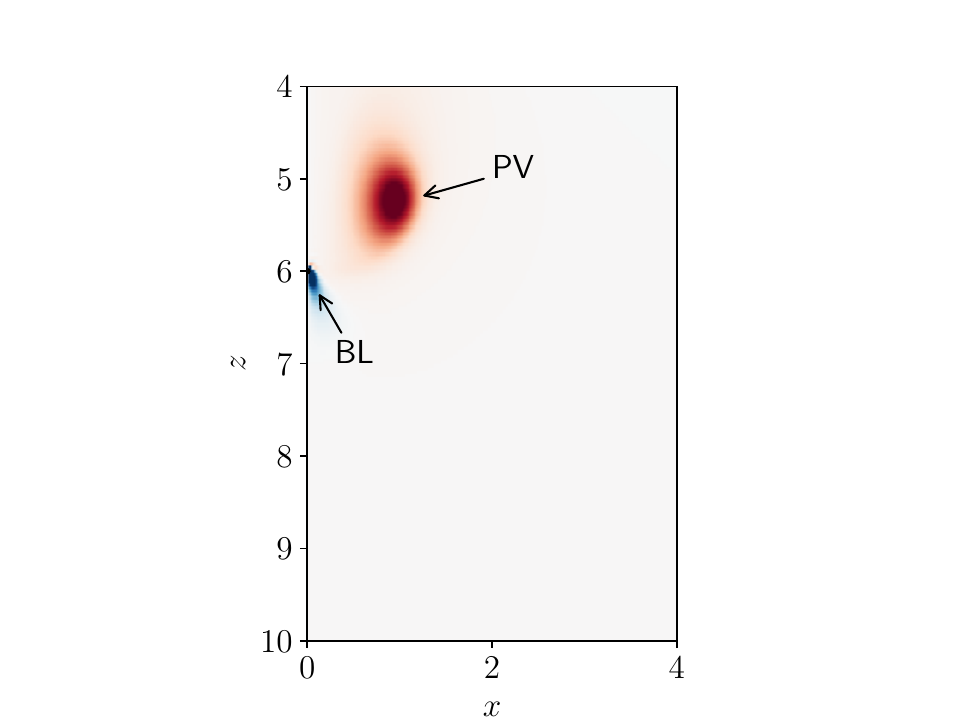}
    \includegraphics[trim={3cm 0 4cm 0},clip,height=0.22\linewidth]{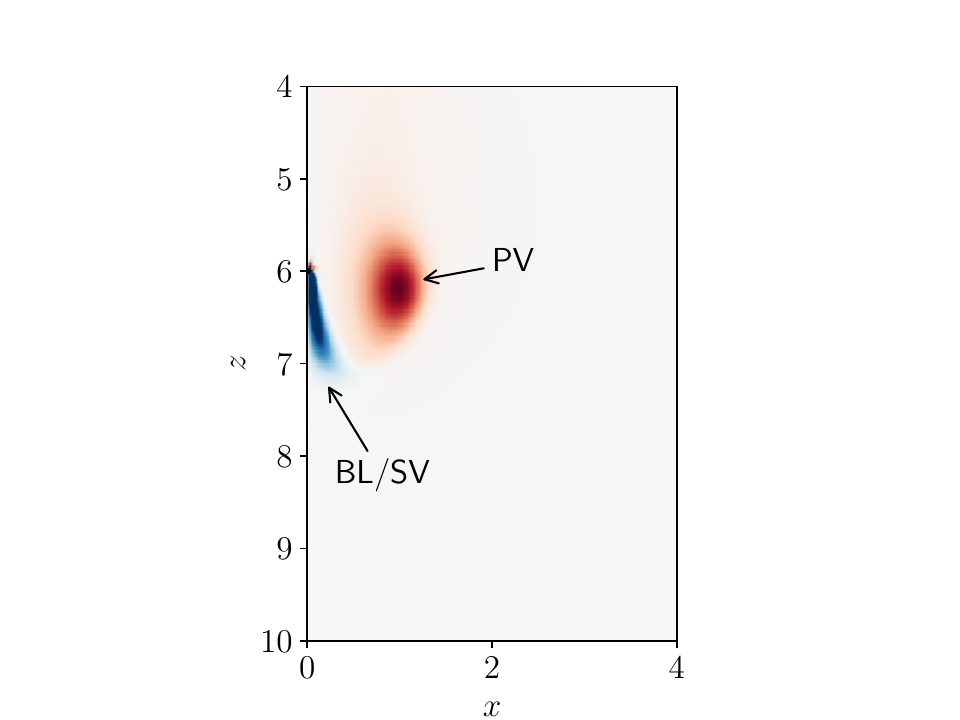}
    \includegraphics[trim={3cm 0 4cm 0},clip,height=0.22\linewidth]{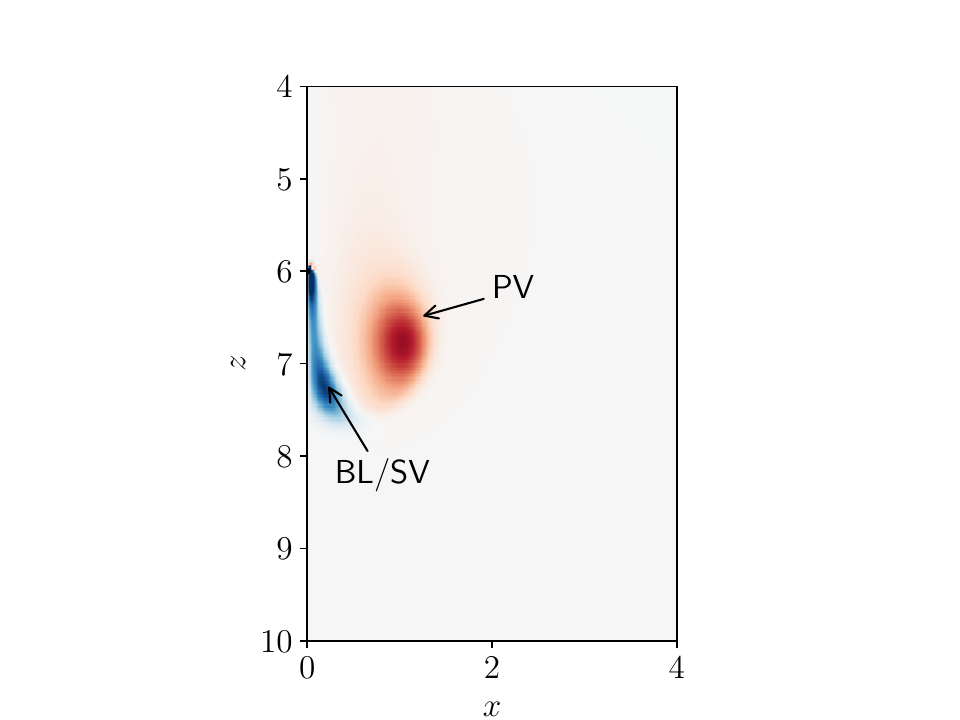}
    \includegraphics[trim={3cm 0 4cm 0},clip,height=0.22\linewidth]{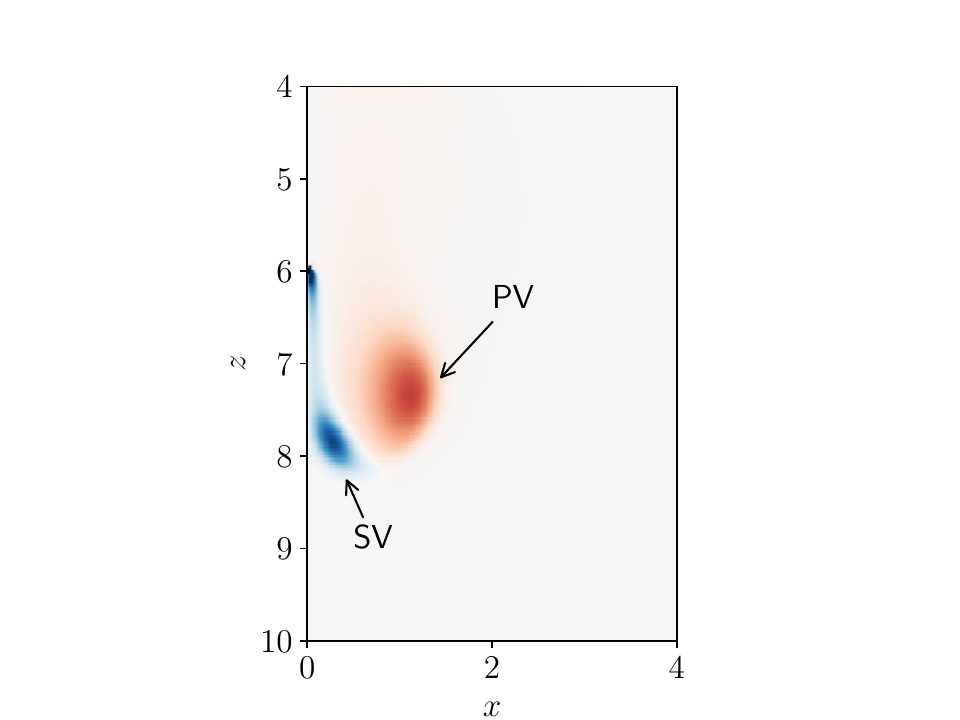}
    \includegraphics[trim={3cm 0 0cm 0},clip,height=0.22\linewidth]{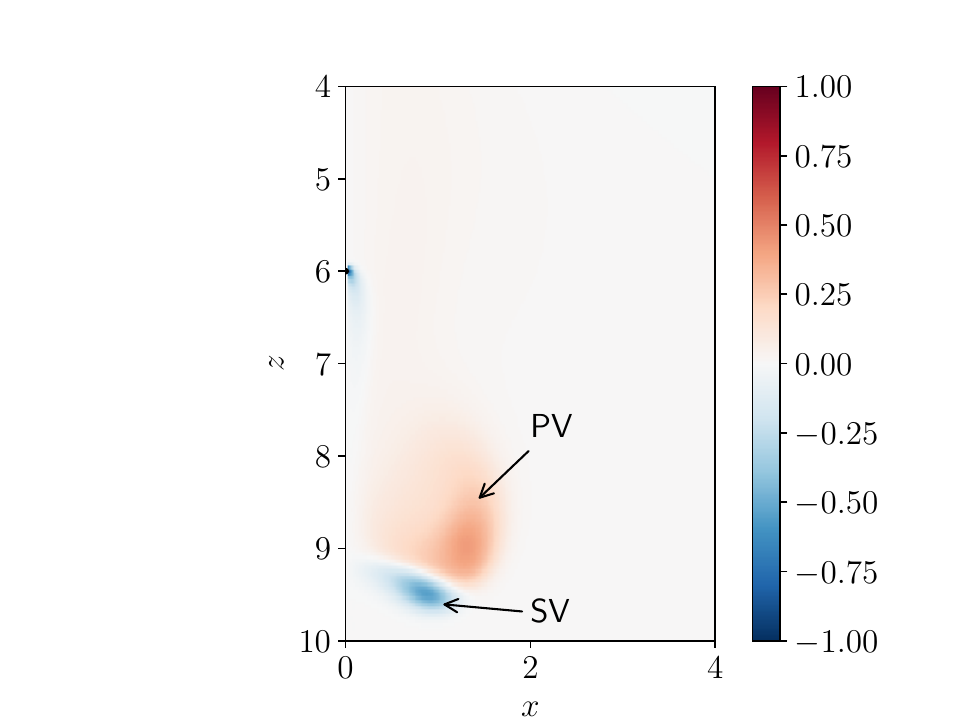}
\\
    \includegraphics[trim={3cm 0 4cm 0},clip,height=0.22\linewidth]{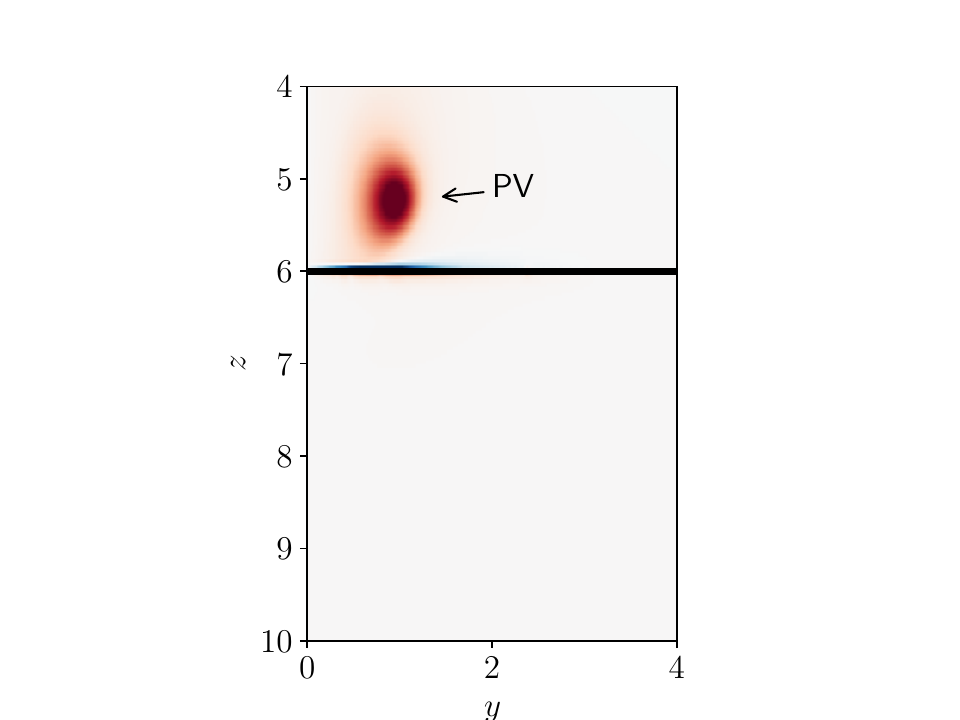}
    \includegraphics[trim={3cm 0 4cm 0},clip,height=0.22\linewidth]{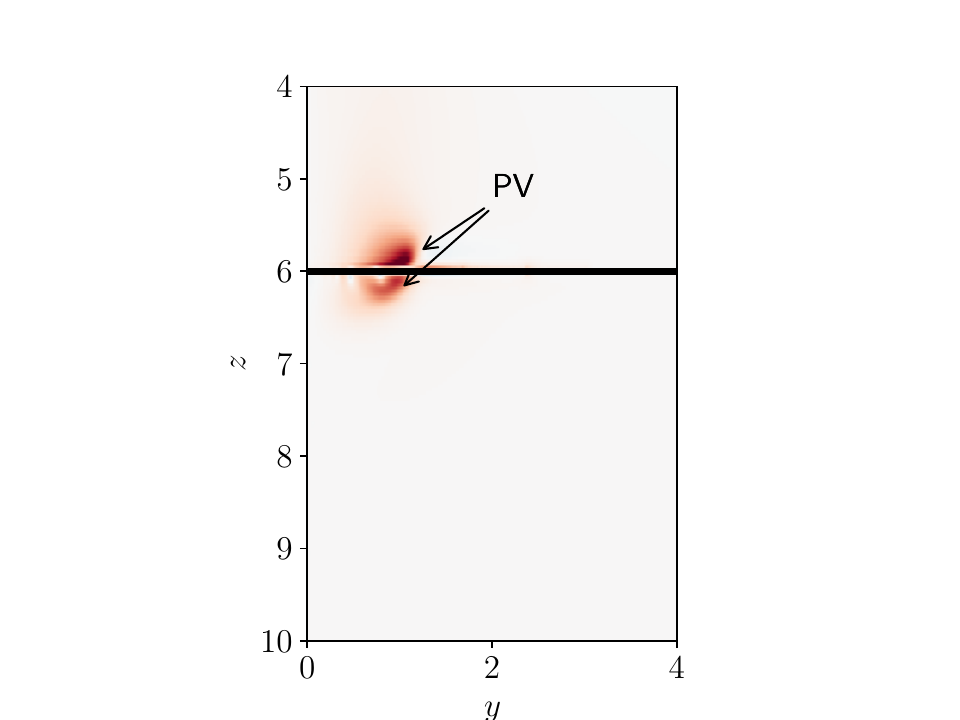}
    \includegraphics[trim={3cm 0 4cm 0},clip,height=0.22\linewidth]{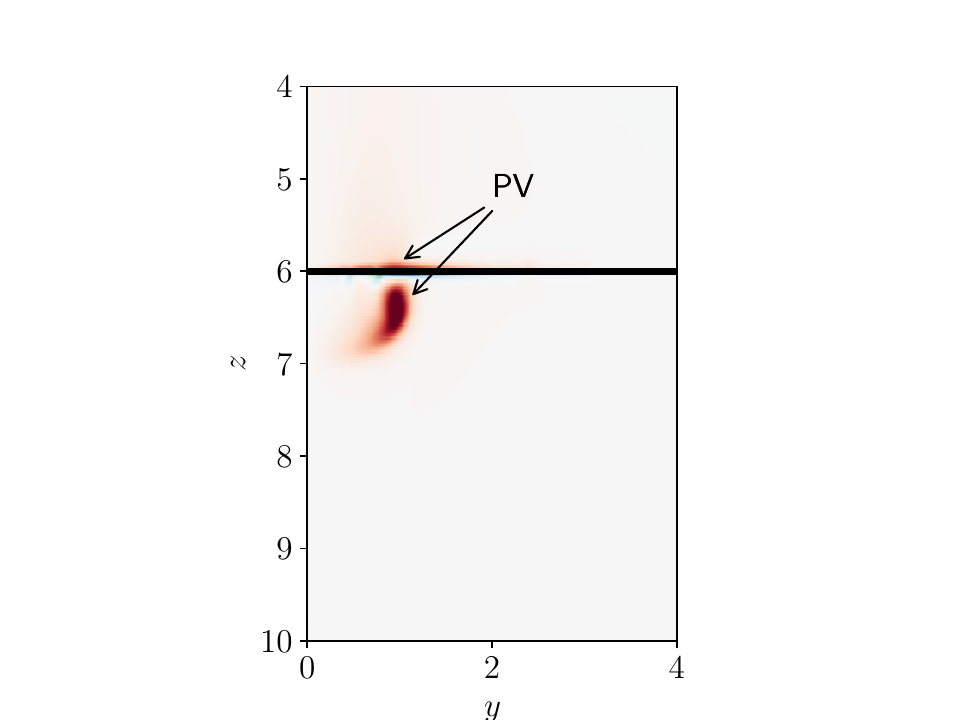}
    \includegraphics[trim={3cm 0 4cm 0},clip,height=0.22\linewidth]{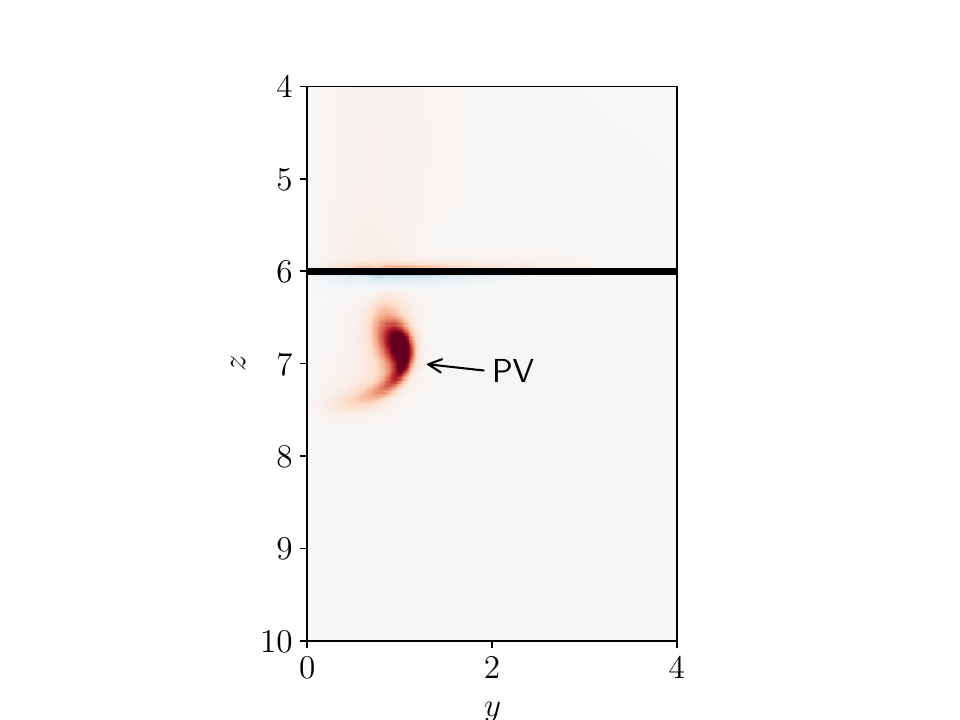}
    \includegraphics[trim={3cm 0 0cm 0},clip,height=0.22\linewidth]{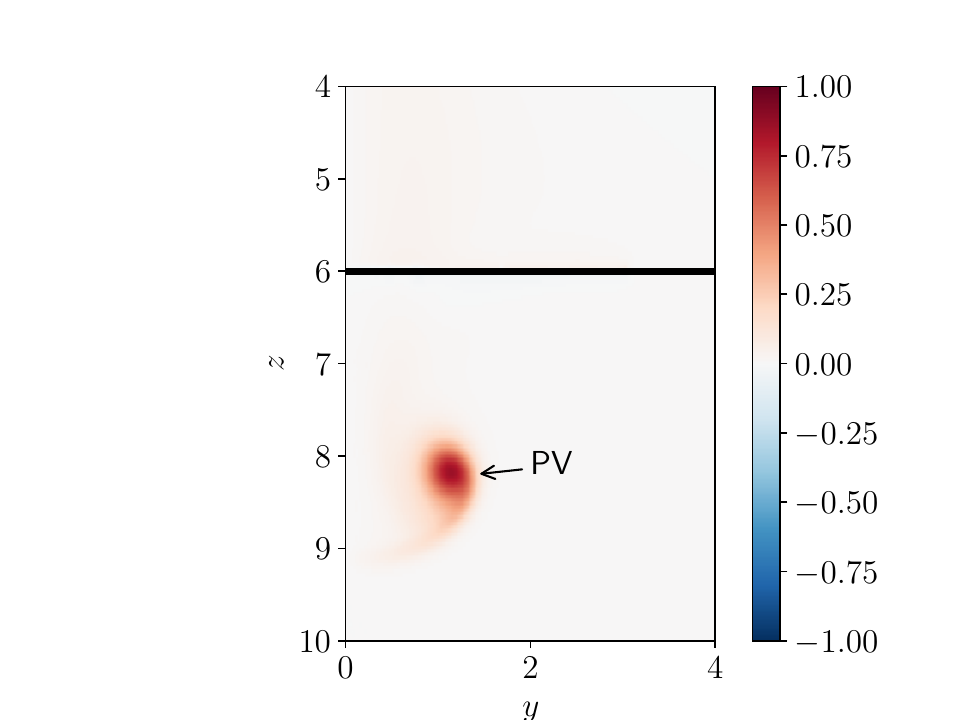}
    \caption{Evolution of the azimuthal vorticity for $T_D=0.025$, $\Lambda=0.2$, $Re_\Gamma=1000$ through the $x$-$z$ plane (top) and $y$-$z$ plane (bottom). From left to right: $t=25$, $32.5$, $37.5$, $42.5$ and $60$. Legend: PV, primary vortex; BL, boundary layer; SV, secondary vorticity. }
    \label{fig:cr0p2-rcyl0p025-wth-re1000}
\end{figure}

Figure~\ref{fig:cr0p2-rcyl0p025-wth-re1000} shows two-dimensional cuts of instantaneous azimuthal vorticity in planes both perpendicular ($x$–$z$) and parallel ($y$–$z$) to the wire. In the $x$–$z$ plane, we can again observe how the primary vortex (PV) forms a boundary layer (BL) around the cylinder that is subsequently lifted and detached as secondary vorticity (SV), slightly weakening the primary ring. This detached layer corresponds to the lobes observed in figure~\ref{fig:cr0p2-rcyls-re1000} (left column) and \ref{fig:cont-cr0p2-rcyl0p025-wth-re1000}, and can be interpreted as a secondary vortex sheet (which later rolls up into a vortex) that travels with the primary structure. In contrast, in the $y$–$z$ plane, the vorticity traverses the wire, disappearing above and reappearing below with minimal morphological changes, although its intensity decreases slightly. This behavior corresponds to the relatively undisturbed regions of the vortex observed in figures~\ref{fig:cr0p2-rcyls-re1000} and \ref{fig:cont-cr0p2-rcyl0p025-wth-re1000}.

\begin{figure}
    \centering
    \includegraphics[trim={3cm 0 4cm 0},clip,height=0.22\linewidth]{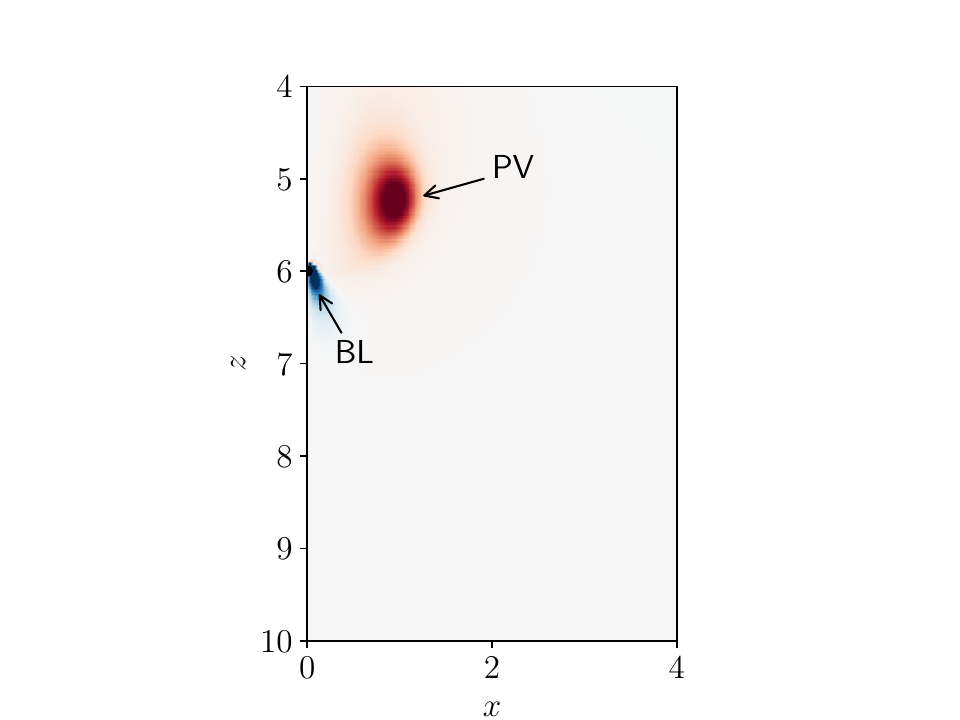}
    \includegraphics[trim={3cm 0 4cm 0},clip,height=0.22\linewidth]{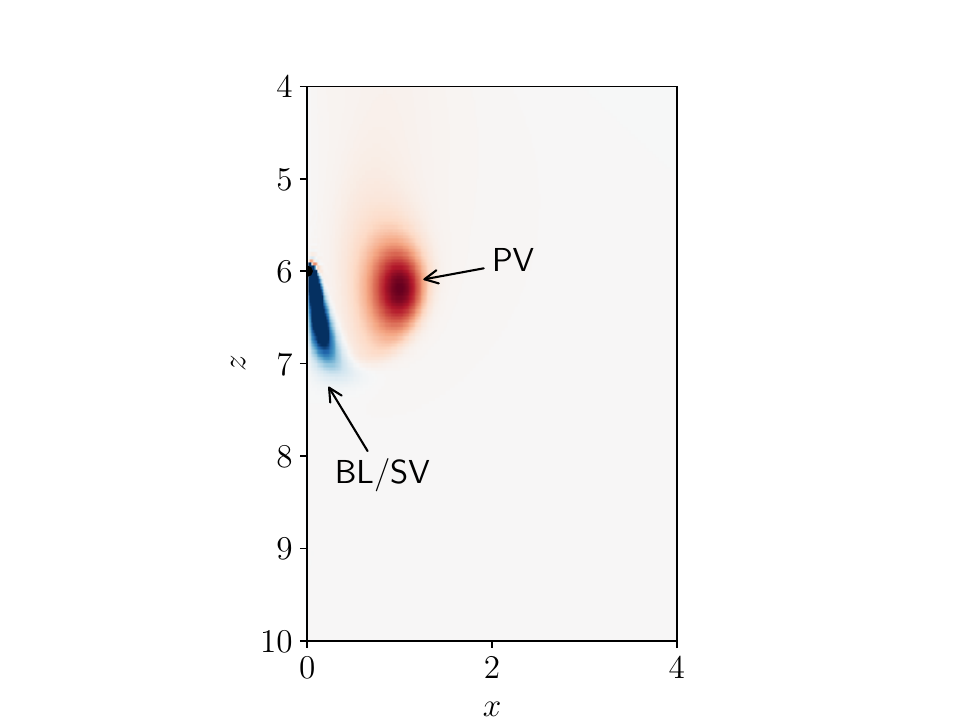}
    \includegraphics[trim={3cm 0 4cm 0},clip,height=0.22\linewidth]{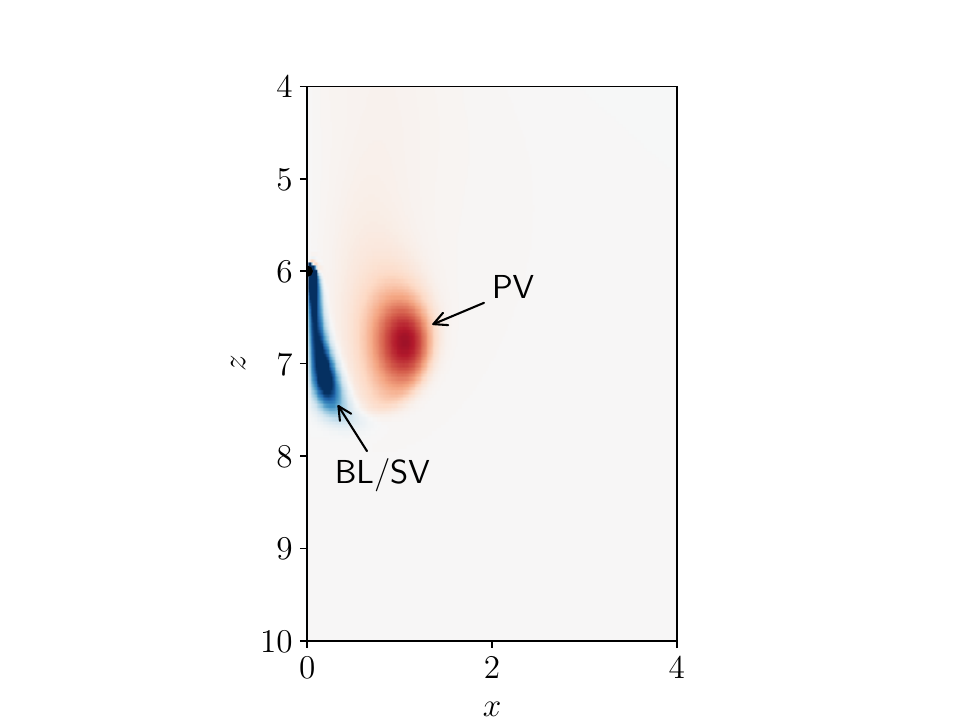}
    \includegraphics[trim={3cm 0 4cm 0},clip,height=0.22\linewidth]{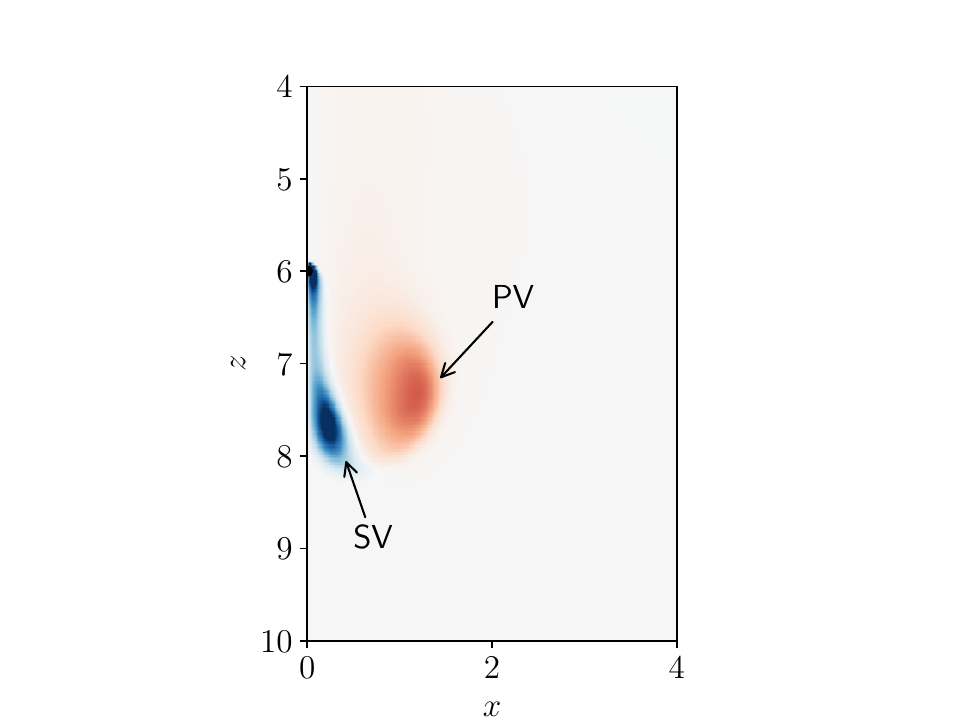}
    \includegraphics[trim={3cm 0 0cm 0},clip,height=0.22\linewidth]{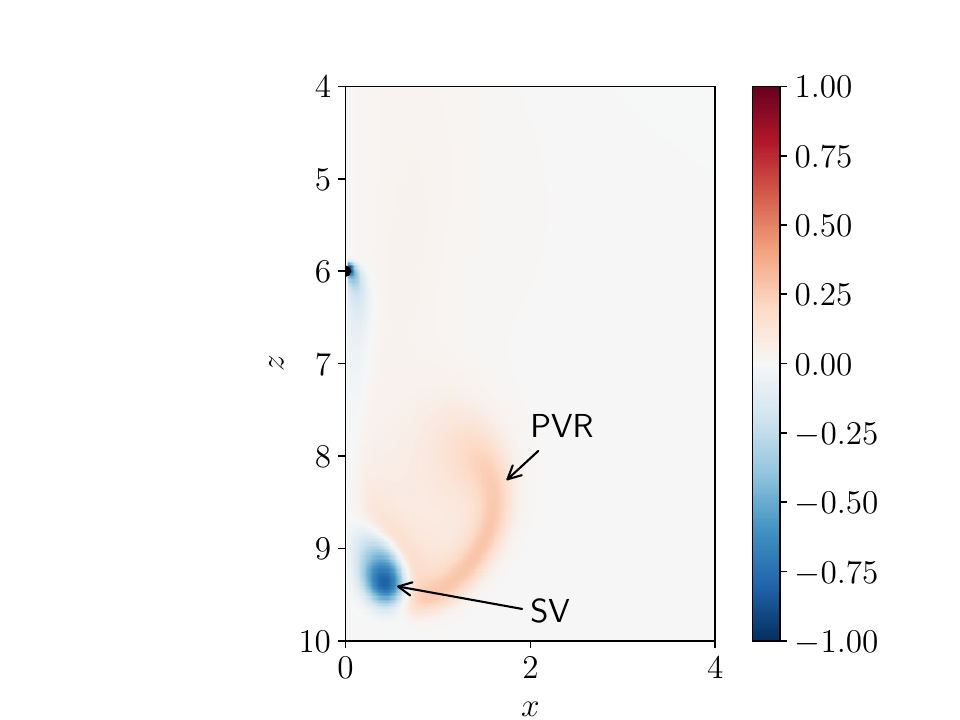}
\\
    \includegraphics[trim={3cm 0 4cm 0},clip,height=0.22\linewidth]{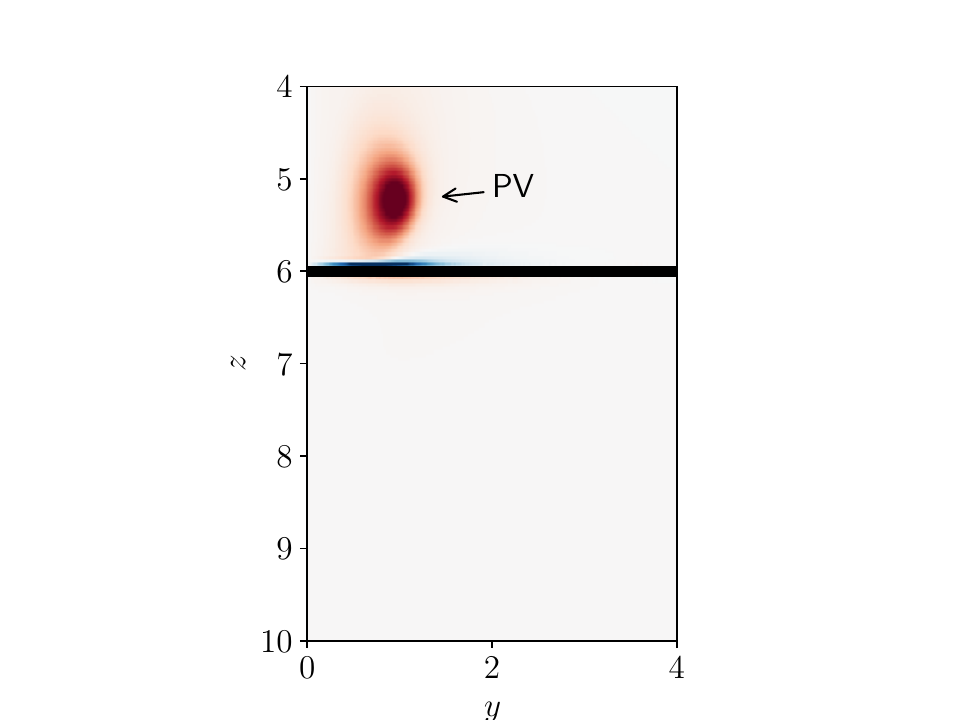}
    \includegraphics[trim={3cm 0 4cm 0},clip,height=0.22\linewidth]{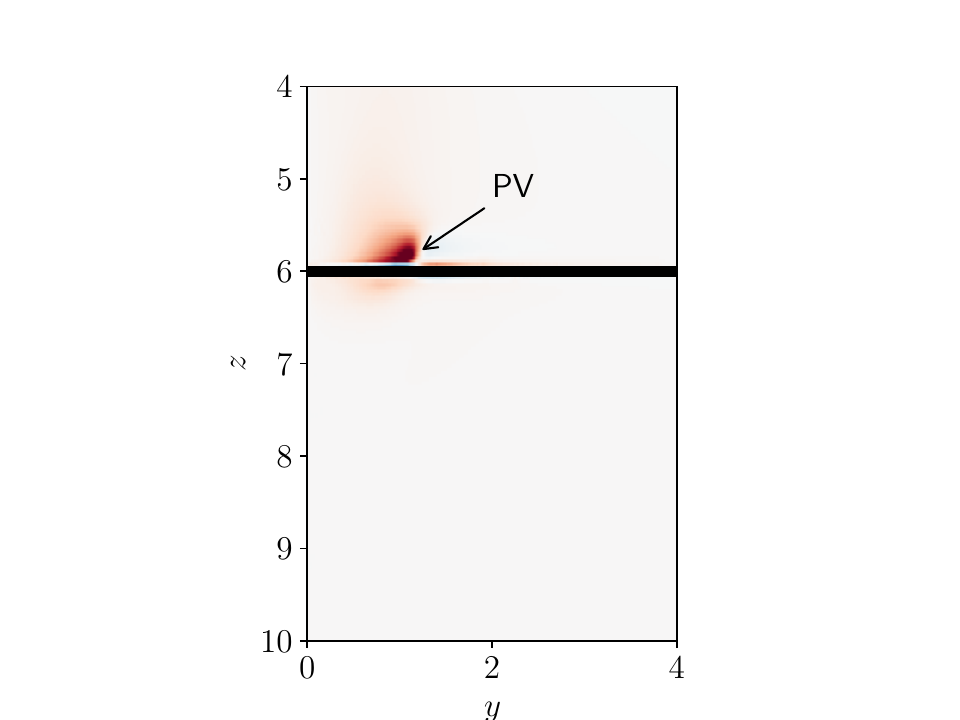}
    \includegraphics[trim={3cm 0 4cm 0},clip,height=0.22\linewidth]{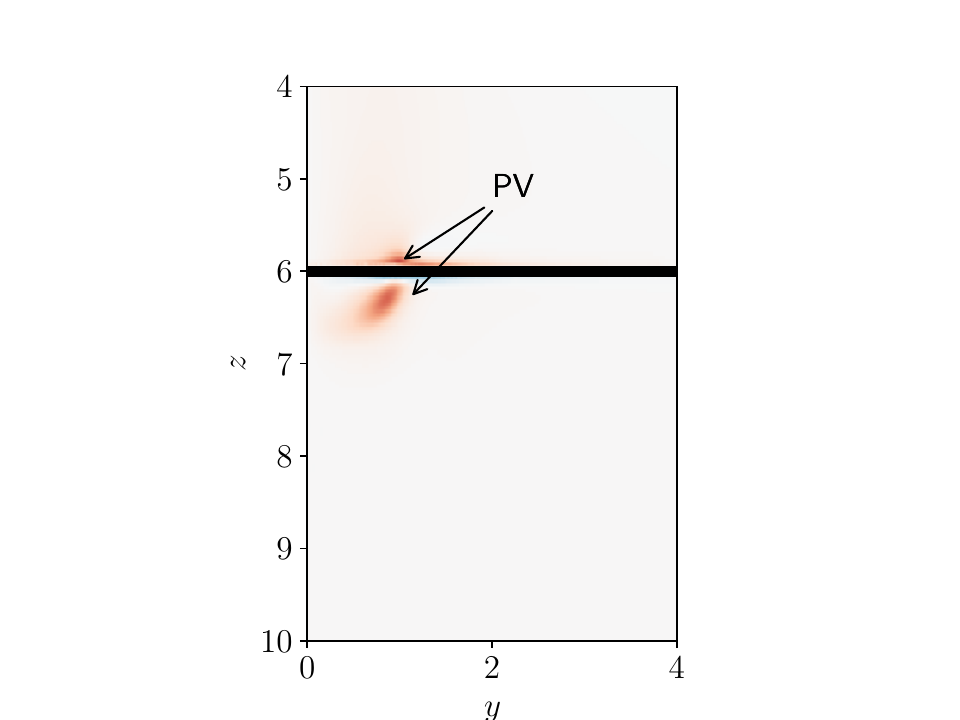}
    \includegraphics[trim={3cm 0 4cm 0},clip,height=0.22\linewidth]{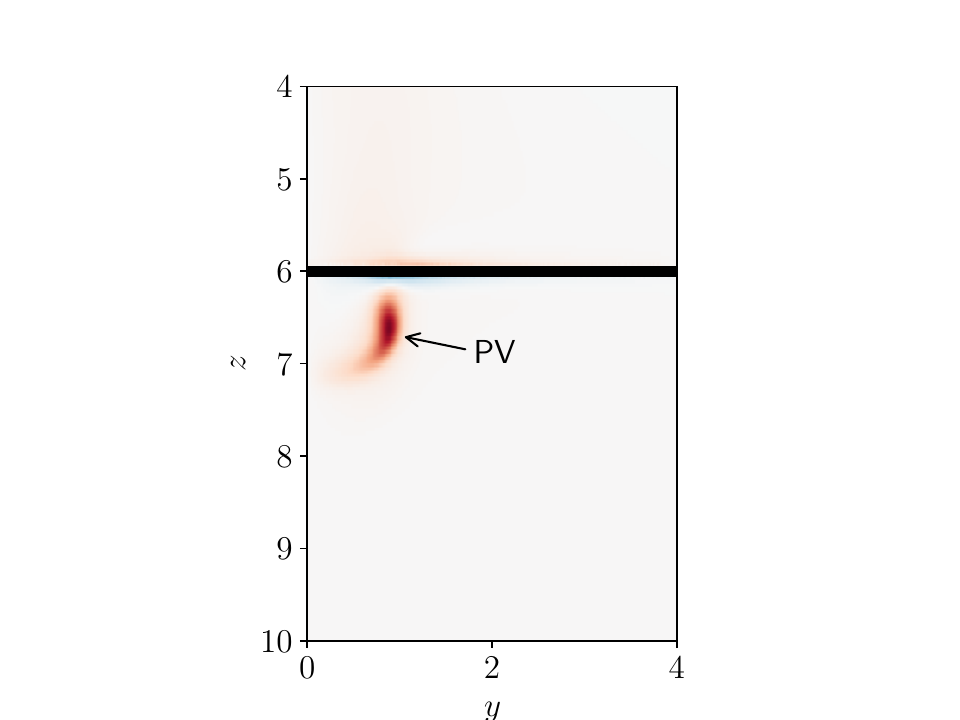}
    \includegraphics[trim={3cm 0 0cm 0},clip,height=0.22\linewidth]{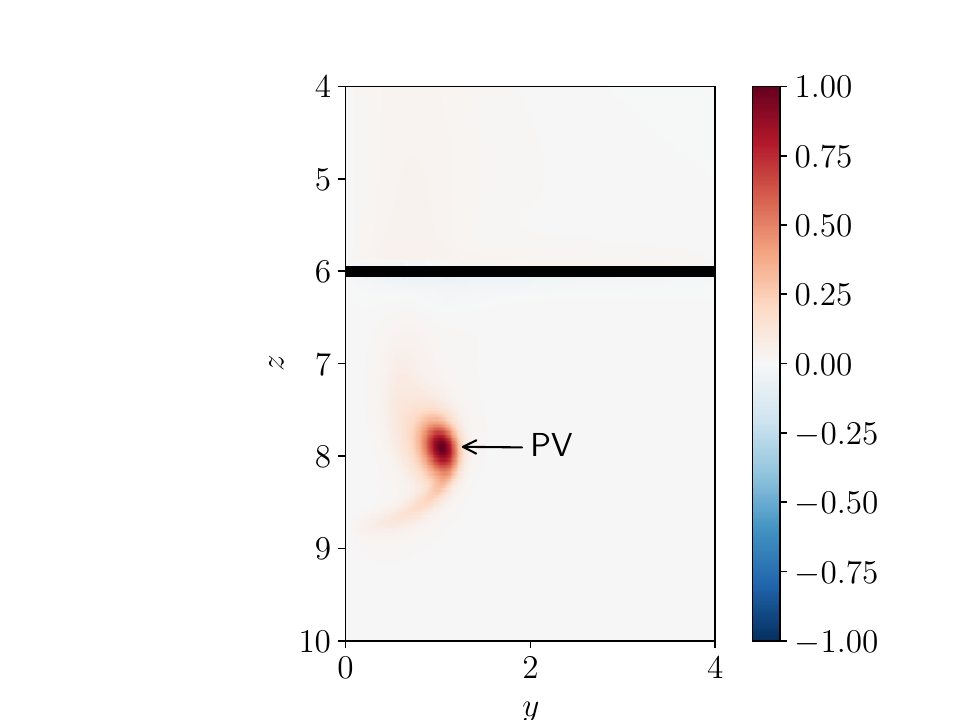}
    \caption{Evolution of the azimuthal vorticity for $T_D=0.05$, $\Lambda=0.2$, $Re_\Gamma=1000$ through the $x$-$z$ plane (top) and $y$-$z$ plane (bottom). From left to right: $t=25$, $32.5$, $37.5$, $42.5$ and $60$. Legend: PV(R), primary vortex (remnants); BL, boundary layer; SV, secondary vorticity.}
    \label{fig:cr0p2-rcyl0p05-wth-re1000}
\end{figure}

Increasing $T_D$ to $0.05$ (figure~\ref{fig:cr0p2-rcyl0p05-wth-re1000}) produces similar qualitative features. In the $x$–$z$ plane, the primary vortex again generates a boundary layer on the wire, which it drags and detaches. The primary ring, however, is more deformed than in the thinner-wire case, adopting a sheet-like structure and spreading out over a larger region. In the $y$–$z$ plane, the primary vortex continues to traverse the wire, vanishing on one side and reappearing on the other, much like the $T_D=0.025$ case. In this plane, the primary vorticity remains a relatively round and compact patch.

\begin{figure}
    \centering
    \includegraphics[width=0.80\linewidth]{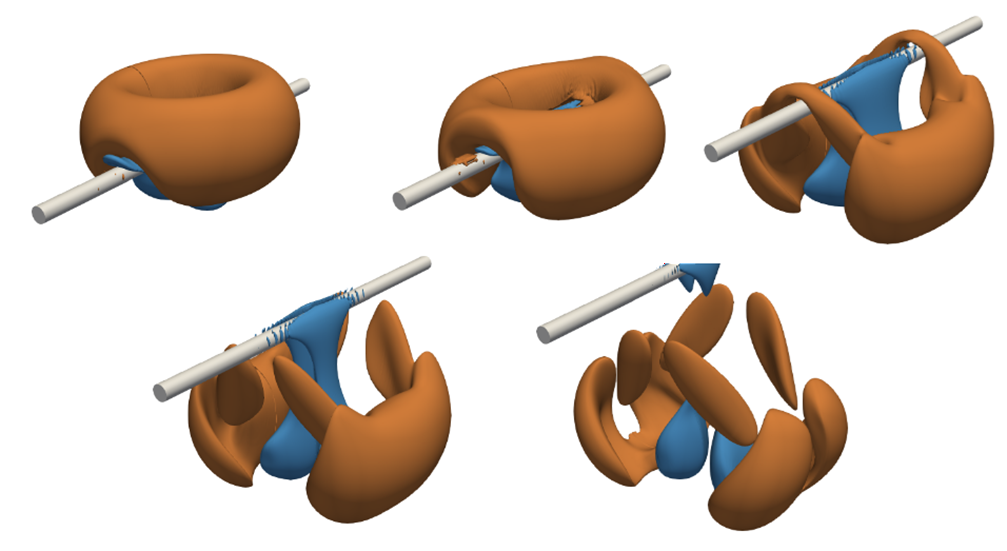}
    \caption{Evolution of the azimuthal vorticity for $T_D=0.1$, $\Lambda=0.2$, $Re_\Gamma=1000$. Isocontours at $\omega_\theta=0.2$ (orangish-red) and $\omega_\theta=-0.15$ (blue).  From top left to bottom right: $t=30$, $35$, $40$, $45$ and $55$. }
    \label{fig:cont-cr0p2-rcyl0p1-wth-re1000}
\end{figure}

\begin{figure}
    \centering
    \includegraphics[trim={3cm 0 4cm 0},clip,height=0.22\linewidth]{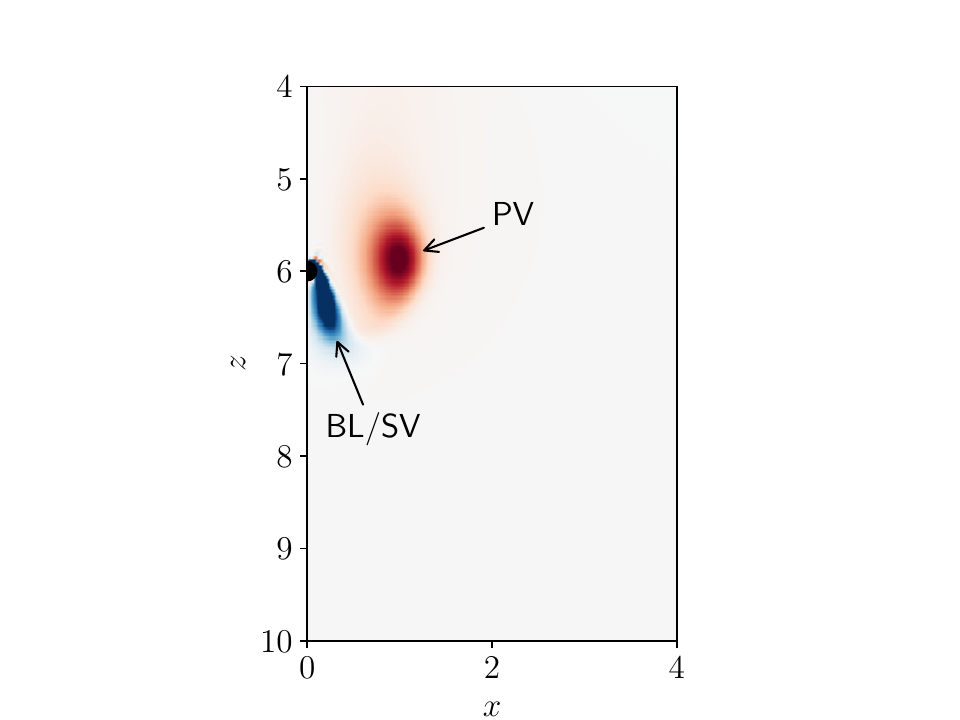}
    \includegraphics[trim={3cm 0 4cm 0},clip,height=0.22\linewidth]{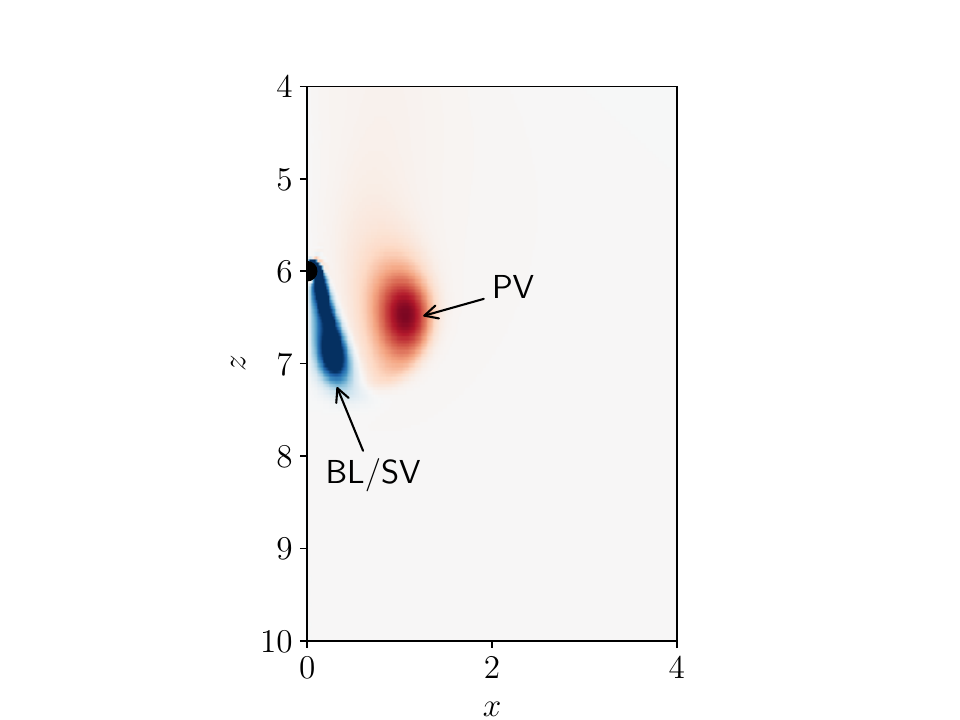}
    \includegraphics[trim={3cm 0 4cm 0},clip,height=0.22\linewidth]{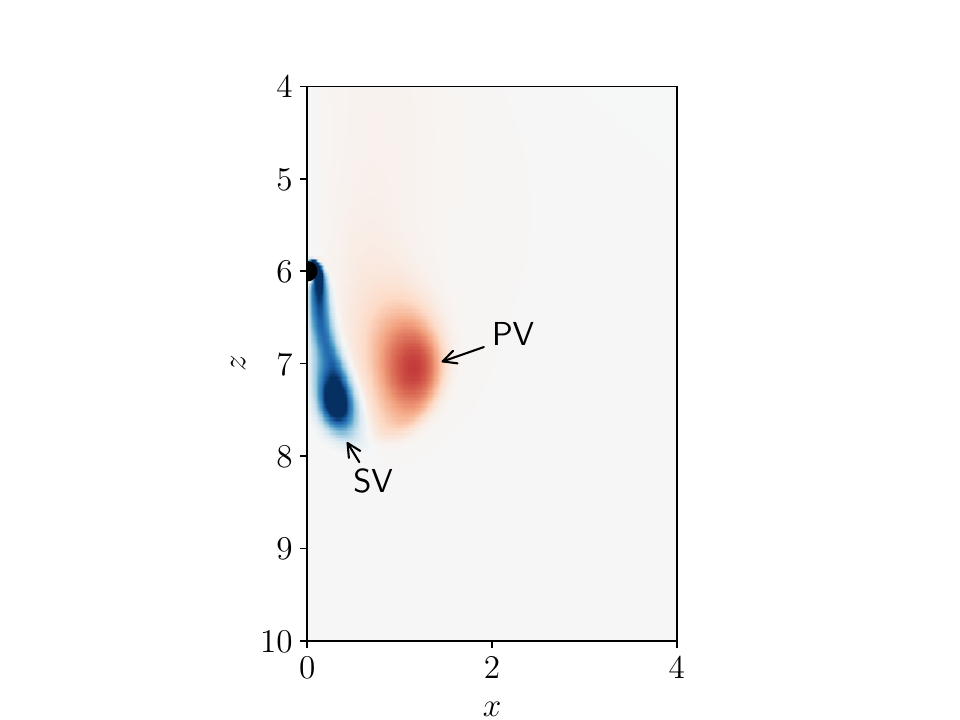}
    \includegraphics[trim={3cm 0 4cm 0},clip,height=0.22\linewidth]{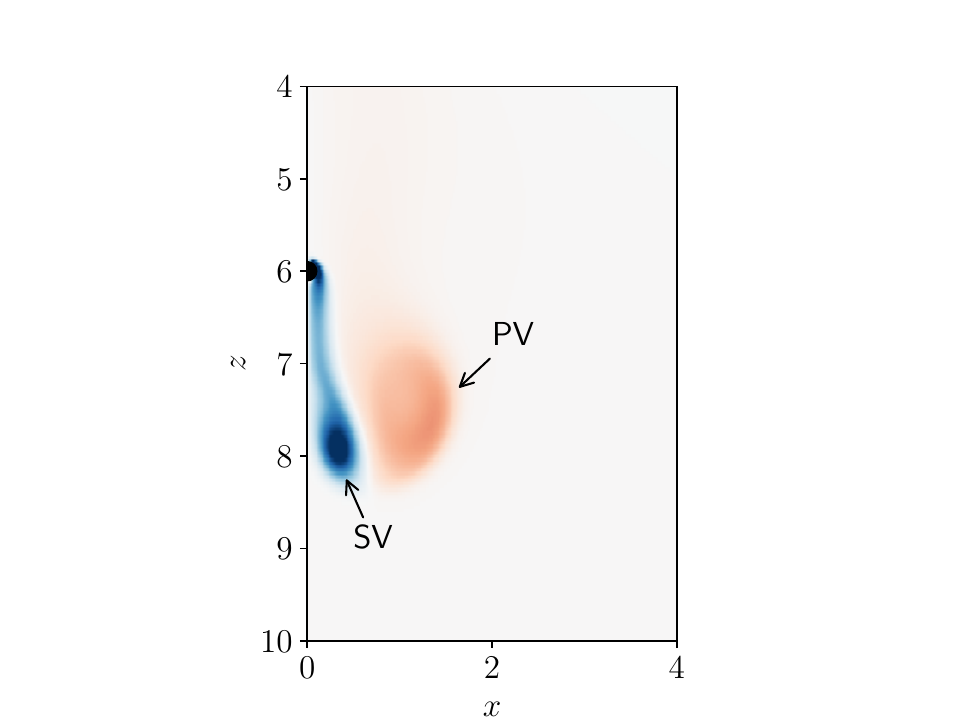}
    \includegraphics[trim={3cm 0 0cm 0},clip,height=0.22\linewidth]{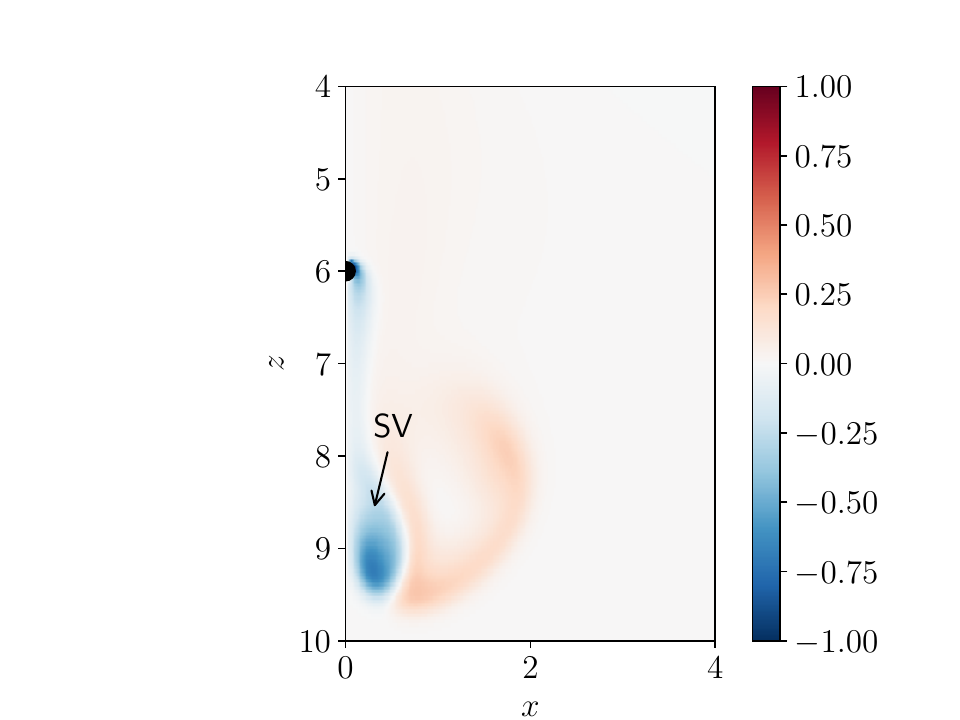}
\\
    \includegraphics[trim={3cm 0 4cm 0},clip,height=0.22\linewidth]{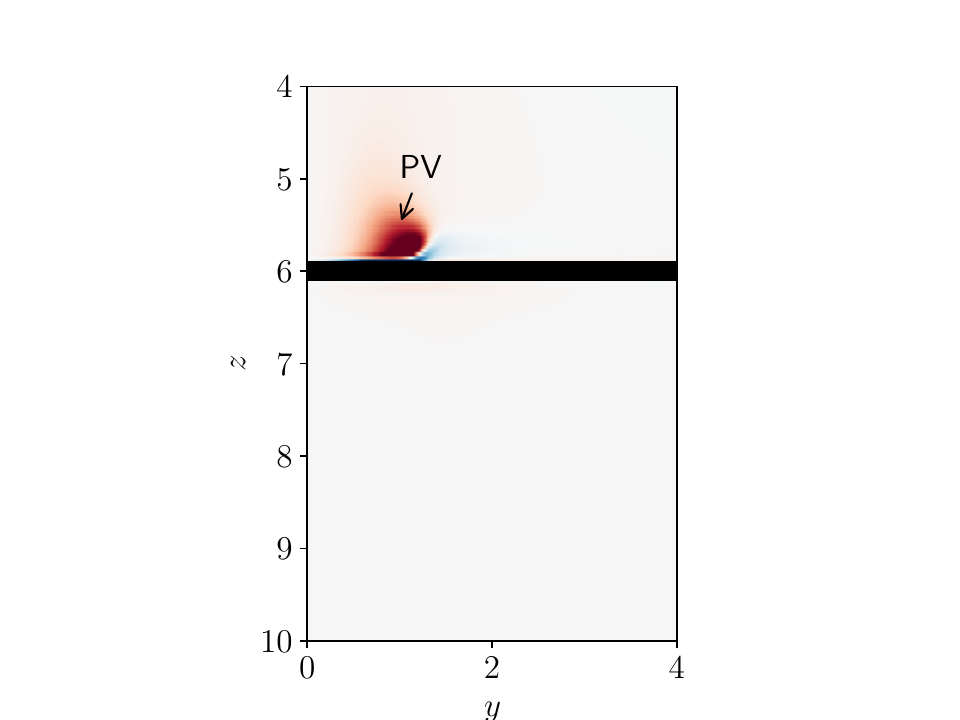}
    \includegraphics[trim={3cm 0 4cm 0},clip,height=0.22\linewidth]{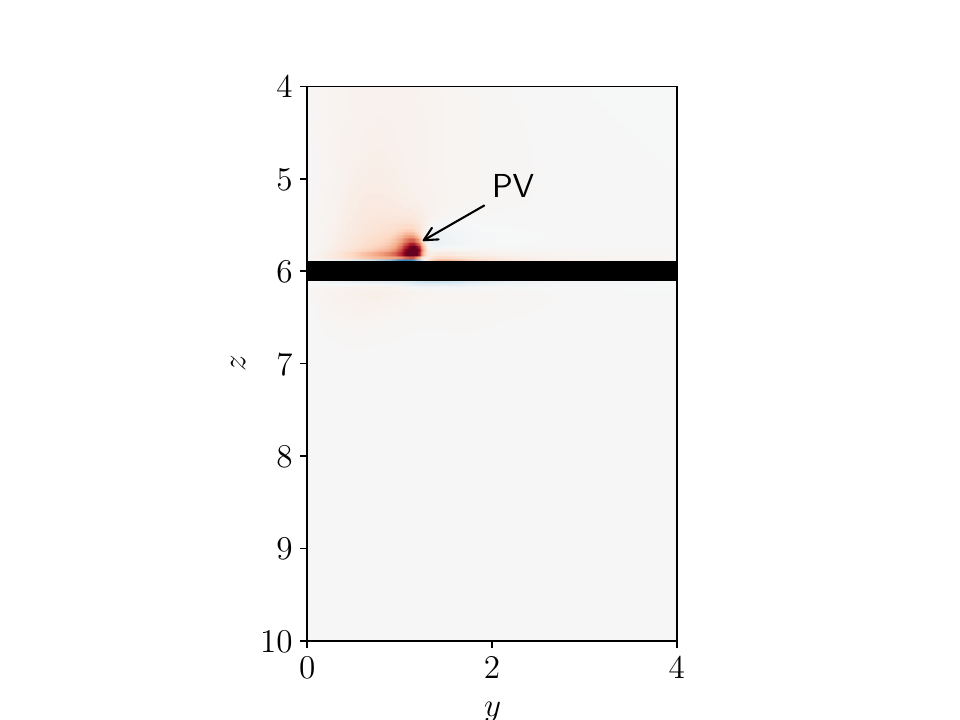}
    \includegraphics[trim={3cm 0 4cm 0},clip,height=0.22\linewidth]{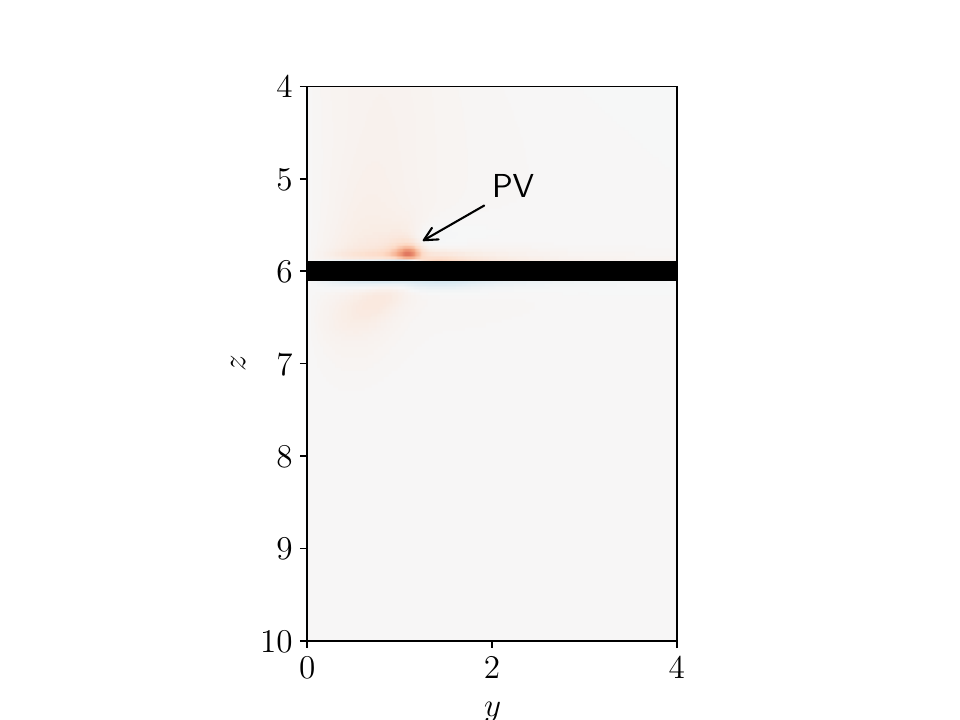}
    \includegraphics[trim={3cm 0 4cm 0},clip,height=0.22\linewidth]{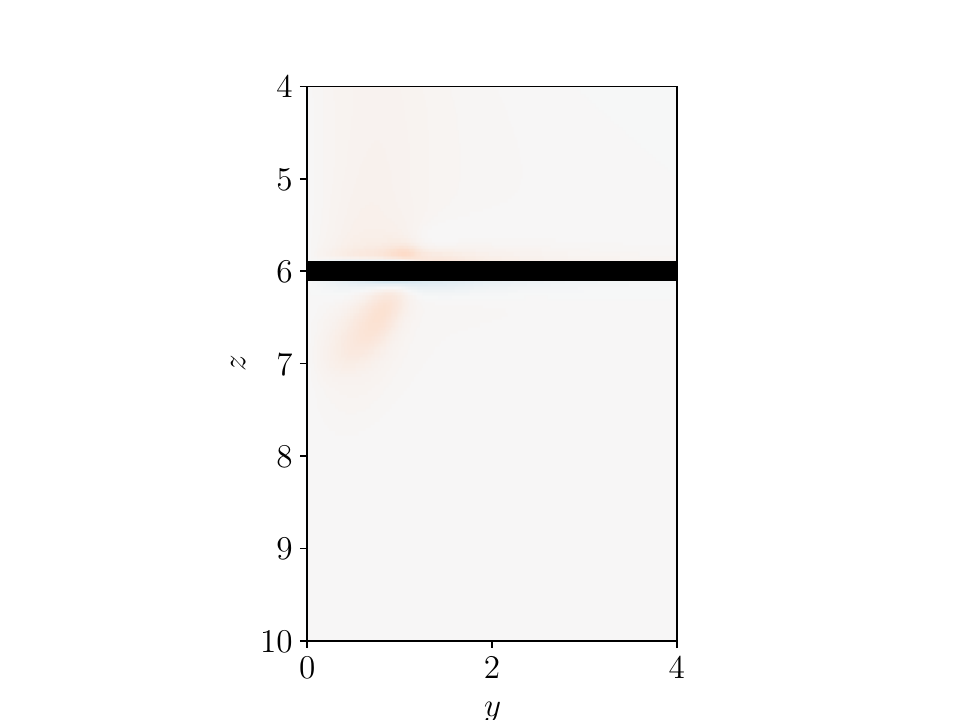}
    \includegraphics[trim={3cm 0 0cm 0},clip,height=0.22\linewidth]{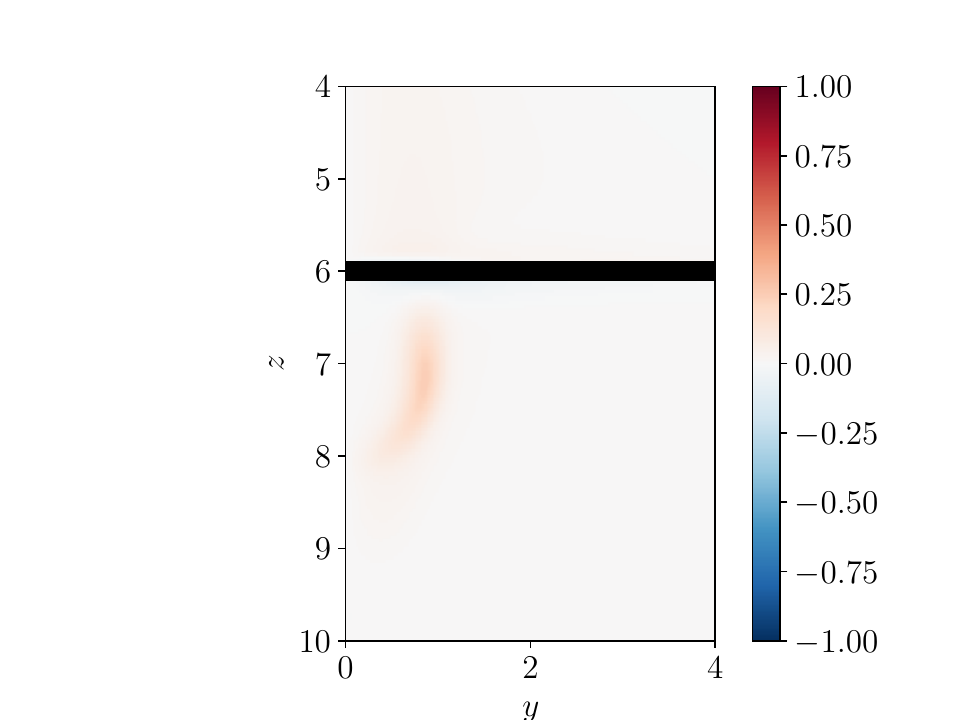}
    \caption{Evolution of the azimuthal vorticity for $T_D=0.1$, $\Lambda=0.2$, $Re_\Gamma=1000$,  through the $x$-$z$ plane (top) and $y$-$z$ plane (bottom). From left to right: $t=30$, $35$, $40$, $45$ and $55$. Legend: PV, primary vortex; BL, boundary layer; SV, secondary vorticity.}
    \label{fig:cr0p2-rcyl0p1-wth-re1000}
\end{figure}

Further increasing the wire diameter to $T_D=0.1$ (Figs.~\ref{fig:cont-cr0p2-rcyl0p1-wth-re1000} and \ref{fig:cr0p2-rcyl0p1-wth-re1000}) leads to a qualitative change in behavior. In the three-dimensional visualizations, we can observe how the primary vorticity is no longer able to transverse the wire, and is instead stretched, then cut by the object. This can be observed in more detail in the two-dimensional cuts: in the $x$–$z$ plane, a secondary vorticity layer forms and detaches alongside the primary vortex as earlier, but the $y$–$z$ plane reveals that the primary vortex can no longer traverse the wire and instead dissipates. This confirms the behavior seen in figure~\ref{fig:crs-rcyls-re1000}: the vortex ring is effectively cut by the obstacle. The two rings ejected from the wire correspond to the vortex dipoles visible in the $x$–$z$ section on both sides of the symmetry plane at $x=0$. These observations indicate that the transition between the wire and cutting regimes is governed by the inability of the primary ring to reform after passing around the object.

\begin{figure}
    \centering
    \includegraphics[width=0.4\linewidth]{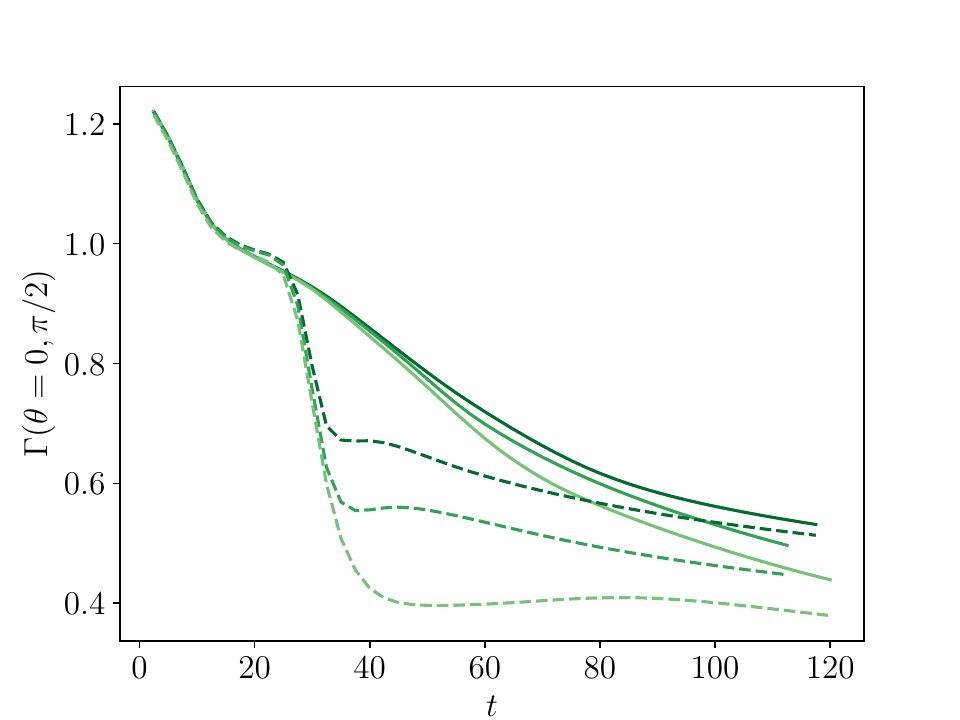}
    \includegraphics[width=0.4\linewidth]{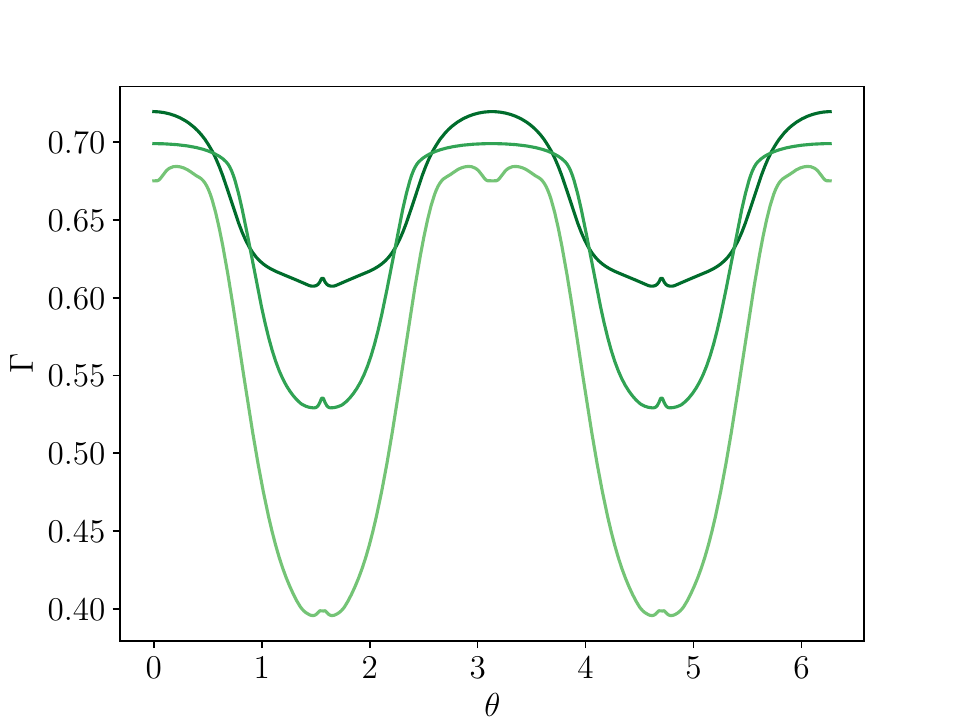}
    \caption{Left: Temporal behaviour of the circulation in the $x$-$z$ (solid) and $y$-$z$ (dashed) planes for $Re_\Gamma=1000$ and $\Lambda=0.2$ for $T_D=0.025$, $0.05$ and $0.1$ (dark to light green). Right: Azimuthal behavior of the circulation at $t=60$, just after the cut for the same cases.  }
    \label{fig:gamma-dep-wire}
\end{figure}

To quantify these effects, we compute the circulation in the $x$–$z$ and $y$–$z$ planes by integrating the positively signed azimuthal vorticity. The left panel of figure~\ref{fig:gamma-dep-wire} shows that, after the startup phase ($t < 10$), the vortex interacts with the wire around $t \approx 25$. The circulation in the $x$–$z$ plane decreases gradually, whereas in the $y$–$z$ plane it drops more sharply but remains non-zero. Interestingly, these smooth temporal variations do not reflect the abrupt qualitative changes observed in the flow visualizations.

The right panel of figure~\ref{fig:gamma-dep-wire} shows the azimuthal distribution of circulation, revealing that the largest values occur in planes perpendicular to the wire and the smallest in parallel planes. Although this trend is consistent with the integrated circulation, it contrasts with the distinctness of the resulting structures. The vortex is more coherent at lower $T_D$, yet the differences of the $\Gamma(\theta)$ profiles between $T_D=0.025$ and $T_D=0.1$ are very small. The reason is that the secondary detached vorticity deforms the primary vortex without significantly altering its net strength. Thus, circulation alone does not fully capture the underlying flow phenomenology.

The present results can be directly compared with the experiments of \citet{naitoh1995vortex,adhikari2009msc}. Our simulations at $Re_\Gamma=1000$ correspond approximately to \cite{naitoh1995vortex}'s reported $Re=680$ once the $1.3$–$1.4\times$ conversion between piston-based and circulation-based Reynolds numbers is applied \citep{mckeown2020turbulence}. Unlike those experiments, however, the simulations here directly visualize vorticity rather than dye, allowing a more detailed interpretation of the underlying dynamics. From this perspective, the transition described by \citet{naitoh1995vortex} between ``laminar'' behavior for thin wires (small $\Pi_3\approx T_\sigma/2$) and ``turbulent'' behavior for thicker wires (large $\Pi_3$) in the range $\Pi_3\in[0.25,0.38]$ corresponds to the regime transition observed here between $T_D=0.05$ ($\Pi_3\approx0.25$) and $T_D=0.1$ ($\Pi_3\approx0.5$). Rather than the onset of turbulence, the simulations show that secondary structures can now detach from the primary vortex, significantly distorting its geometry, and this corresponds to the change of behaviour observed. Furthermore, as discussed in \S\ref{sec:res1}, this transition is controlled primarily by $T_D$ rather than $T_\sigma$ (or $\Pi_3$) as reported in the manuscript.

When compared to \cite{adhikari2009msc}, more similarities arise: what this author denotes as the ``cut-and-reconnect'' regime can be understood here as a regime where the vorticity is able to transverse the wire on some planes. However, we do not observe the vorticity to be fully ``cut'' until $T_D=0.1$ where the ring is no longer able to reform on the other side of the interaction. Therefore, we choose the name ``wire''-regime for the interactions at low $T_D$ where the ring is never effectively cut. 

In addition, the dye visualizations in \cite{naitoh1995vortex} showed an approximately elliptical ring after the interaction, but the experiments in \cite{adhikari2009msc} show a highly deformed vorticity field. The present results fields reveal an intermediate picture: the ring does not deform into a simple ellipse but instead carries a lobe of secondary vorticity generated by its interaction with the wire. At higher Reynolds numbers, as shown in figure~\ref{fig:crs-rcyls-re2000} and in the following section, these lobes can detach and be ejected as secondary rings, while the primary ring remains largely intact.

\subsection{Results for $Re_\Gamma=2000$}

\begin{figure}
    \centering
    \includegraphics[width=0.80\linewidth]{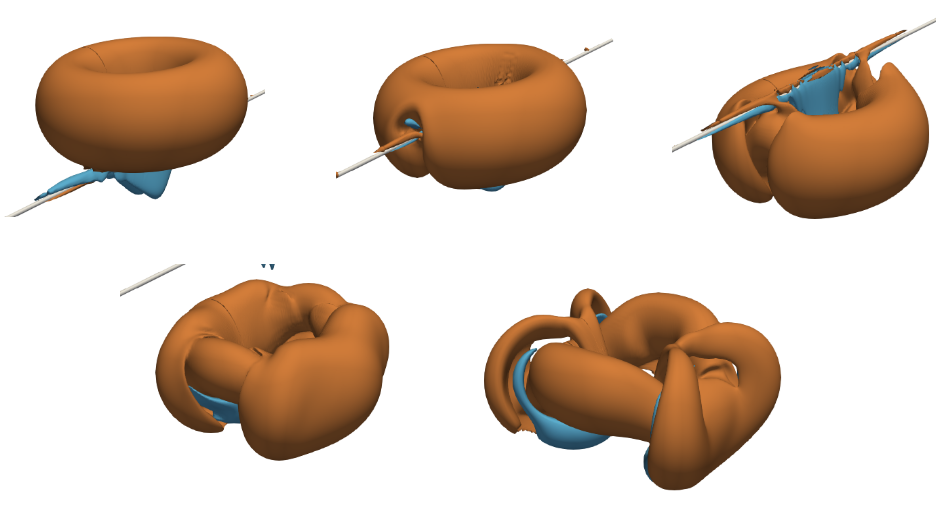}
    \caption{Evolution of the azimuthal vorticity for $T_D=0.025$, $\Lambda=0.2$, $Re_\Gamma=2000$. Isocontours at $\omega_\theta=0.2$ (orangish-red) and $\omega_\theta=-0.15$ (blue).  From top left to bottom right: $t=25$, $30$, $35$, $45$ and $60$. }
    \label{fig:cont-cr0p2-rcyl0p025-wth-re2000}
\end{figure}

\begin{figure}
    \centering
    \includegraphics[trim={3.5cm 0 4.5cm 0},clip,height=0.25\linewidth]{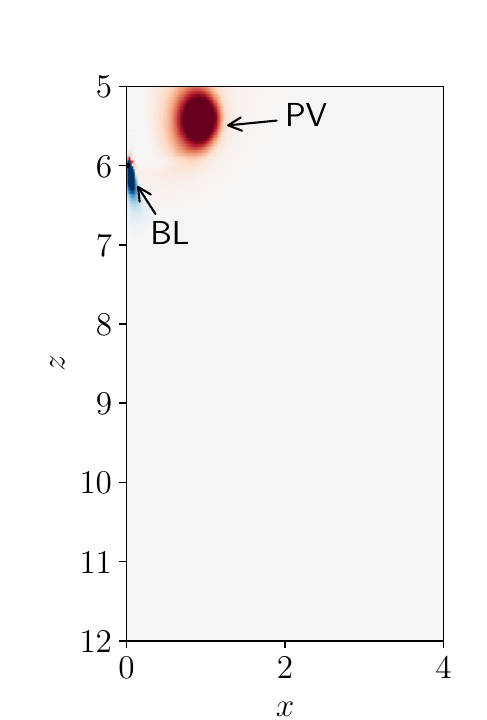}
    \includegraphics[trim={3.5cm 0 4.5cm 0},clip,height=0.25\linewidth]{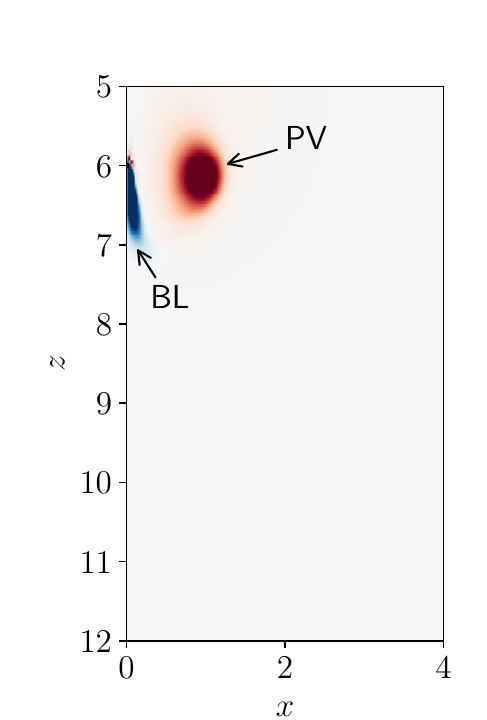}
    \includegraphics[trim={3.5cm 0 4.5cm 0},clip,height=0.25\linewidth]{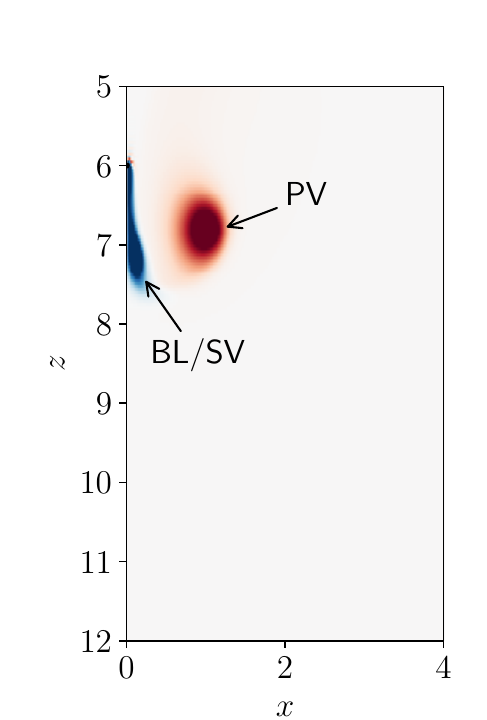}
    \includegraphics[trim={3.5cm 0 4.5cm 0},clip,height=0.25\linewidth]{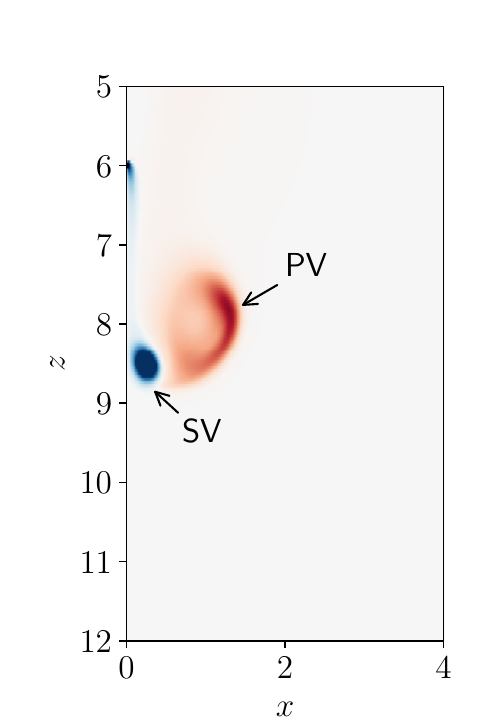}
    \includegraphics[trim={3.5cm 0 1cm 0},clip,height=0.25\linewidth]{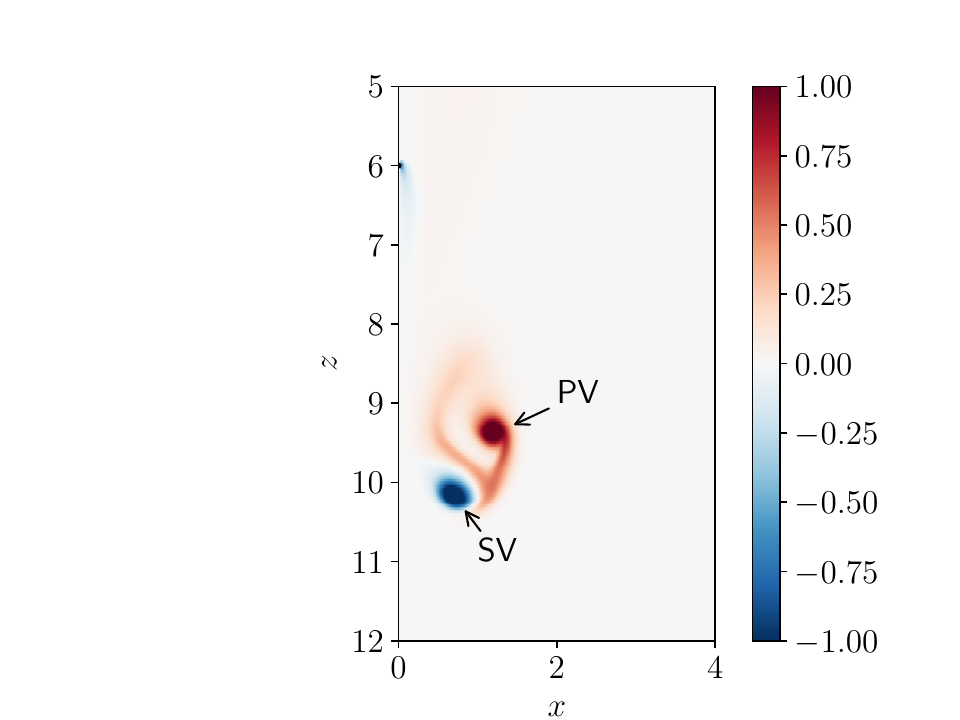}
\\
    \includegraphics[trim={3.5cm 0 4.5cm 0},clip,height=0.25\linewidth]{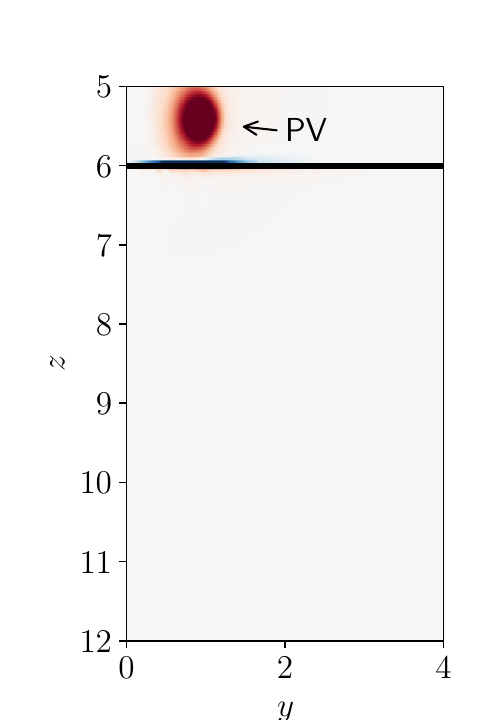}
    \includegraphics[trim={3.5cm 0 4.5cm 0},clip,height=0.25\linewidth]{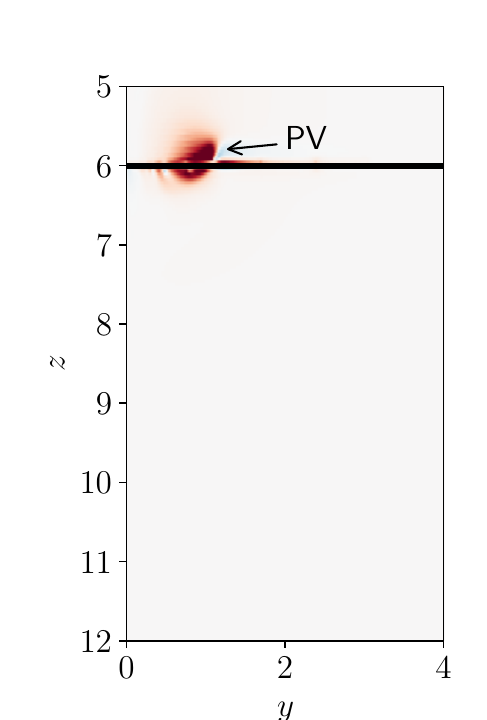}
    \includegraphics[trim={3.5cm 0 4.5cm 0},clip,height=0.25\linewidth]{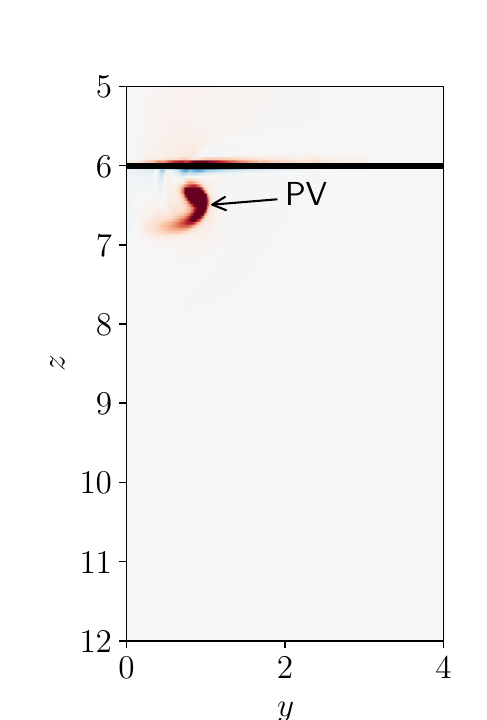}
    \includegraphics[trim={3.5cm 0 4.5cm 0},clip,height=0.25\linewidth]{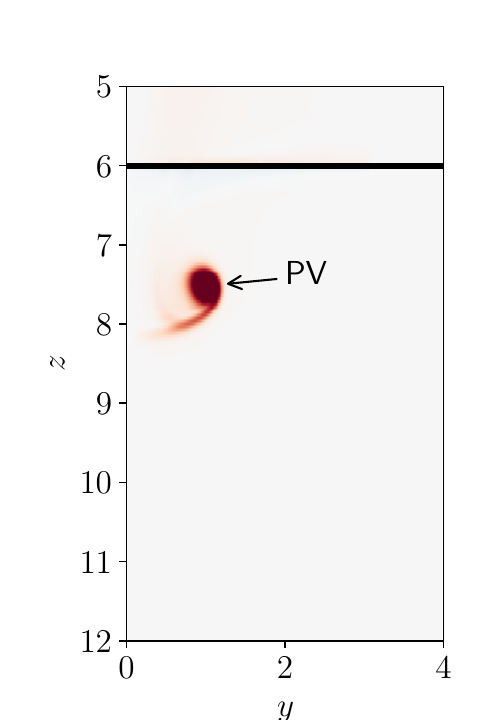}
    \includegraphics[trim={3.5cm 0 1cm 0},clip,height=0.25\linewidth]{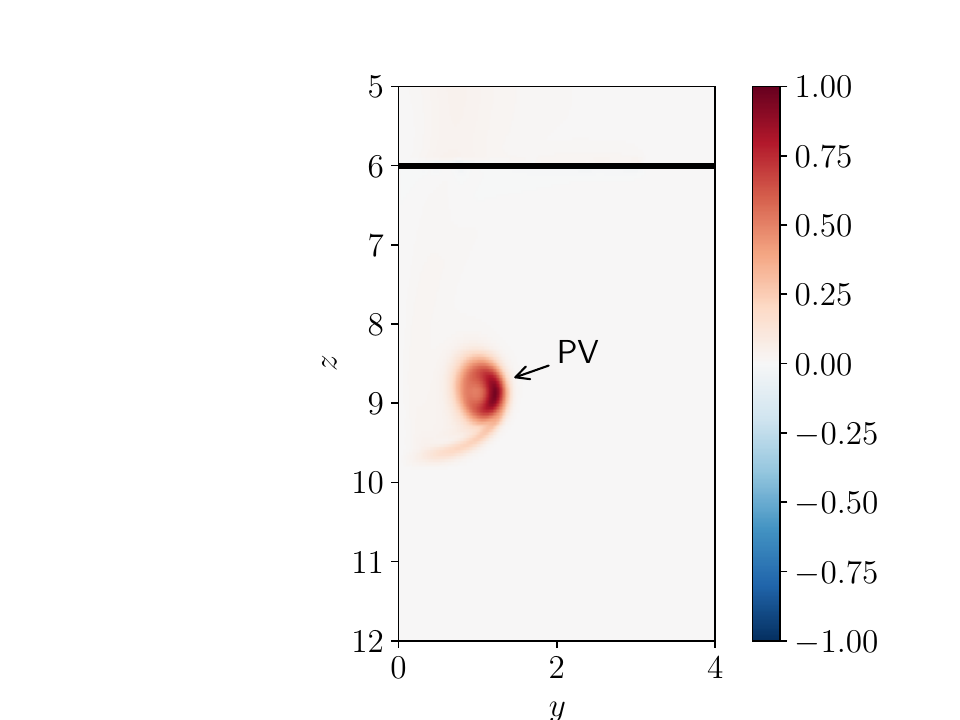}
    \caption{Evolution of the azimuthal vorticity for $T_D=0.025$, $\Lambda=0.2$, $Re_\Gamma=2000$, through the $x$-$z$ plane (top) and $y$-$z$ plane (bottom). From left to right: $t=25$, $35$, $45$, $60$ and $80$. Legend: PV, primary vortex; BL, boundary layer; SV, secondary vorticity.}
    \label{fig:cr0p2-rcyl0p025-wth-re2000}
\end{figure}

To examine the effect of Reynolds number, figures \ref{fig:cont-cr0p2-rcyl0p025-wth-re2000} and \ref{fig:cr0p2-rcyl0p025-wth-re2000} show iso-contours and two-dimensional cuts of $\omega_\theta$ for $T_D=0.025$ at $Re_\Gamma=2000$. The overall phenomenology closely resembles that observed at lower $Re_\Gamma$: in the $x$–$z$ plane, the primary vortex drags a secondary vorticity layer that travels with it, while in the $y$–$z$ plane, the primary vortex reforms after traversing the wire. The vortex is deformed but not cut. At later times, the secondary vorticity is drawn from the wire and interacts with the primary vortex, producing stronger deformation. Eventually, this lobe is slowly dissipated, but it may be ejected at higher $Re_\Gamma$ as described in \cite{adhikari2009msc}. 

\begin{figure}
    \centering
    \includegraphics[width=0.80\linewidth]{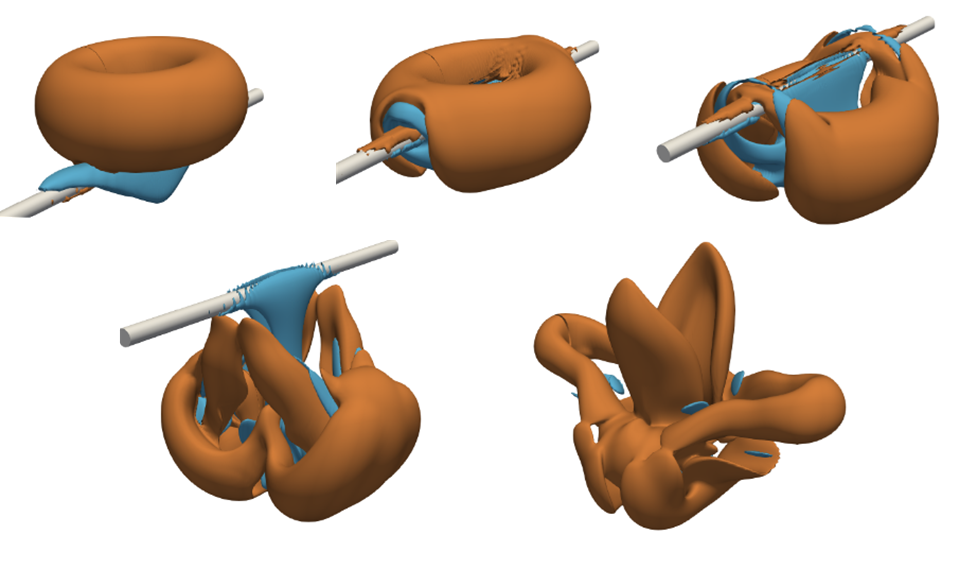}
    \caption{Evolution of the azimuthal vorticity for $T_D=0.1$, $\Lambda=0.2$, $Re_\Gamma=2000$. Isocontours at $\omega_\theta=0.2$ (orangish-red) and $\omega_\theta=-0.15$ (blue).  From top left to bottom right: $t=25$, $30$, $35$, $45$ and $60$. }
    \label{fig:cont-cr0p2-rcyl0p1-wth-re2000}
\end{figure}

\begin{figure}
    \centering
    \includegraphics[trim={3.5cm 0 4.5cm 0},clip,height=0.25\linewidth]{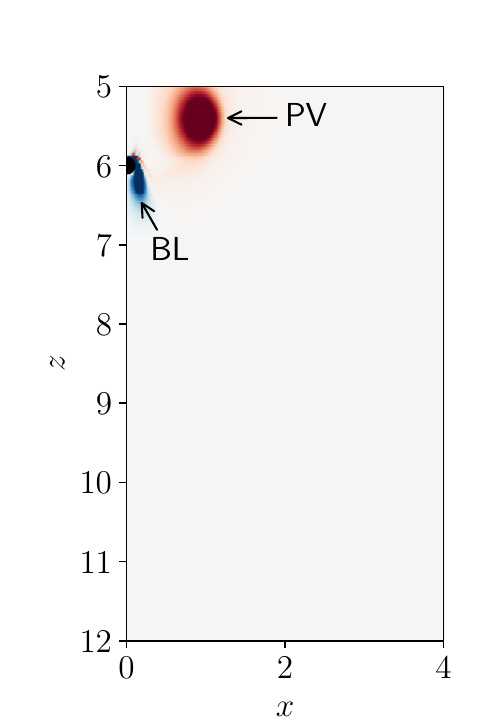}
    \includegraphics[trim={3.5cm 0 4.5cm 0},clip,height=0.25\linewidth]{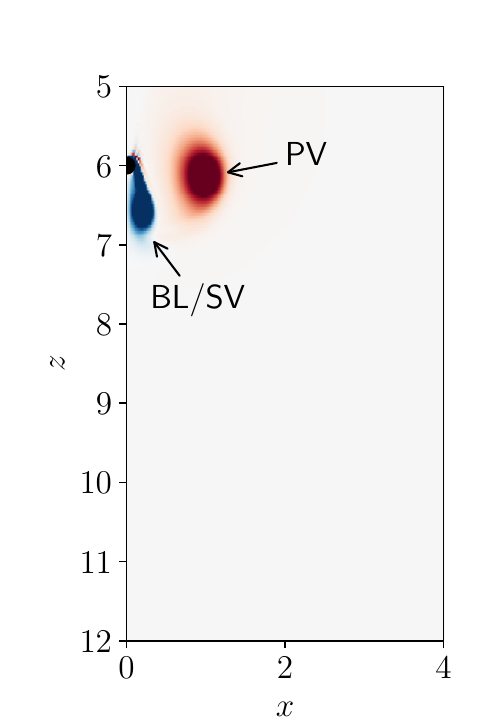}
    \includegraphics[trim={3.5cm 0 4.5cm 0},clip,height=0.25\linewidth]{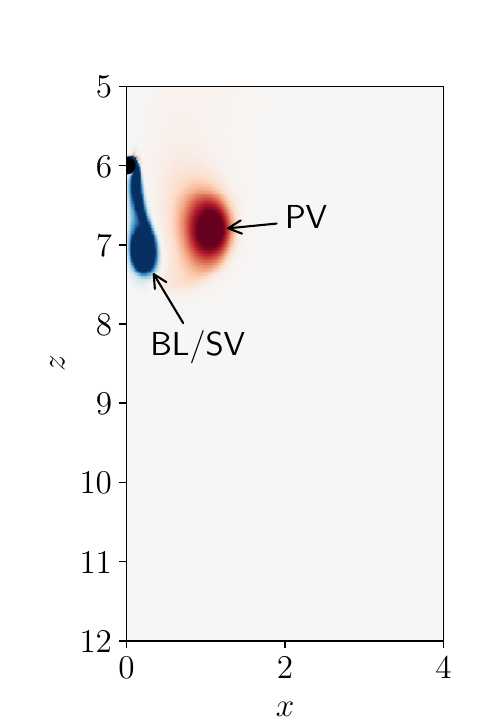}
    \includegraphics[trim={3.5cm 0 4.5cm 0},clip,height=0.25\linewidth]{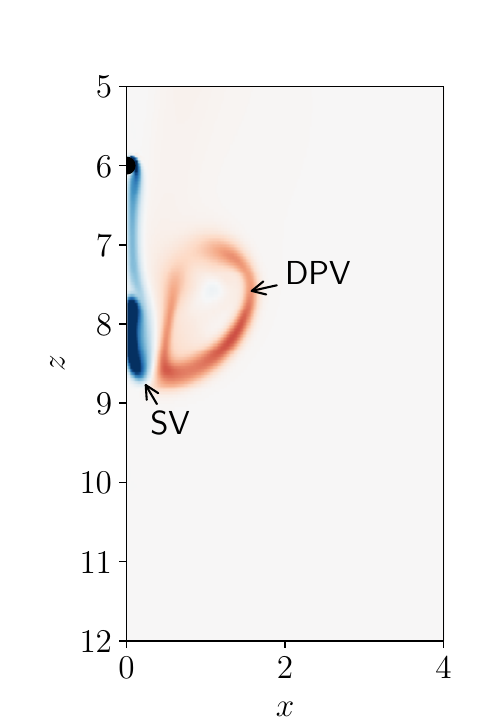}
    \includegraphics[trim={3.5cm 0 1cm 0},clip,height=0.25\linewidth]{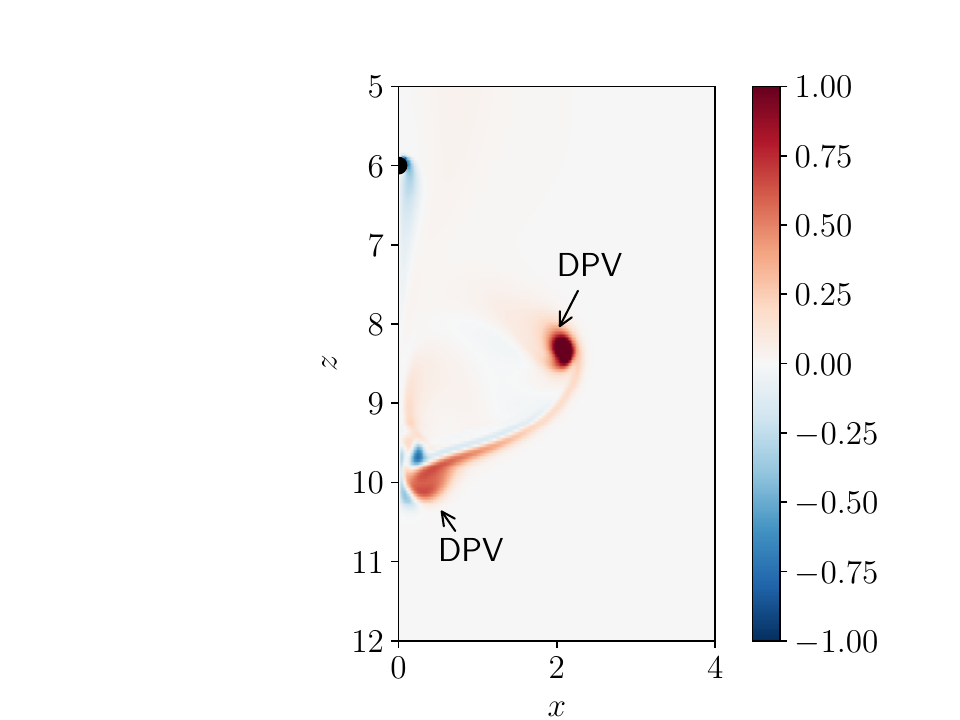}
\\
    \includegraphics[trim={3.5cm 0 4.5cm 0},clip,height=0.25\linewidth]{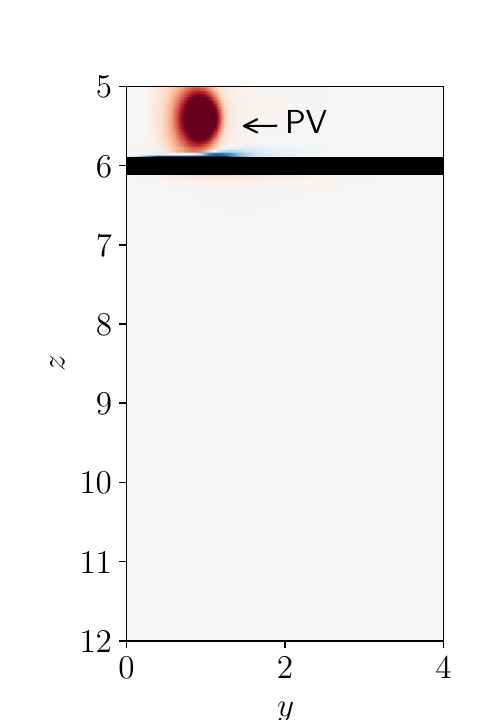}
    \includegraphics[trim={3.5cm 0 4.5cm 0},clip,height=0.25\linewidth]{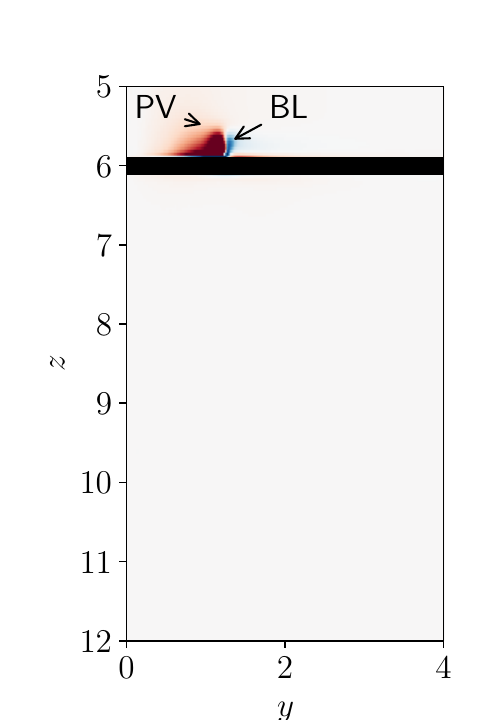}
    \includegraphics[trim={3.5cm 0 4.5cm 0},clip,height=0.25\linewidth]{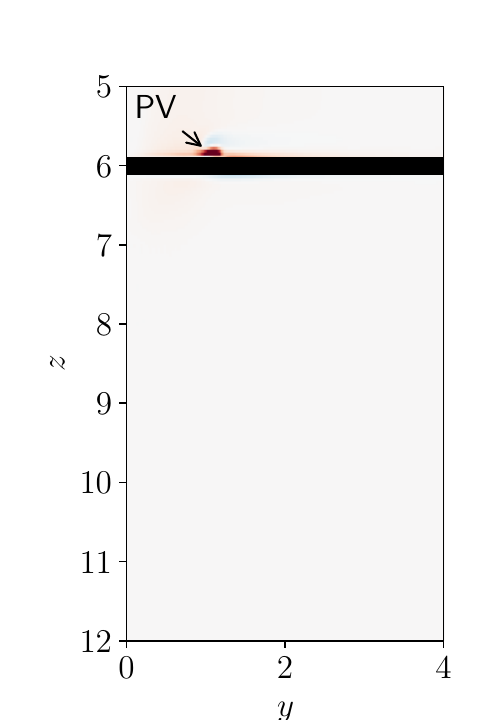}
    \includegraphics[trim={3.5cm 0 4.5cm 0},clip,height=0.25\linewidth]{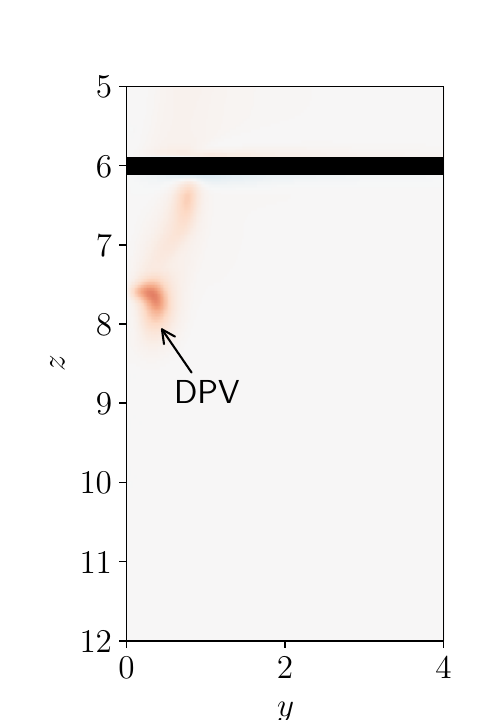}
    \includegraphics[trim={3.5cm 0 1cm 0},clip,height=0.25\linewidth]{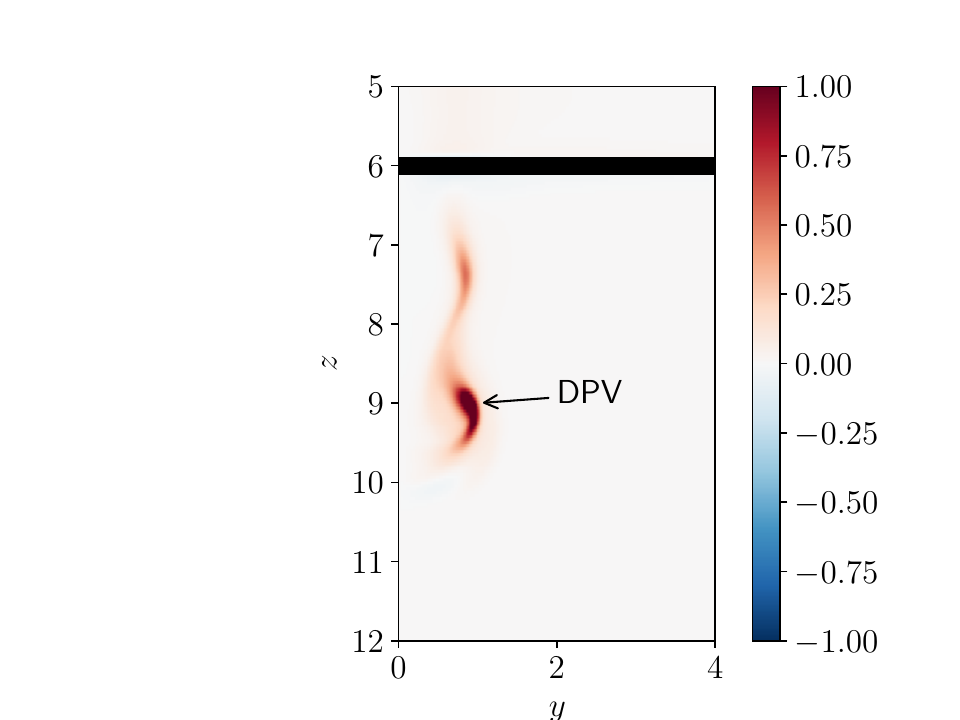}
    \caption{Evolution of the azimuthal vorticity for $T_D=0.1$, $\Lambda=0.2$, $Re_\Gamma=2000$,  through the $x$-$z$ plane (top) and $y$-$z$ plane (bottom).  From left to right: $t=25$, $35$, $45$, $60$ and $80$.  Legend: (D)PV, (deformed) primary vortex; BL, boundary layer; SV, secondary vorticity.}
    \label{fig:cr0p2-rcyl0p1-wth-re2000}
\end{figure}

For intermediate $T_D$, the phenomenology becomes increasingly complex. Figures~\ref{fig:cont-cr0p2-rcyl0p1-wth-re2000} and \ref{fig:cr0p2-rcyl0p1-wth-re2000} show the behavior for $T_D=0.1$ at $Re_\Gamma=2000$. In this simulation, the primary vortex sustains significant deformation. When looking at the two-dimensional cuts, it initially disappears from the $y$–$z$ plane but later reforms in the final frame. This contrasts with the lower-$Re_\Gamma$ case in figure~\ref{fig:cr0p2-rcyl0p1-wth-re1000}, where no such reformation occurs. At higher $Re_\Gamma$, the cutting process becomes more intricate: vortex sheets may roll up into tubes or stretch into elongated structures, leading to reconnections and partial recovery of the primary ring. This behavior is consistent with the complex phenomenology observed in figure~\ref{fig:cr0p2-rcyls-re2000}. We note that this is not the same ``cut-and-reconnect'' process observed in \cite{adhikari2009msc}, as it occurs at higher values of $T_D$, but instead one where the vortex dynamics is highliy intricate.

In summary, increasing $Re_\Gamma$ enhances the interaction between primary and secondary vorticity, promoting deformation and reconnection. The ``wire'' regime thus persists at higher Reynolds numbers, but with greater structural complexity and less clearly identifiable surviving rings.

\section{The ``cutting'' regime}
\label{sec:res3}

The results presented above demonstrate that, for sufficiently thin wires, the primary vortex ring largely survives the interaction, although it may incorporate within it secondary vorticity and experience moderate deformation. As the wire diameter increases, the interaction becomes progressively stronger, eventually preventing the primary ring from reforming after passage around the obstacle. This marks the onset of the ``cutting'' regime, in which the ring is divided into multiple coherent structures that evolve independently. In what follows, we examine this transition in detail and characterize the flow phenomenology associated with the destruction of the primary vortex ring.

\subsection{Formation and ejection of secondary structures}

\begin{figure}
    \centering
    \includegraphics[width=0.80\linewidth]{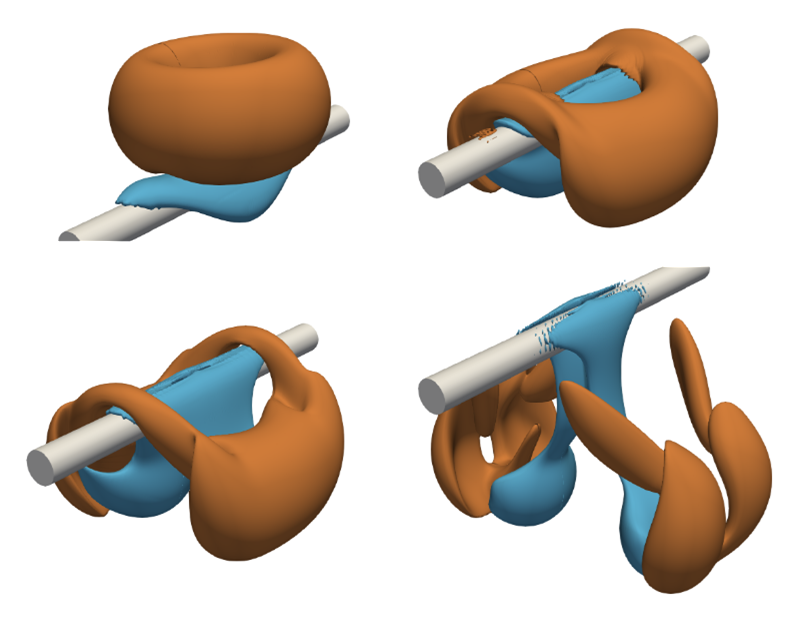}
    \caption{Evolution of the azimuthal vorticity for $T_D=0.2$, $\Lambda=0.2$, $Re_\Gamma=1000$. Isocontours at $\omega_\theta=0.2$ (orangish-red) and $\omega_\theta=-0.15$ (blue).  From top left to bottom right: $t=25$, $35$, $45$ and $55$. }
    \label{fig:cont-cr0p2-rcyl0p2-wth-re1000}
\end{figure}

\begin{figure}
    \centering
    \includegraphics[trim={3cm 0 4cm 0},clip,height=0.25\linewidth]{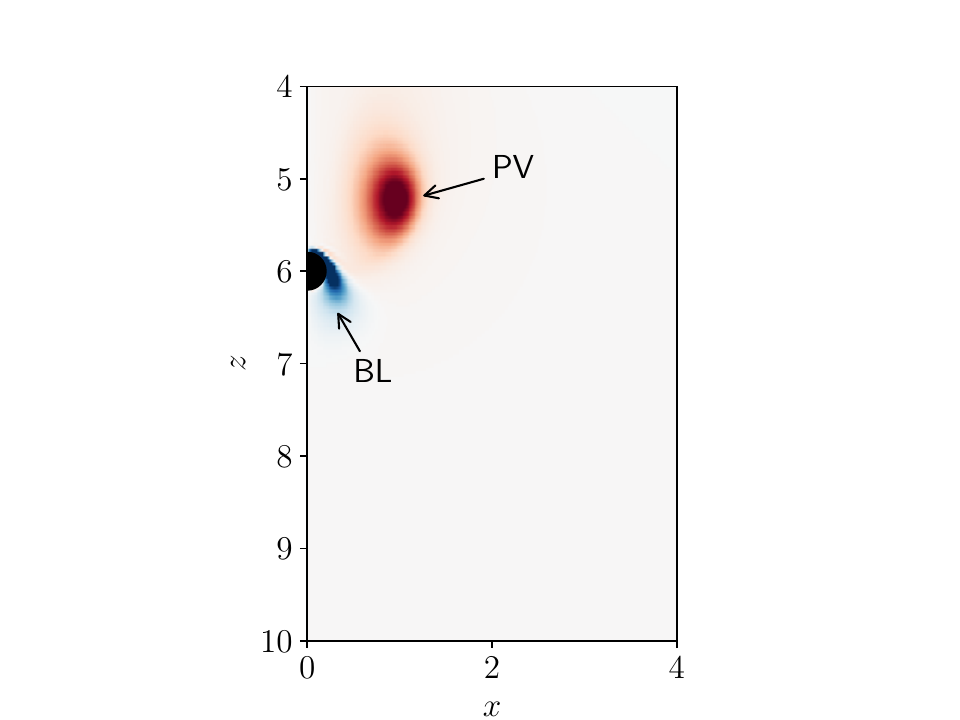}
    \includegraphics[trim={3cm 0 4cm 0},clip,height=0.25\linewidth]{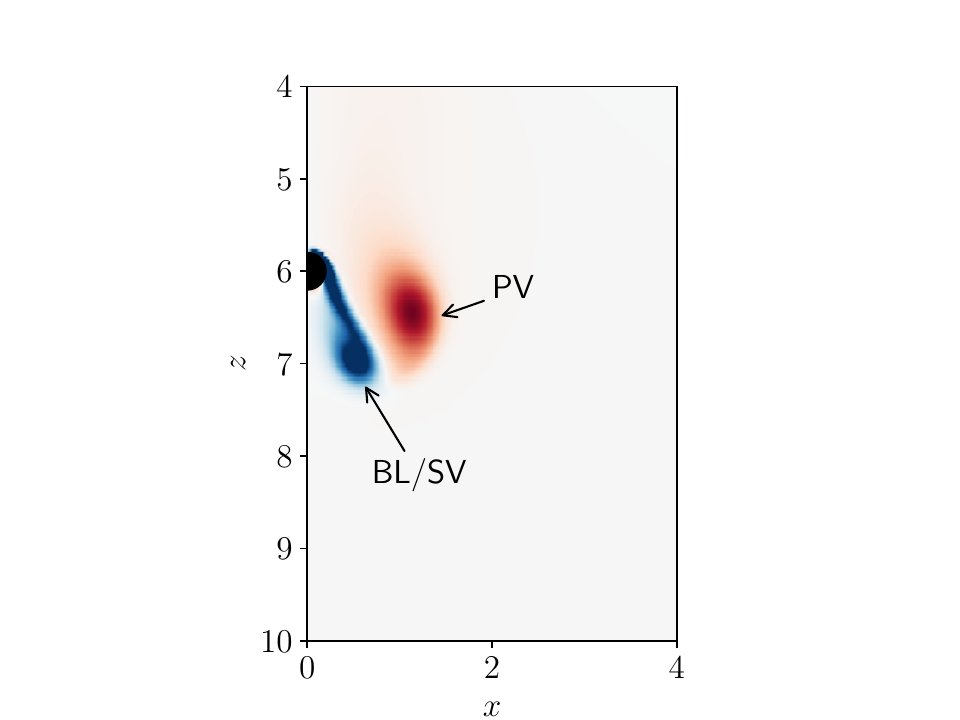}
    \includegraphics[trim={3cm 0 4cm 0},clip,height=0.25\linewidth]{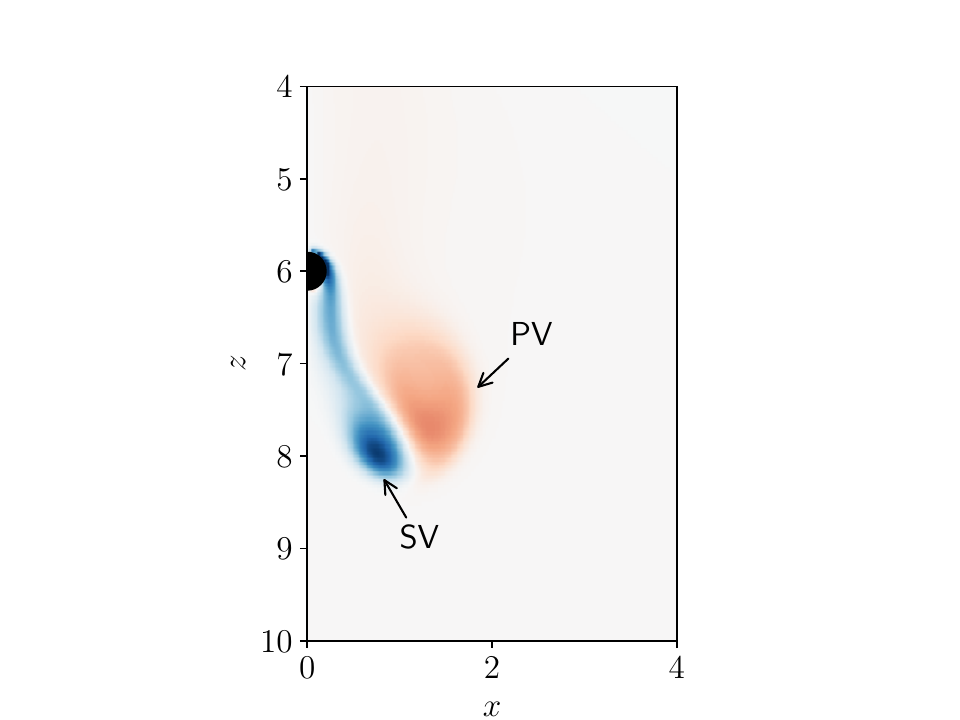}
    \includegraphics[trim={3cm 0 0cm 0},clip,height=0.25\linewidth]{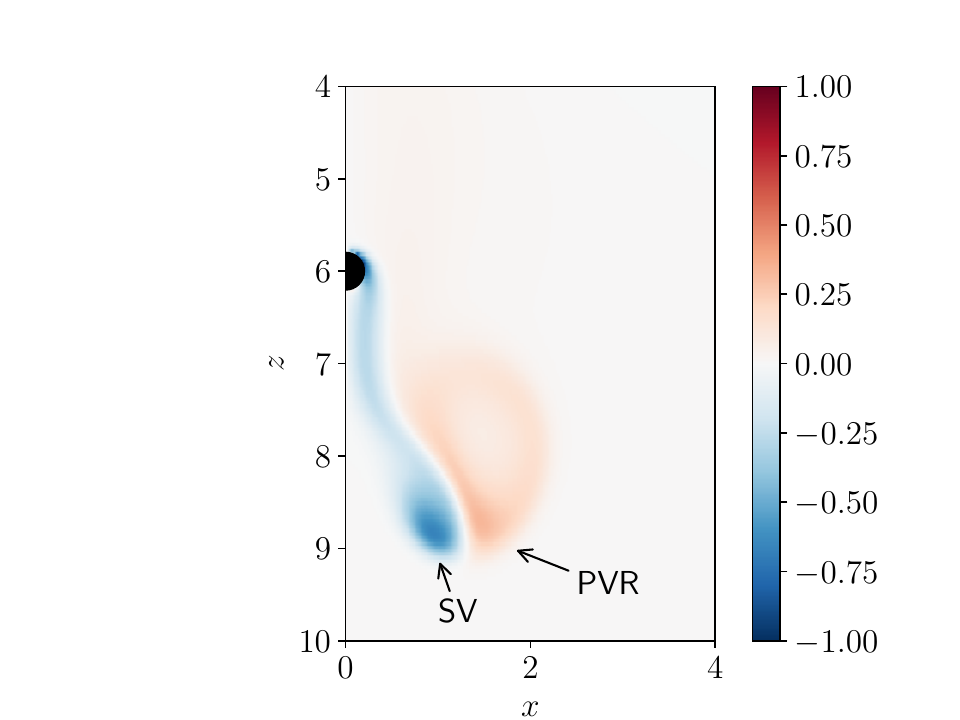}
\\
    \includegraphics[trim={3cm 0 4cm 0},clip,height=0.25\linewidth]{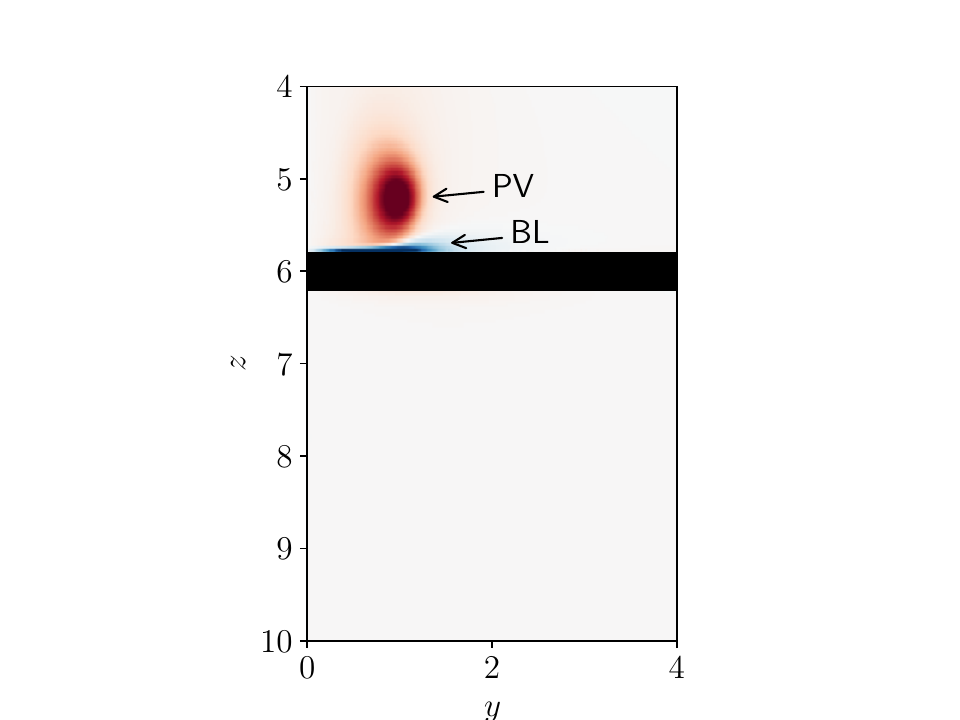}
    \includegraphics[trim={3cm 0 4cm 0},clip,height=0.25\linewidth]{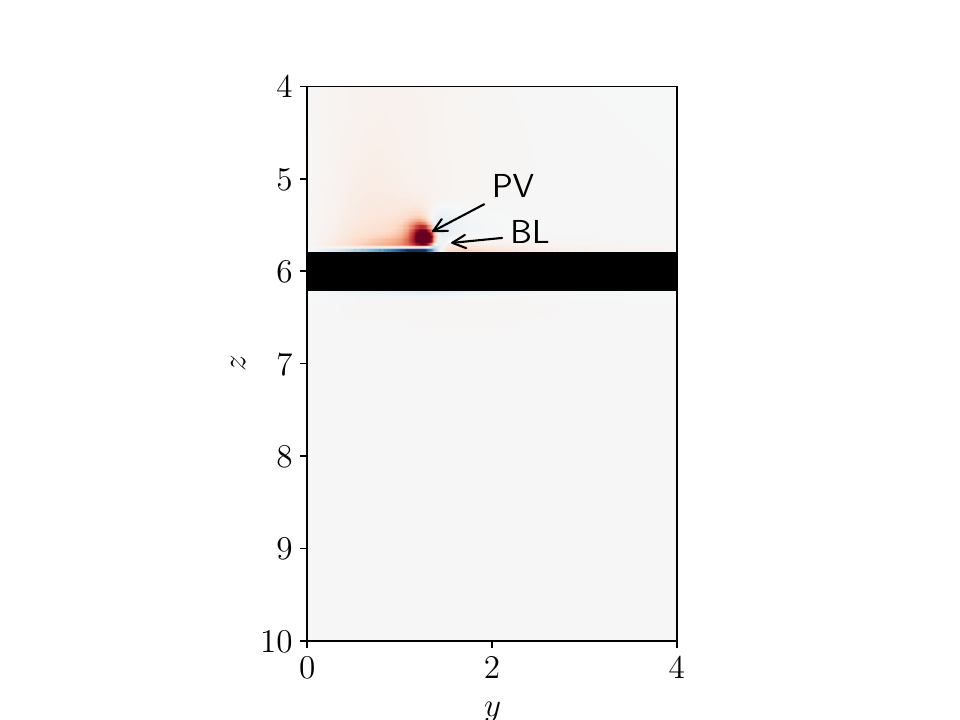}
    \includegraphics[trim={3cm 0 4cm 0},clip,height=0.25\linewidth]{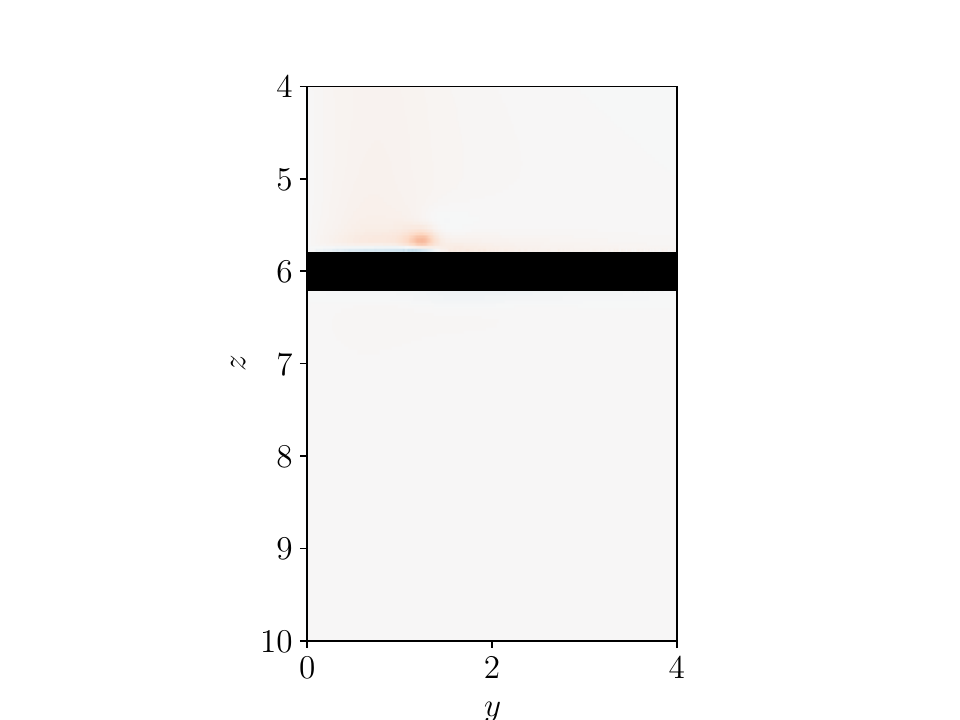}
    \includegraphics[trim={3cm 0 0cm 0},clip,height=0.25\linewidth]{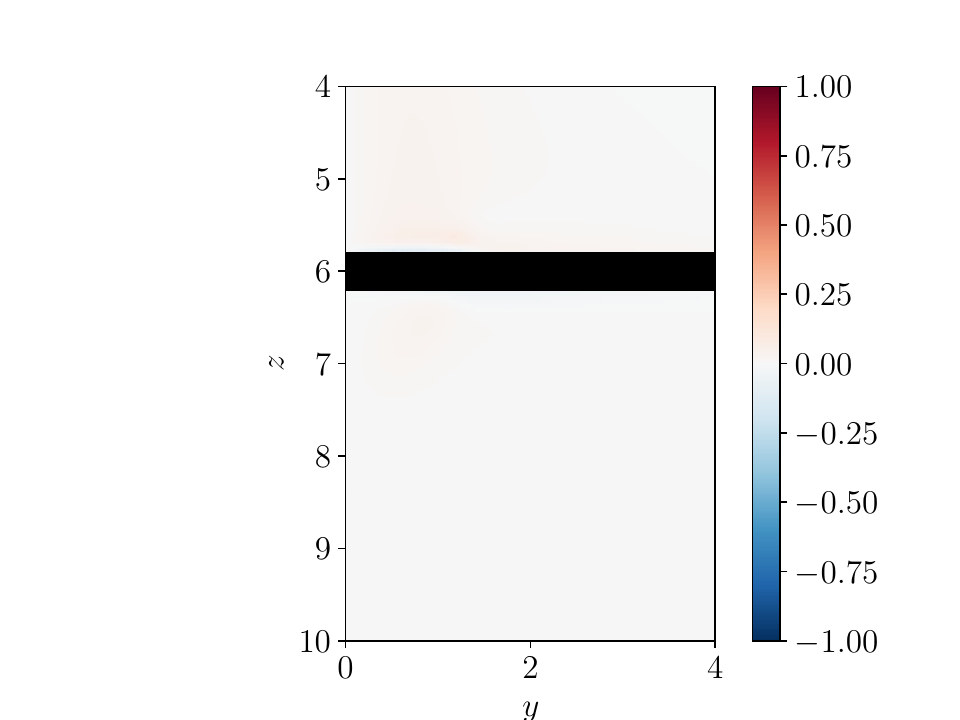}
    \caption{Temporal evolution of the azimuthal vorticity for $T_D=0.2$, $\Lambda=0.2$, $Re_\Gamma=1000$,  through the $x$-$z$ plane (top) and $y$-$z$ plane (bottom). From left to right: $t=25$, $35$, $45$ and $55$. Legend: PV(R), primary vortex (remnants); BL, boundary layer; SV, secondary vorticity.}
    \label{fig:cr0p2-rcyl0p2-wth-re1000}
\end{figure}

Figures\ref{fig:cont-cr0p2-rcyl0p2-wth-re1000} and \ref{fig:cr0p2-rcyl0p2-wth-re1000} show the evolution of the azimuthal vorticity for the case $T_D=0.2$ and $Re_\Gamma=1000$ in three-dimensional iso-contours and in two-dimensional cuts along the $x$-$z$ and $y$-$z$ planes. Compared with figures \ref{fig:cont-cr0p2-rcyl0p1-wth-re1000} and \ref{fig:cr0p2-rcyl0p1-wth-re1000}, only minor changes are observed especially in the three-dimensional iso-contours. When observing the two-dimensional cuts, we can observe that in the $x$–$z$ plane the secondary vorticity originating from the cylinder becomes increasingly well defined and now departs at a finite angle to the original trajectory rather than parallel to the $z$–axis. This vorticity forms the bottom half of the rings which are ejected from the interaction surface. In the $y$–$z$ plane, the primary vorticity can be seen to slightly slide across the cylinder surface. The increased wire thickness causes this interaction to begin resembling that of a vortex with a flat wall. However, in the $y$-$z$ plane the secondary vorticity generated in the boundary layer remains weak and does not significantly affect the primary vortex.

\begin{figure}
    \centering
    \includegraphics[width=0.80\linewidth]{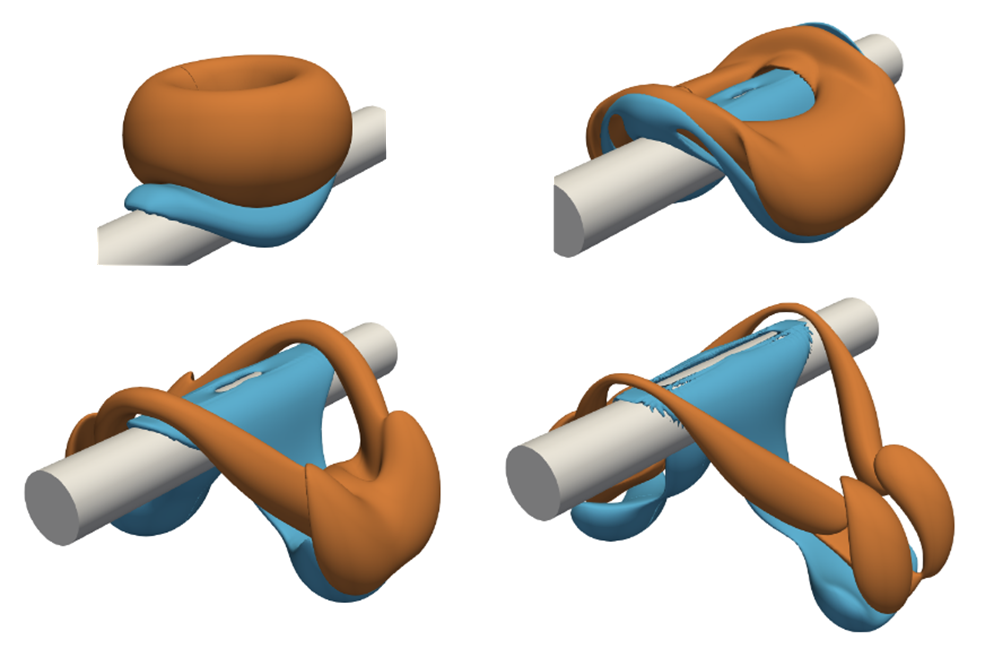}
    \caption{Evolution of the azimuthal vorticity for $T_D=0.4$, $\Lambda=0.2$, $Re_\Gamma=1000$. Isocontours at $\omega_\theta=0.2$ (orangish-red) and $\omega_\theta=-0.15$ (blue).  From top left to bottom right: $t=25$, $35$, $45$ and $55$. }
    \label{fig:cont-cr0p2-rcyl0p4-wth-re1000}
\end{figure}

\begin{figure}
    \centering
    \includegraphics[trim={3cm 0 4cm 0},clip,height=0.25\linewidth]{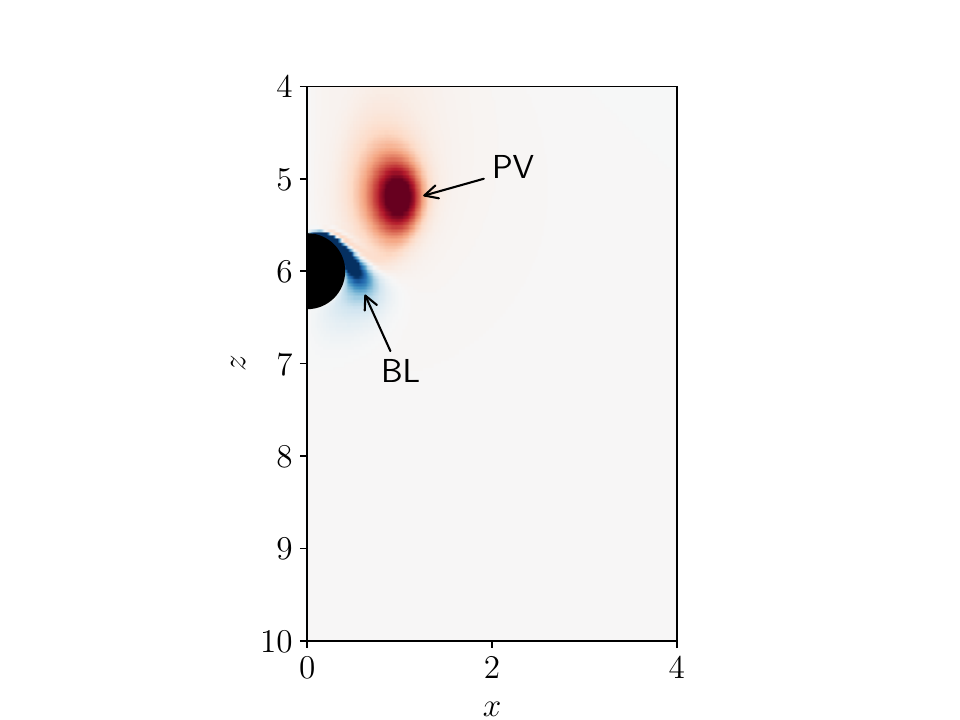}
    \includegraphics[trim={3cm 0 4cm 0},clip,height=0.25\linewidth]{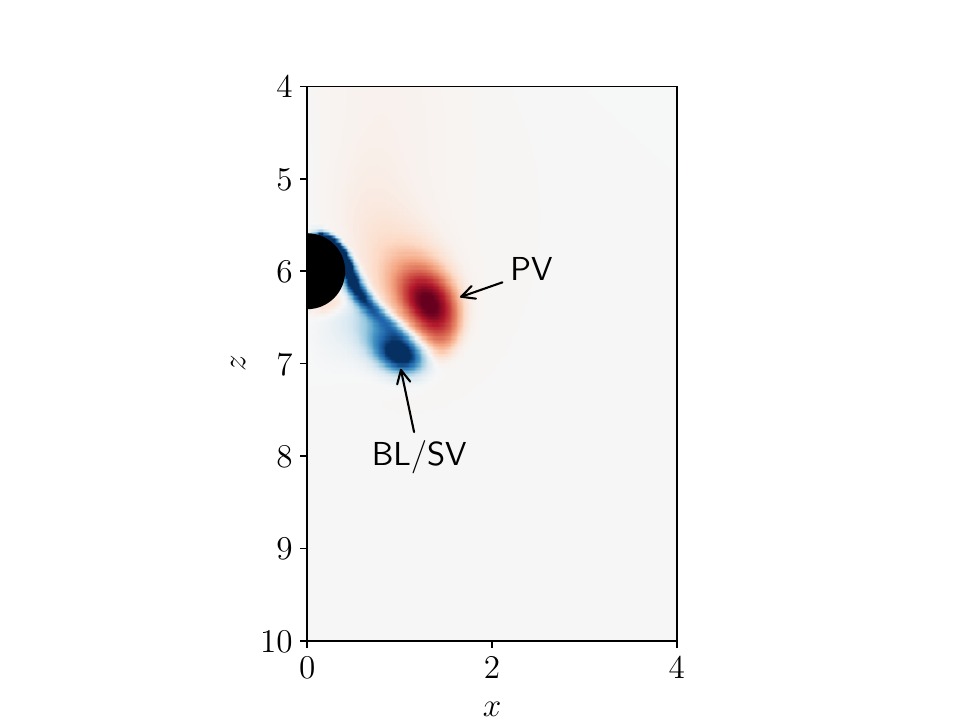}
    \includegraphics[trim={3cm 0 4cm 0},clip,height=0.25\linewidth]{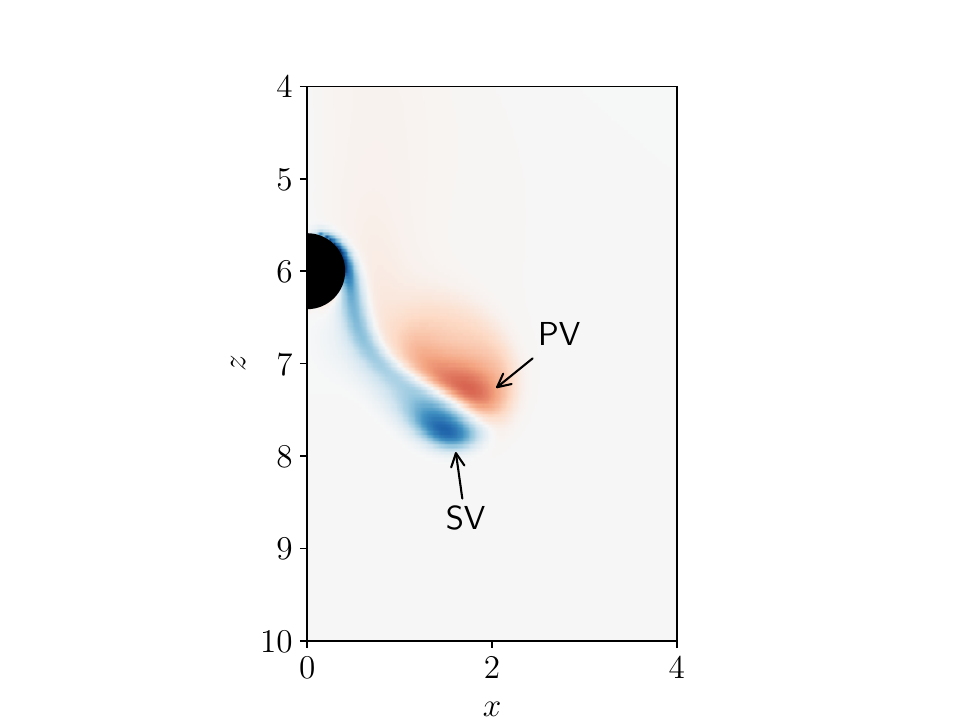}
    \includegraphics[trim={3cm 0 0cm 0},clip,height=0.25\linewidth]{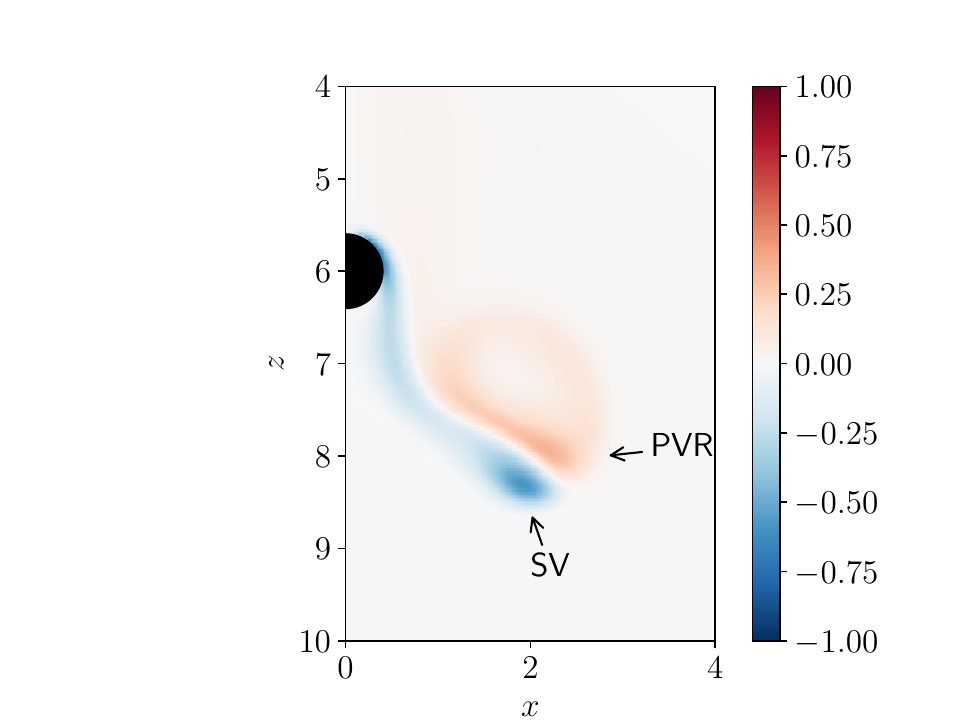}
\\
    \includegraphics[trim={3cm 0 4cm 0},clip,height=0.25\linewidth]{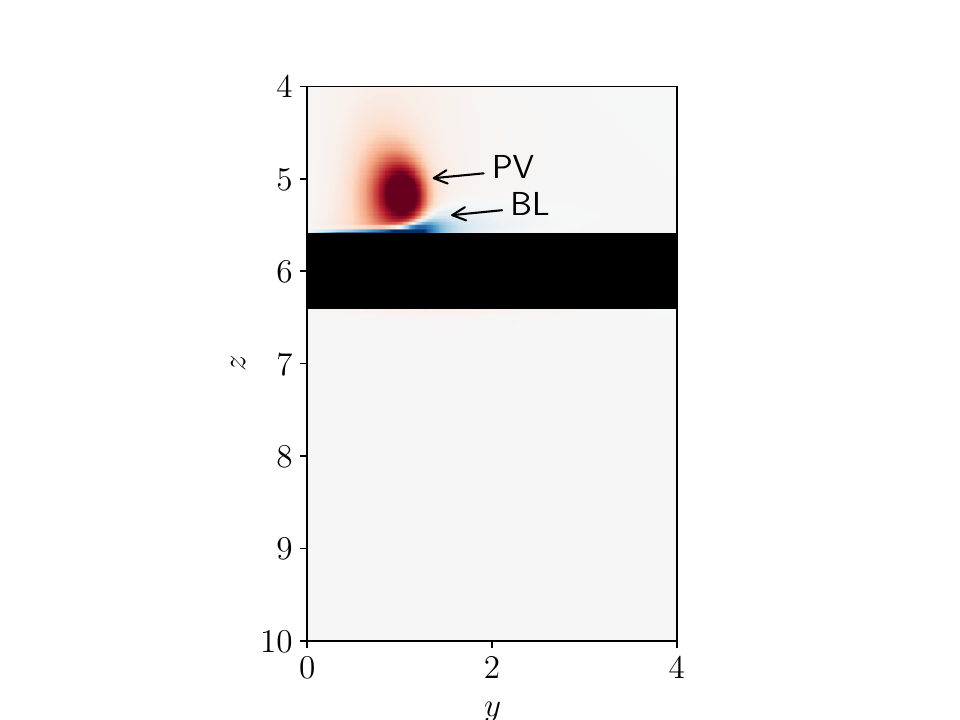}
    \includegraphics[trim={3cm 0 4cm 0},clip,height=0.25\linewidth]{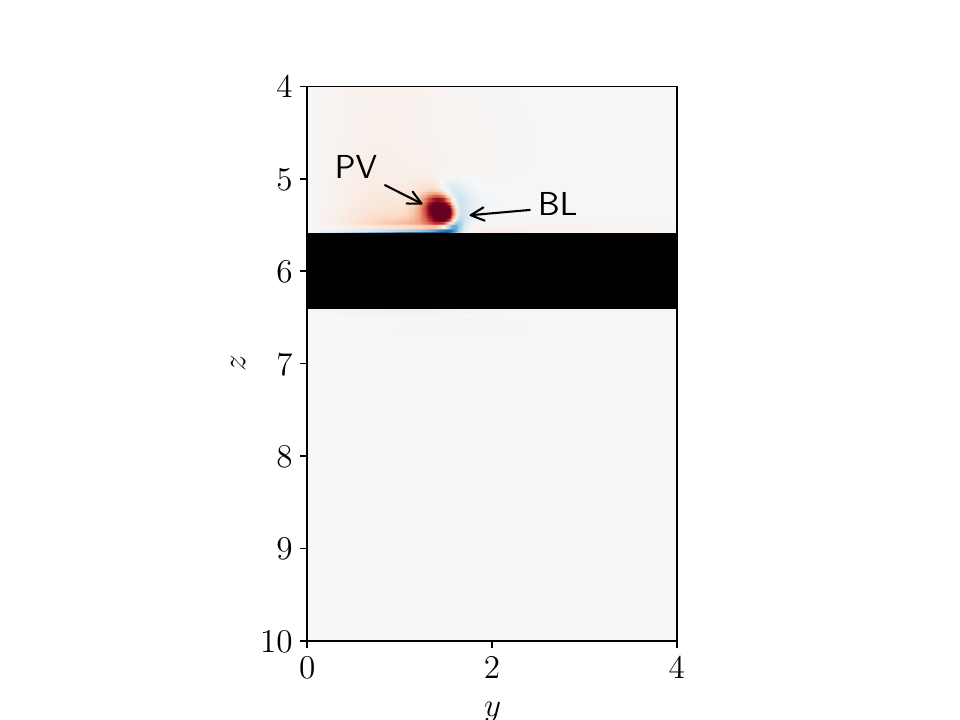}
    \includegraphics[trim={3cm 0 4cm 0},clip,height=0.25\linewidth]{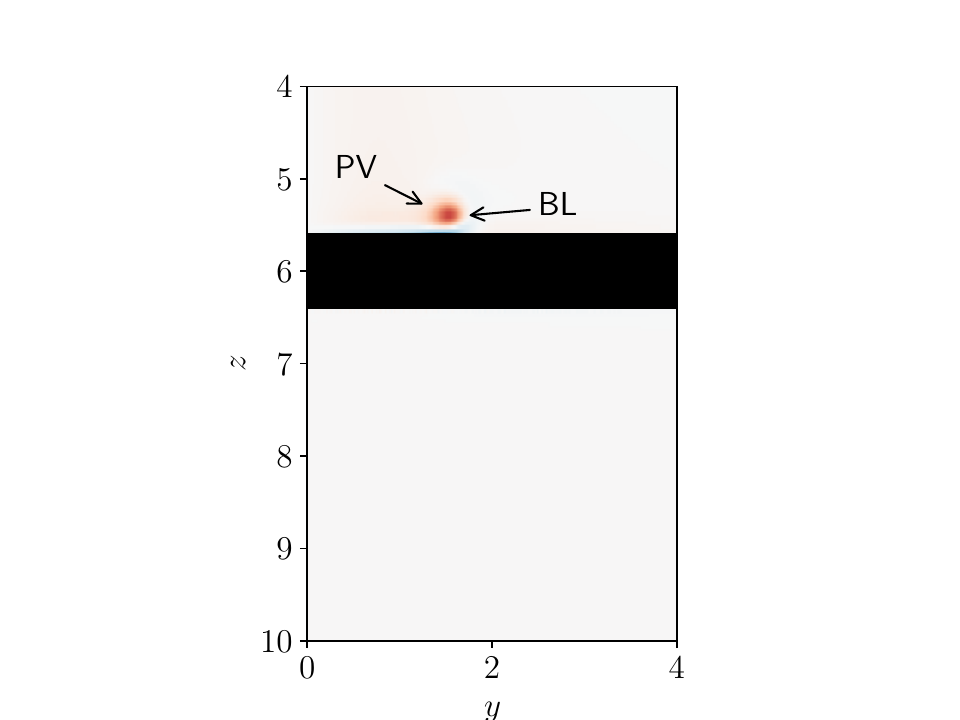}
    \includegraphics[trim={3cm 0 0cm 0},clip,height=0.25\linewidth]{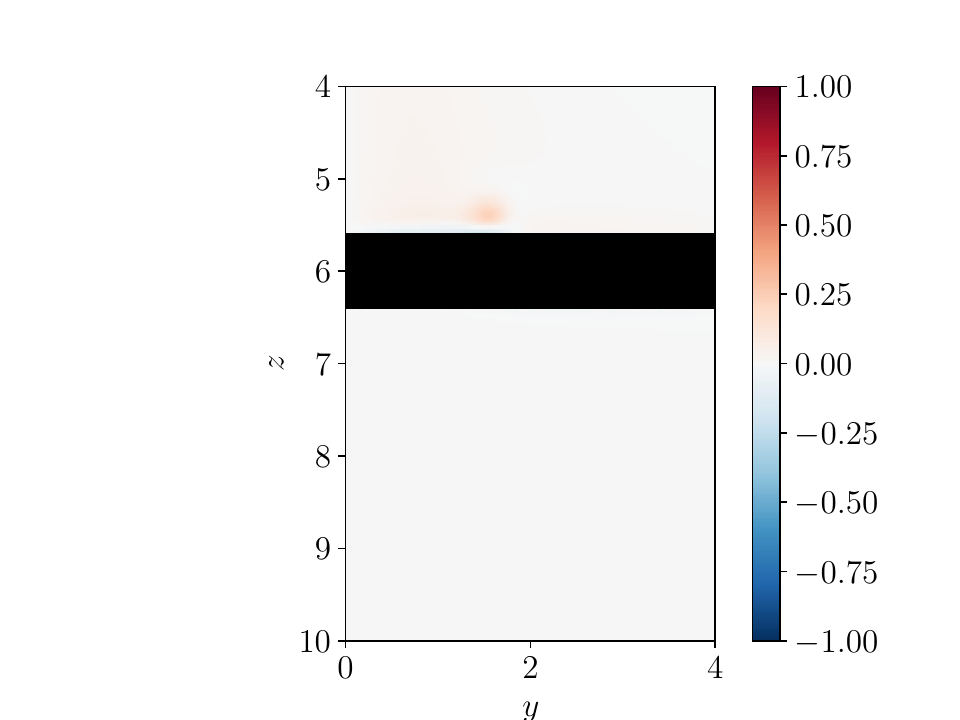}
    \caption{Temporal evolution of the azimuthal vorticity for $T_D=0.4$, $\Lambda=0.2$, $Re_\Gamma=1000$,  through the $x$-$z$ plane (top) and $y$-$z$ plane (bottom). From left to right: $t=25$, $35$, $45$ and $55$. Legend: PV(R), primary vortex (remnants); BL, boundary layer; SV, secondary vorticity.}
    \label{fig:cr0p2-rcyl0p4-wth-re1000}
\end{figure}

Further increasing the wire diameter to $T_D=0.4$ (figures~\ref{fig:cont-cr0p2-rcyl0p4-wth-re1000} and \ref{fig:cr0p2-rcyl0p4-wth-re1000}) yields a qualitatively similar picture. In the $x$–$z$ plane, the secondary vorticity is now clearly defined and leaves at a larger angle than for $T_D=0.2$. A small amount of vorticity is dragged along the $z$–axis, as also observed earlier in figure~\ref{fig:cr0p2-rcyls-re1000}, but it dissipates rapidly. Meanwhile, in the $y$–$z$ plane, the primary vorticity interacts more strongly with the cylinder surface, forming a boundary layer, although the overall interaction between ring and boundary layer remains relatively weak when compared to that in the $x$-$z$ plane.

\begin{figure}
    \centering
    \includegraphics[width=0.80\linewidth]{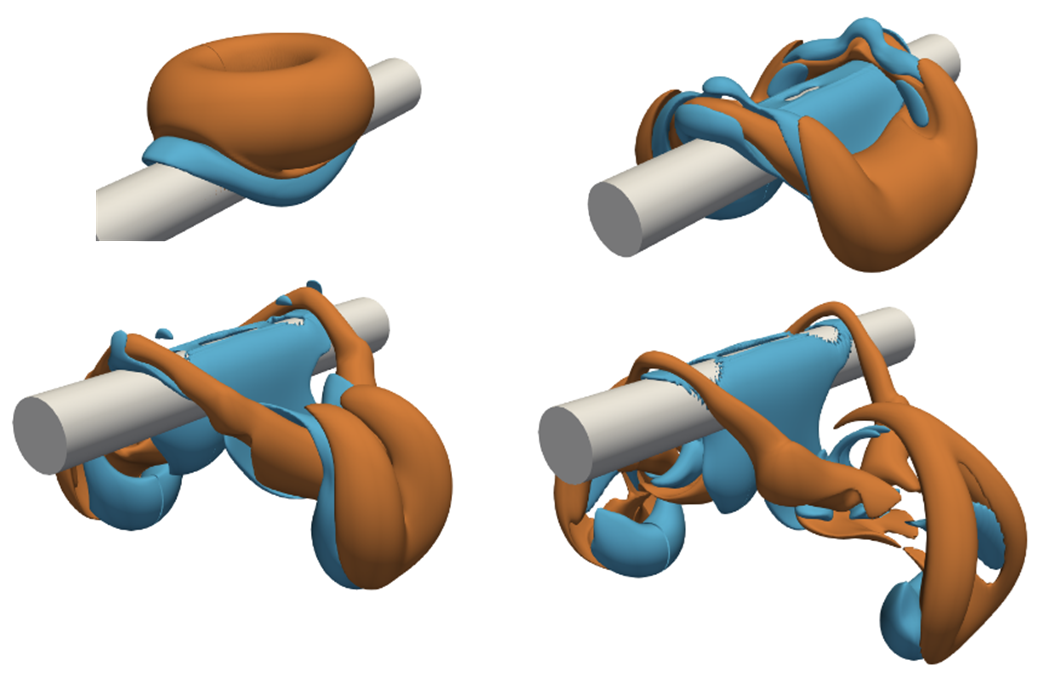}
    \caption{Evolution of the azimuthal vorticity for $T_D=0.4$, $\Lambda=0.2$, $Re_\Gamma=2000$. Isocontours at $\omega_\theta=0.2$ (orangish-red) and $\omega_\theta=-0.15$ (blue).  From top left to bottom right: $t=25$, $35$, $45$ and $55$. }
    \label{fig:cont-cr0p2-rcyl0p4-wth-re2000}
\end{figure}

\begin{figure}
    \centering
    \includegraphics[trim={3.5cm 0 4.5cm 0},clip,height=0.25\linewidth]{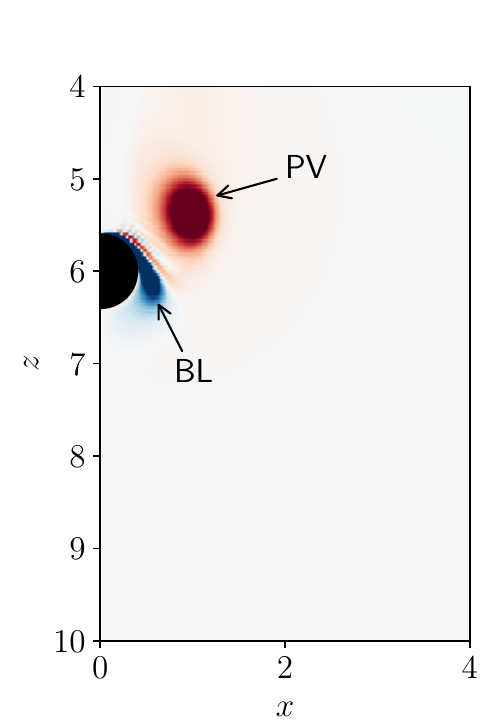}
    \includegraphics[trim={3.5cm 0 4.5cm 0},clip,height=0.25\linewidth]{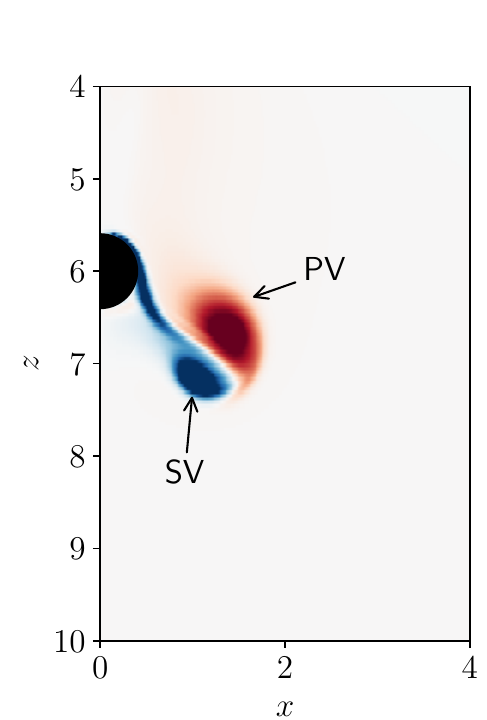}
    \includegraphics[trim={3.5cm 0 4.5cm 0},clip,height=0.25\linewidth]{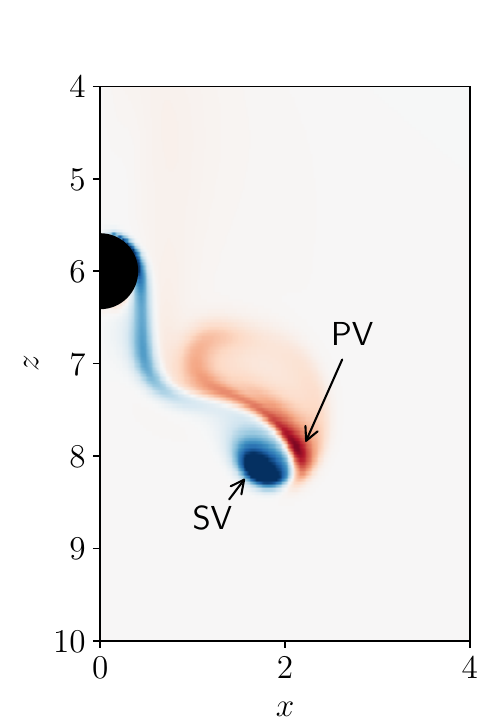}
    \includegraphics[trim={3.5cm 0 1cm 0},clip,height=0.25\linewidth]{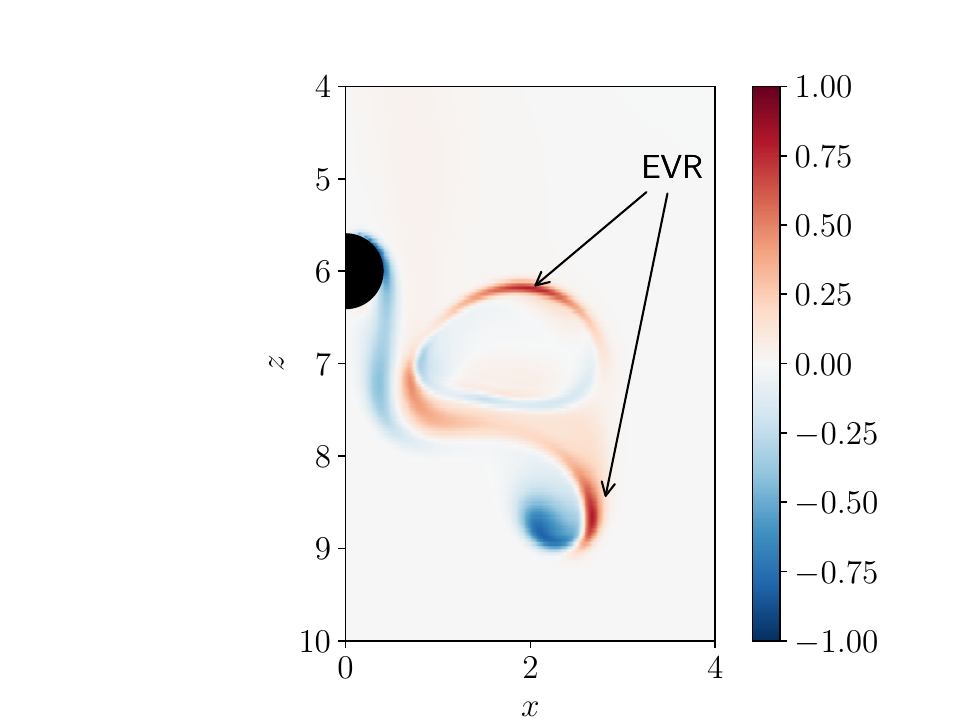}
\\
    \includegraphics[trim={3.5cm 0 4.5cm 0},clip,height=0.25\linewidth]{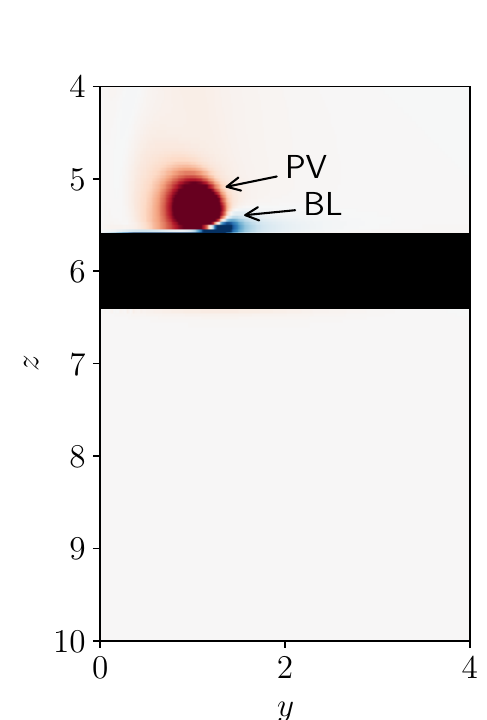}
    \includegraphics[trim={3.5cm 0 4.5cm 0},clip,height=0.25\linewidth]{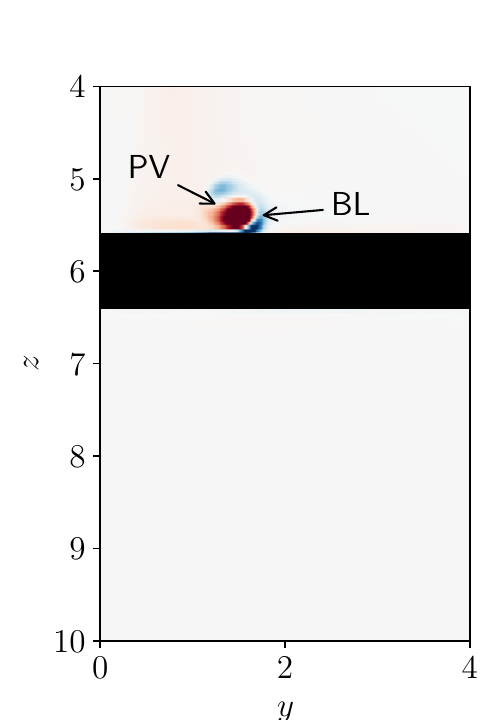}
    \includegraphics[trim={3.5cm 0 4.5cm 0},clip,height=0.25\linewidth]{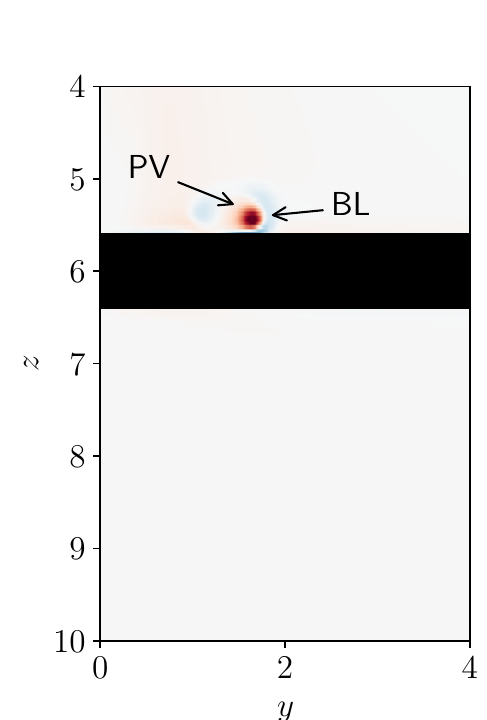}
    \includegraphics[trim={3.5cm 0 1cm 0},clip,height=0.25\linewidth]{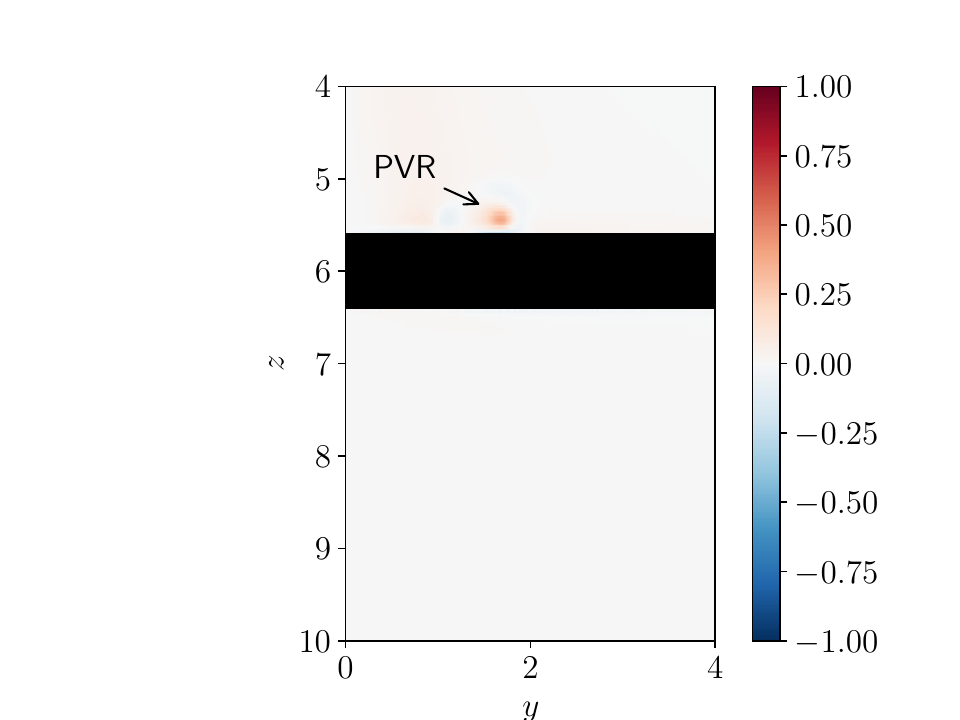}
    \caption{Temporal evolution of the azimuthal vorticity for $T_D=0.4$, $\Lambda=0.2$, $Re_\Gamma=2000$,  through the $x$-$z$ plane (top) and $y$-$z$ plane (bottom).  From left to right: $t=25$, $35$, $45$ and $55$. Legend: PV, primary vortex; BL, boundary layer; SV, secondary vorticity; EVR, ejected vortex ring.}
    \label{fig:cr0p2-rcyl0p4-wth-re2000}
\end{figure}

As in the previous section, increasing $Re_\Gamma$ complicates the flow phenomenology. Figures~\ref{fig:cont-cr0p2-rcyl0p4-wth-re2000} and \ref{fig:cr0p2-rcyl0p4-wth-re2000} show the evolution of the azimuthal vorticity for $T_D=0.4$ and $Re_\Gamma=2000$. Initially, the behaviour resembles that at lower $Re_\Gamma$: the primary vortex forms a boundary layer of oppositely signed secondary vorticity around the cylinder, which can detach to form a secondary structure. However, the increased stretching results in the secondary ring emerging at a much larger angle and with a greater diameter. 

Furthermore, as observed in figure~\ref{fig:cr0p2-rcyls-re2000}. a weaker ring structure that continues to propagate in the $z$ direction can also form from the interaction. This ring is barely visible in \ref{fig:cont-cr0p2-rcyl0p4-wth-re2000} and \ref{fig:cr0p2-rcyl0p4-wth-re2000} as it is much weaker than the other structures. It is a tertiary structure of the process, and contains only a small fraction of the original circulation: slightly less than 10\% in the $y$–$z$ plane, where it is strongest. It is therefore much weaker than the surviving primary rings observed in the wire regime. We can still state that $T_D=0.4$ remains within the cutting regime at higher Reynolds numbers, as the remnant ring is phenomenologically distinct from a surviving primary ring.

\subsection{From cutting to the wall regime}
\label{subsec:cutting-to-wall}

\begin{figure}
    \centering
    \includegraphics[width=0.80\linewidth]{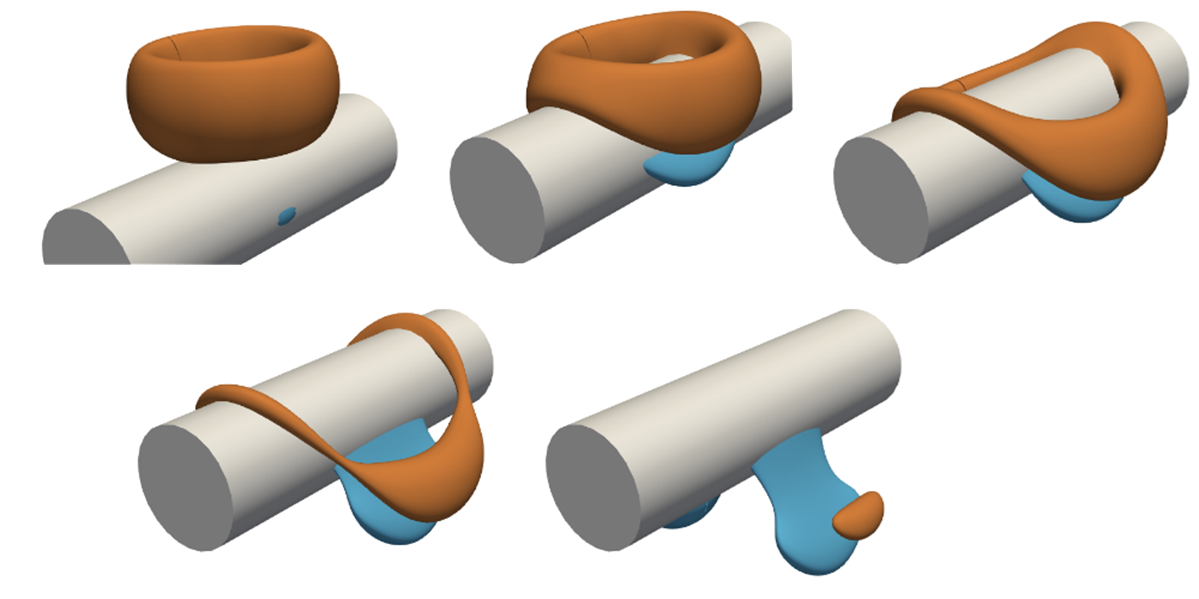}
    \caption{Evolution of the azimuthal vorticity for $T_D=0.7$, $\Lambda=0.2$, $Re_\Gamma=1000$. Isocontours at $\omega_\theta=0.2$ (orangish-red) and $\omega_\theta=-0.15$ (blue).  From top left to bottom right: $t=25$, $35$, $45$, $55$ and $65$. }
    \label{fig:cont-cr0p2-rcyl0p7-wth-re1000}
\end{figure}

\begin{figure}
    \centering
    \includegraphics[trim={3cm 0 4cm 0},clip,height=0.22\linewidth]{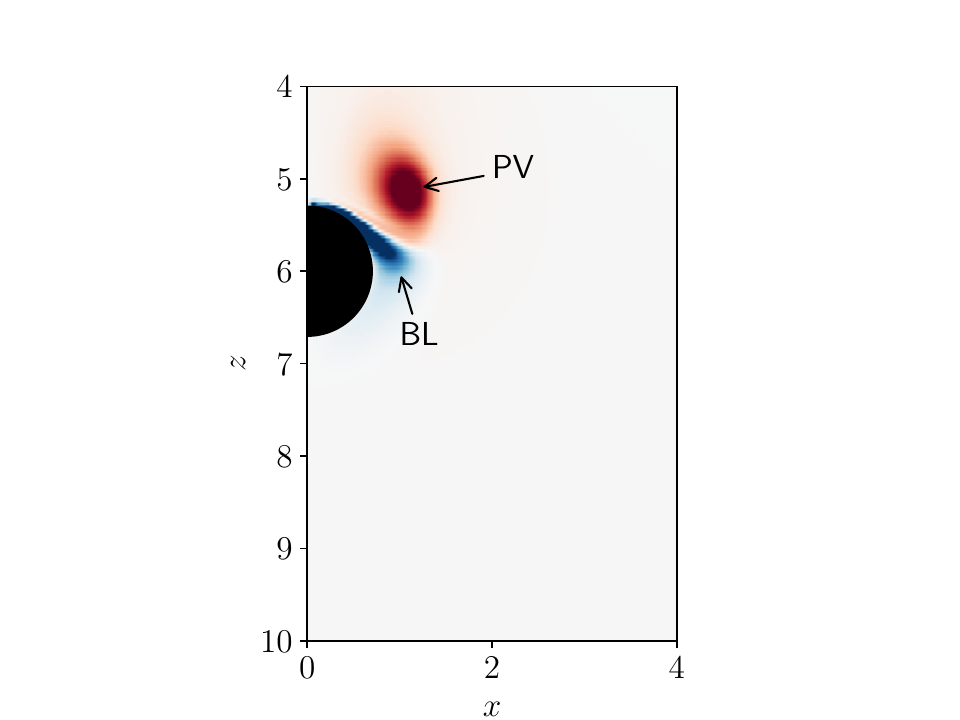}
    \includegraphics[trim={3cm 0 4cm 0},clip,height=0.22\linewidth]{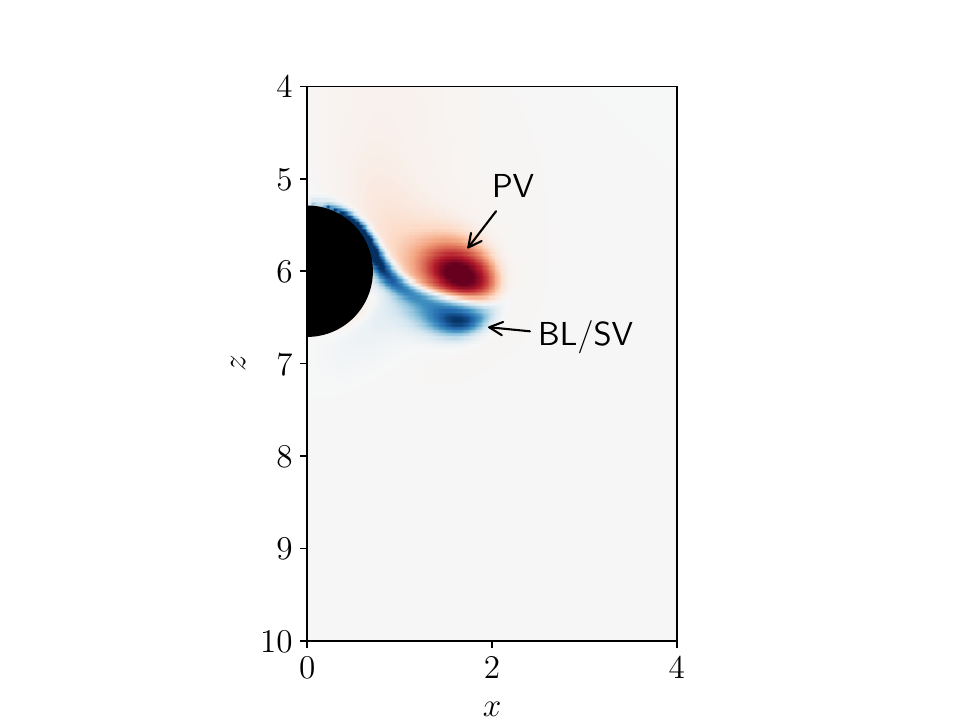}
    \includegraphics[trim={3cm 0 4cm 0},clip,height=0.22\linewidth]{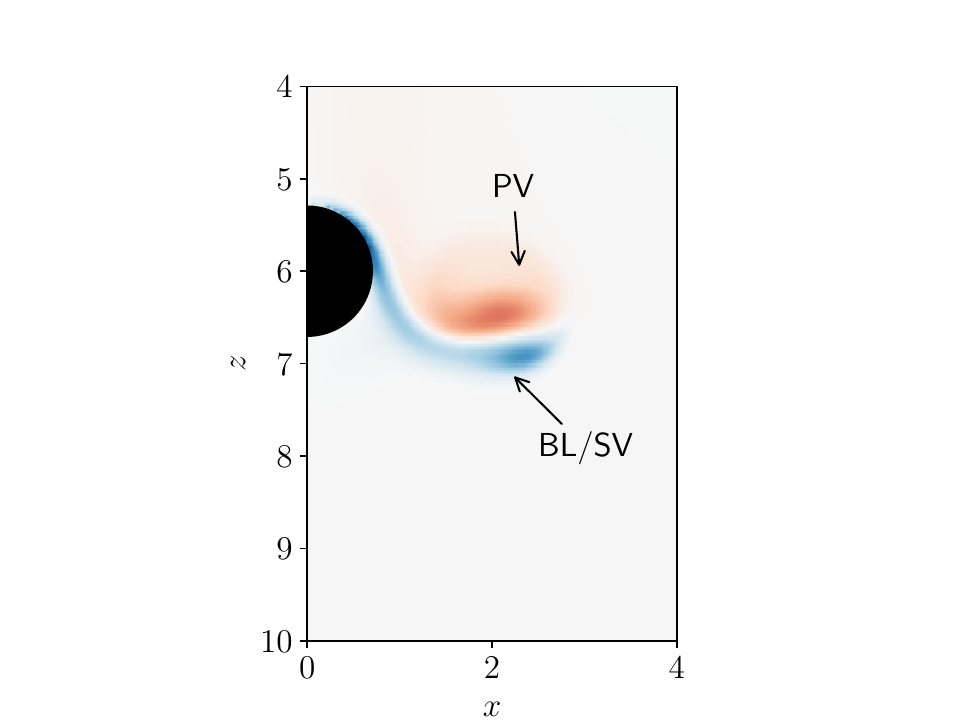}
    \includegraphics[trim={3cm 0 4cm 0},clip,height=0.22\linewidth]{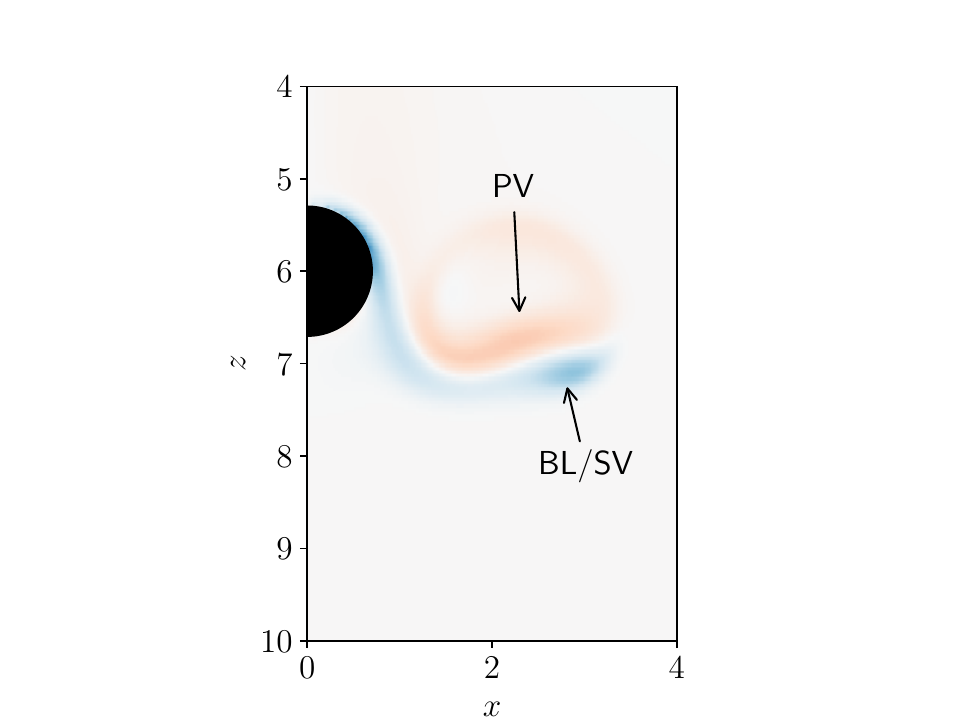}
    \includegraphics[trim={3cm 0 0cm 0},clip,height=0.22\linewidth]{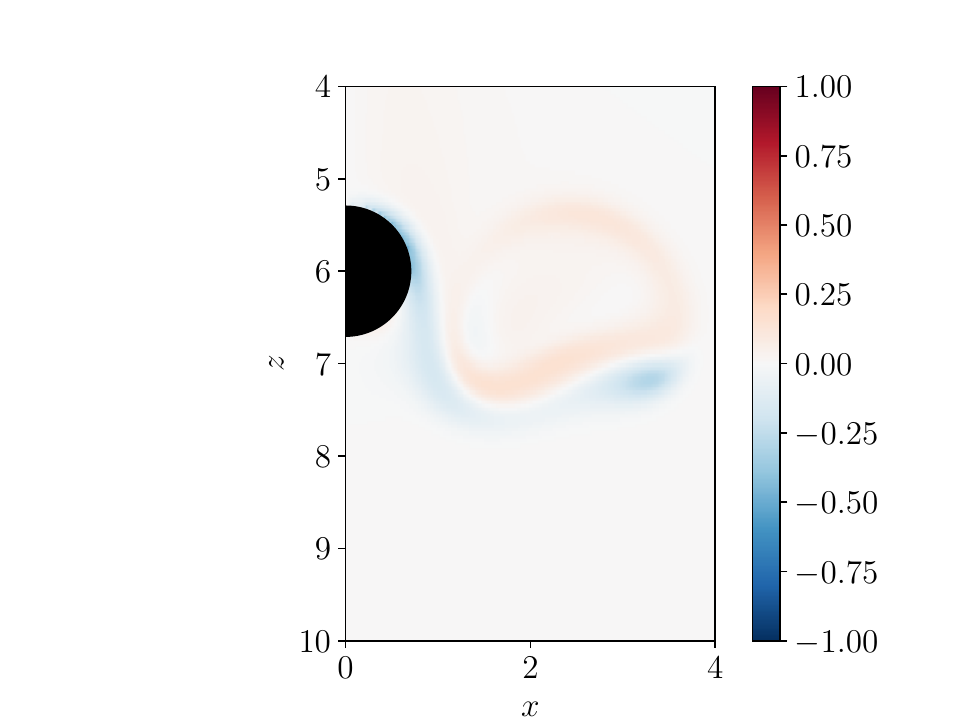}
\\
    \includegraphics[trim={3cm 0 4cm 0},clip,height=0.22\linewidth]{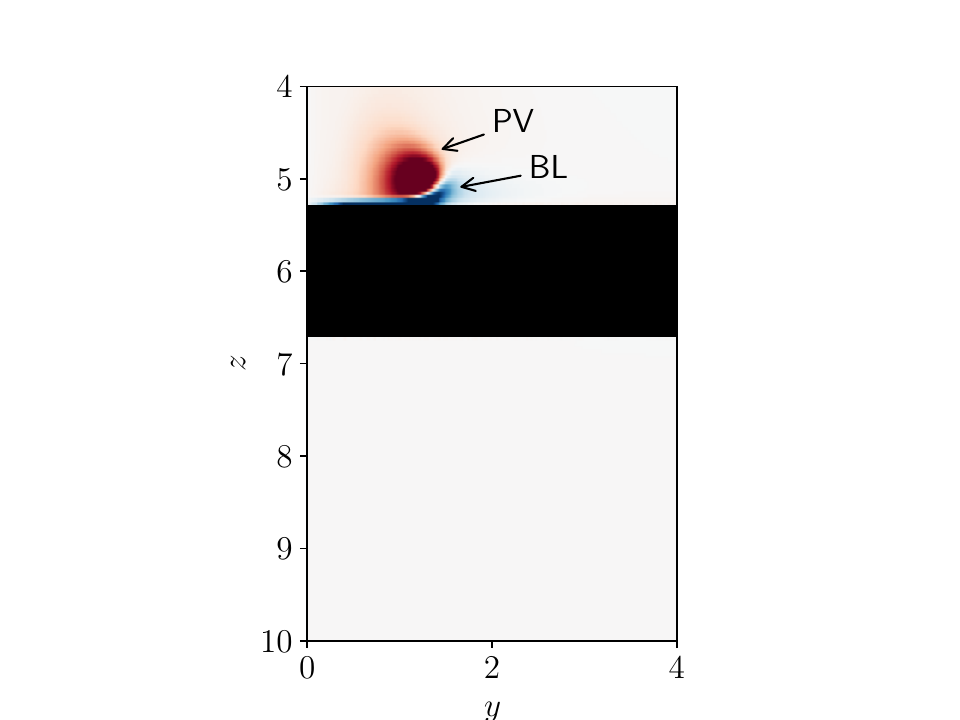}
    \includegraphics[trim={3cm 0 4cm 0},clip,height=0.22\linewidth]{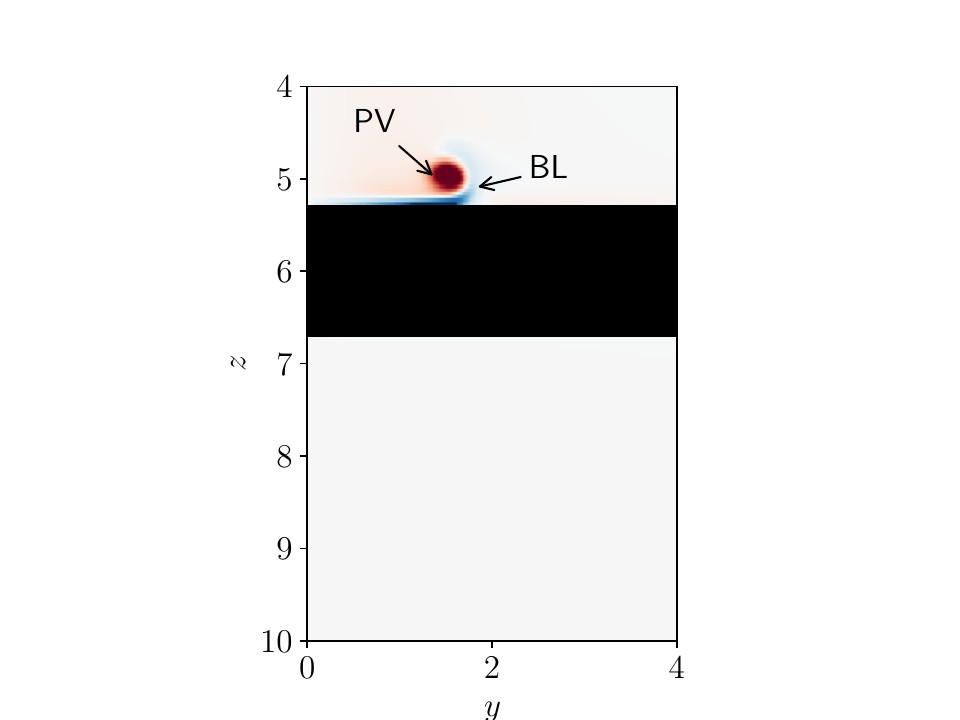}
    \includegraphics[trim={3cm 0 4cm 0},clip,height=0.22\linewidth]{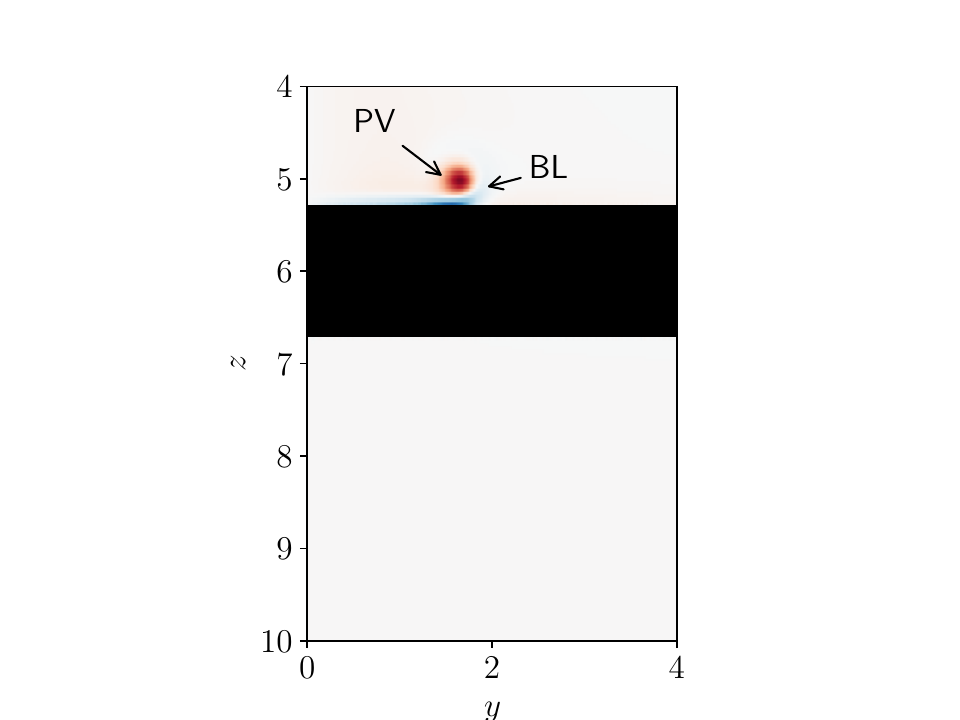}
    \includegraphics[trim={3cm 0 4cm 0},clip,height=0.22\linewidth]{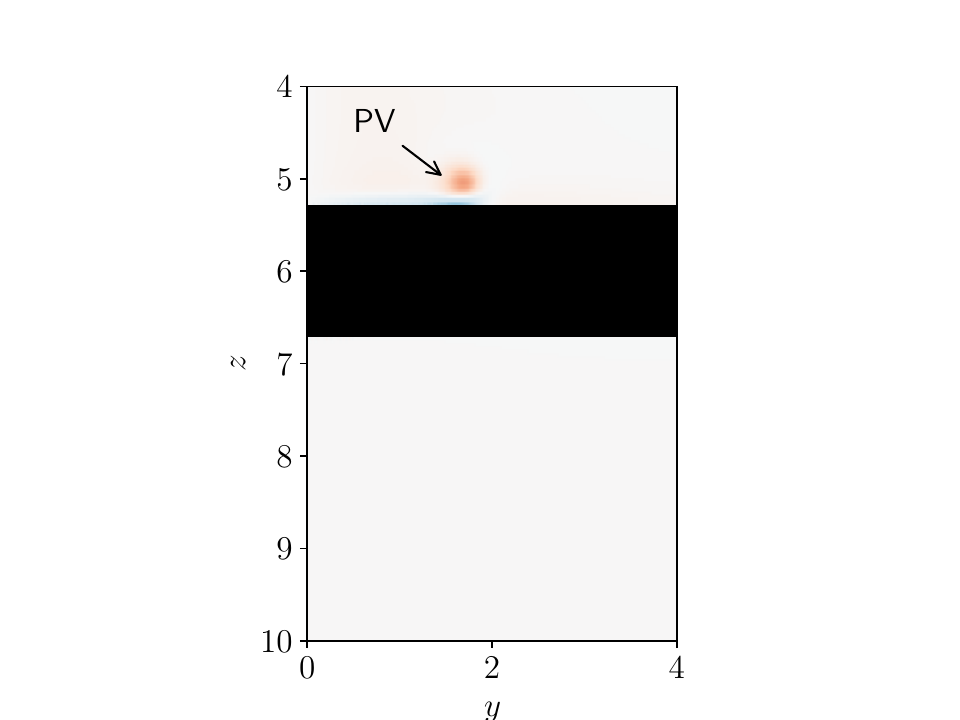}
    \includegraphics[trim={3cm 0 0cm 0},clip,height=0.22\linewidth]{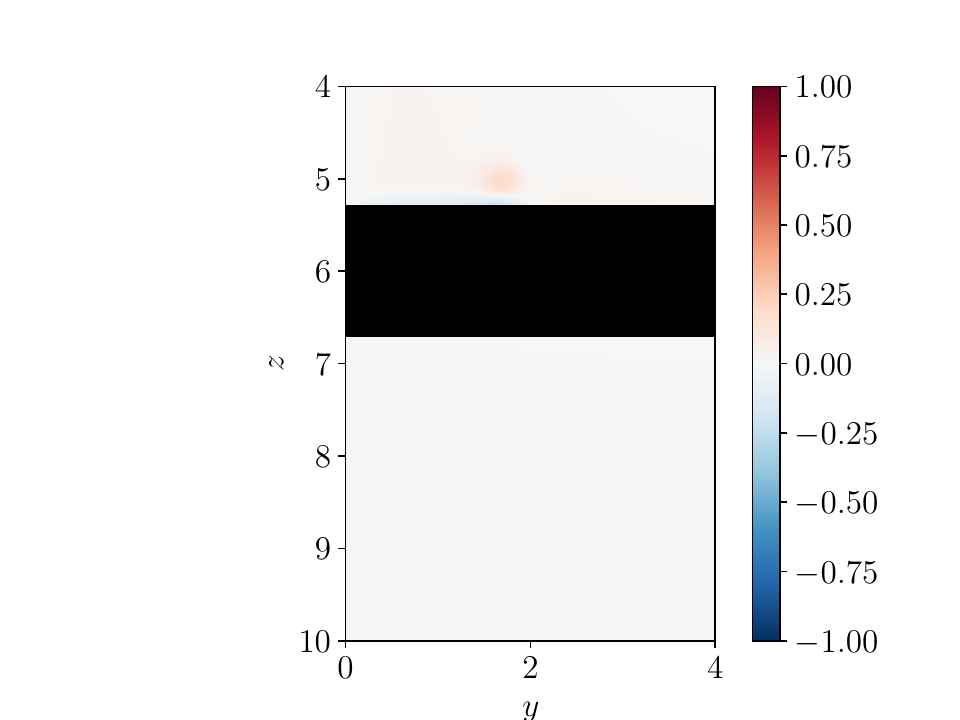}
    \caption{Temporal evolution of the azimuthal vorticity for $T_D=0.7$, $\Lambda=0.2$, $Re_\Gamma=1000$, through the $x$-$z$ plane (top) and $y$-$z$ plane (bottom). From left to right: $t=25$, $35$, $45$, $55$ and $65$. Legend: PV, primary vortex; BL, boundary layer; SV, secondary vorticity. }
    \label{fig:cr0p2-rcyl0p7-wth-re1000}
\end{figure}

Returning to $Re_\Gamma=1000$ and further increasing $T_D$ to $T_D=0.7$ (figures~\ref{fig:cont-cr0p2-rcyl0p7-wth-re1000} and \ref{fig:cr0p2-rcyl0p7-wth-re1000}), we begin to observe the characteristic signatures of the wall regime. The ring interacts with the wire and is stretched in both directions. In the $x$–$z$ plane, the angle at which the secondary vortex structures are ejected increases further and becomes nearly perpendicular to the original trajectory, as the secondary vorticity is now strong enough to substantially deflect the primary vortex. In the $y$–$z$ plane, the secondary vorticity also intensifies and interacts more directly with the primary ring, slightly lifting it upward. Due to vortex stretching, in the three-dimensional iso-contours the ring appears to dissipate. However, when seen in the in the rightmost columns of the volumetric visualizations of figure~\ref{fig:crs-rcyls-re1000}: the primary ring is much weaker even if the topology remains intact.

\begin{figure}
    \centering
    \includegraphics[width=0.80\linewidth]{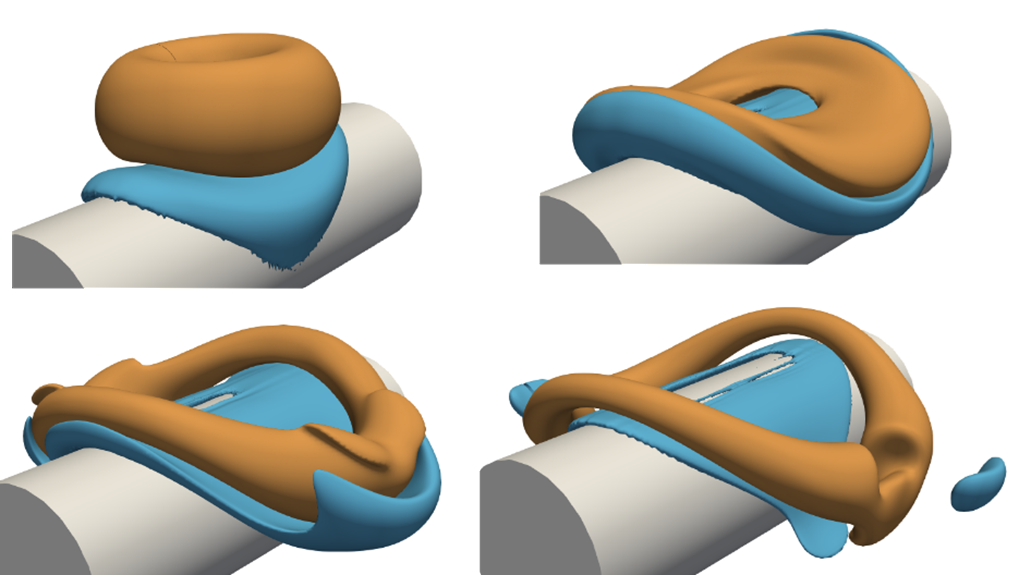}
    \caption{Evolution of the azimuthal vorticity for $T_D=1$, $\Lambda=0.2$, $Re_\Gamma=1000$. Isocontours at $\omega_\theta=0.15$ (orangish-red) and $\omega_\theta=-0.15$ (blue).  From top left to bottom right: $t=20$, $30$, $40$ and $60$. }
    \label{fig:cont-cr0p2-rcyl1-wth-re1000}
\end{figure}

\begin{figure}
    \centering
    \includegraphics[trim={3cm 0 4cm 0},clip,height=0.25\linewidth]{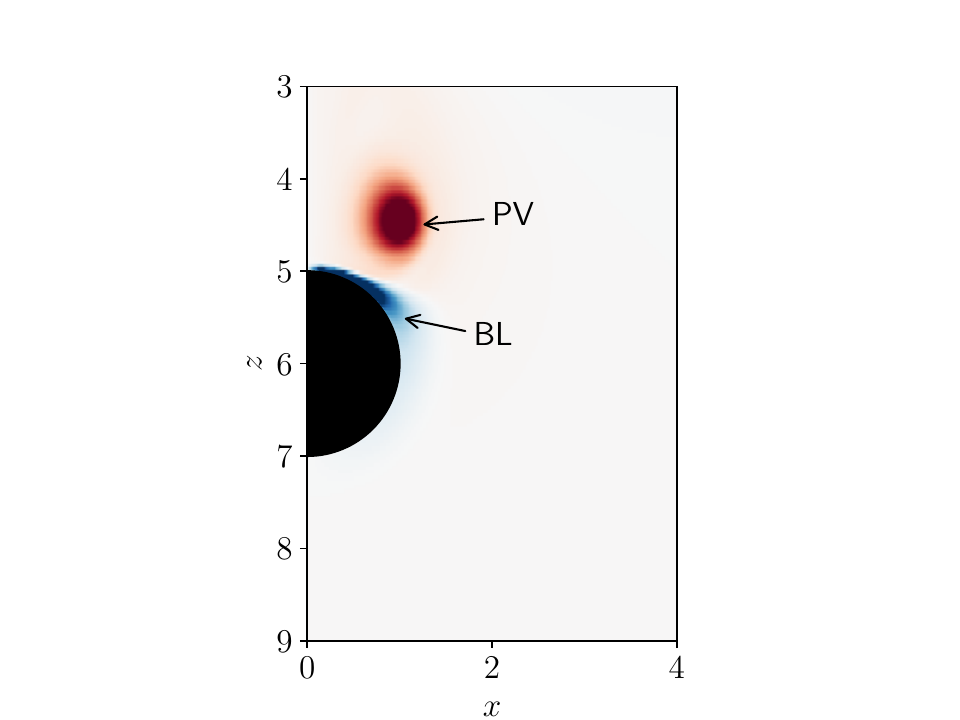}
    \includegraphics[trim={3cm 0 4cm 0},clip,height=0.25\linewidth]{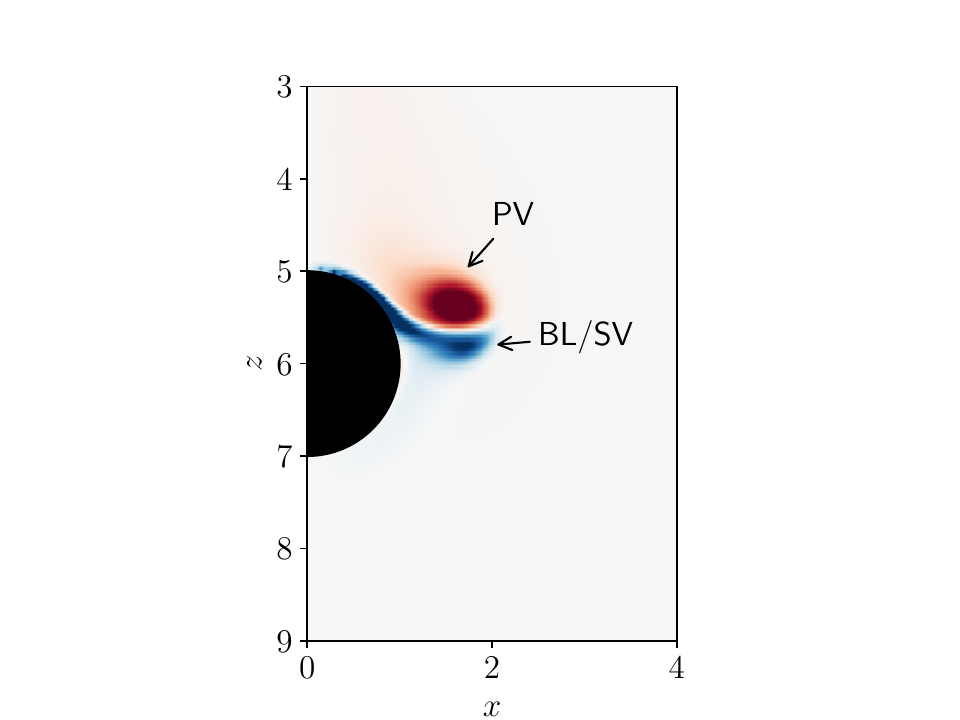}
    \includegraphics[trim={3cm 0 4cm 0},clip,height=0.25\linewidth]{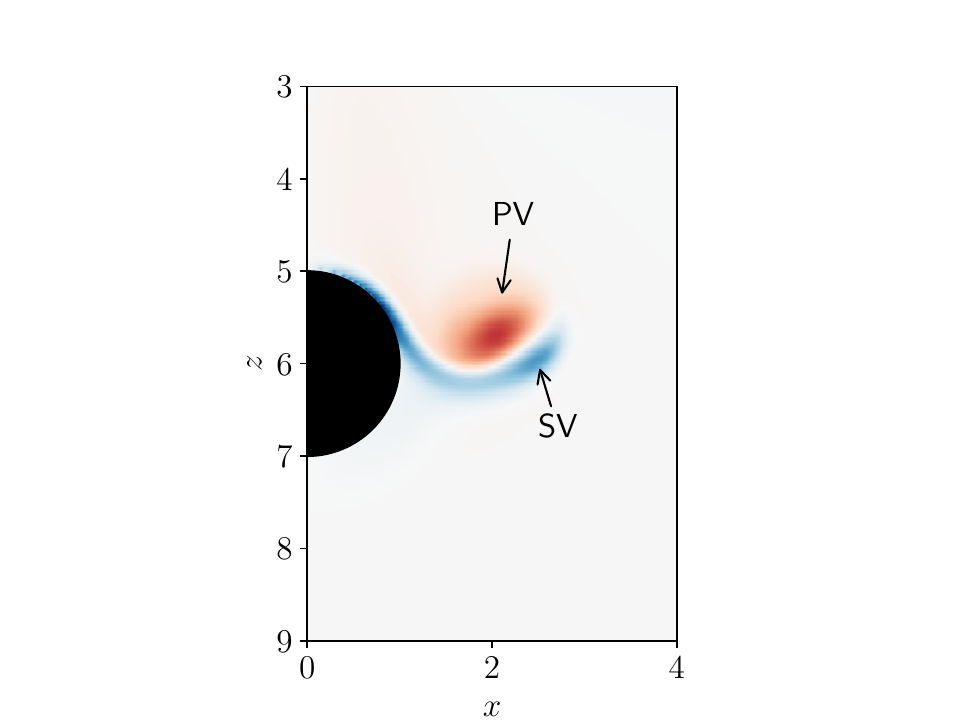}
    \includegraphics[trim={3cm 0 0cm 0},clip,height=0.25\linewidth]{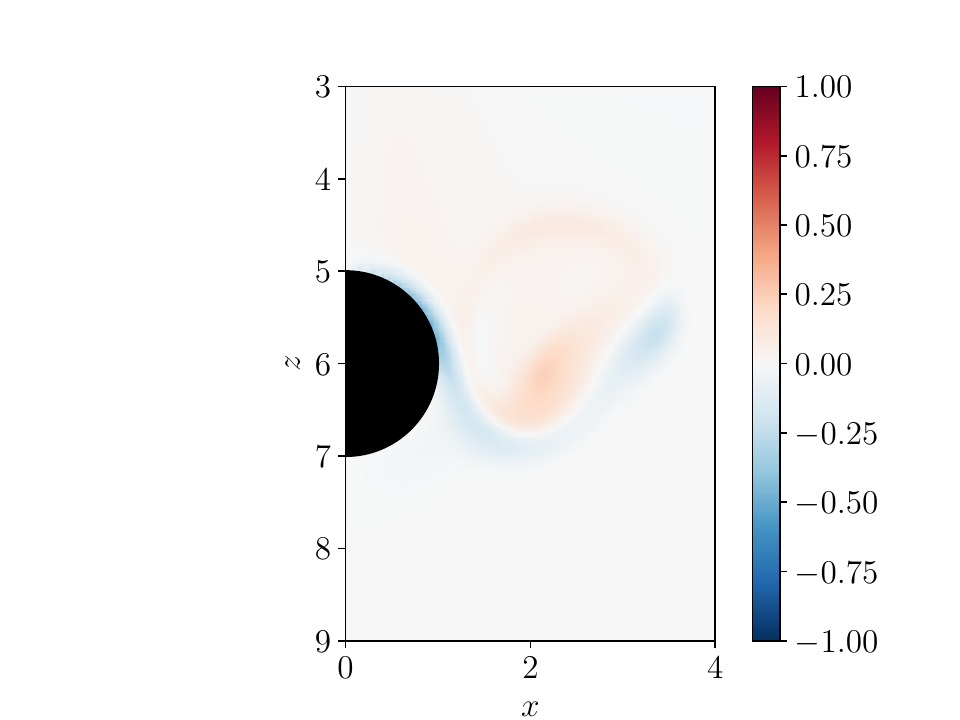}
\\
    \includegraphics[trim={3cm 0 4cm 0},clip,height=0.25\linewidth]{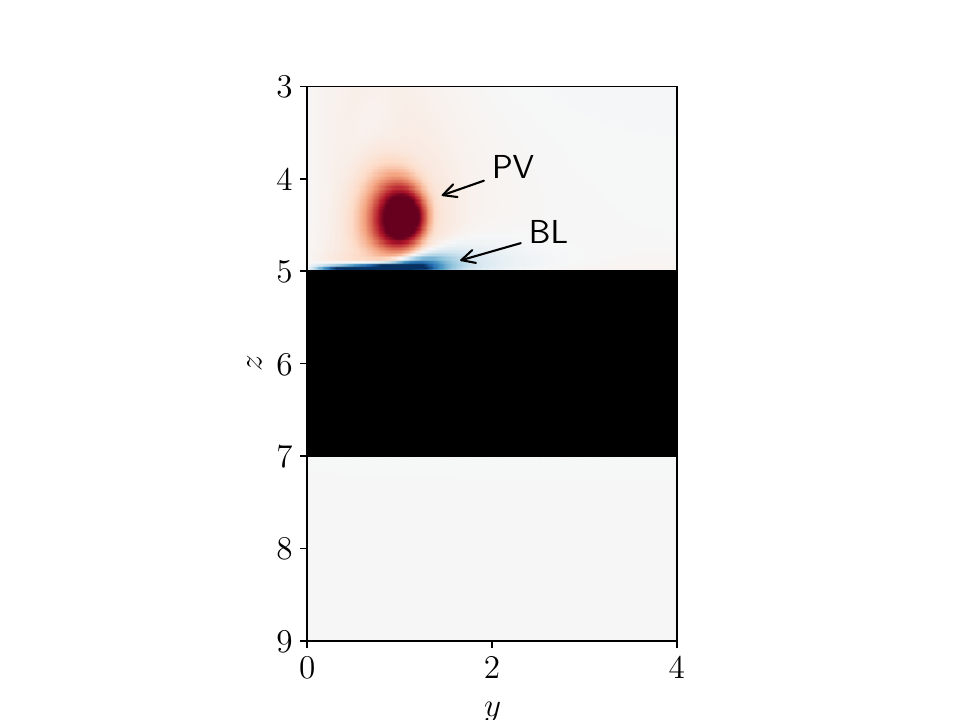}
    \includegraphics[trim={3cm 0 4cm 0},clip,height=0.25\linewidth]{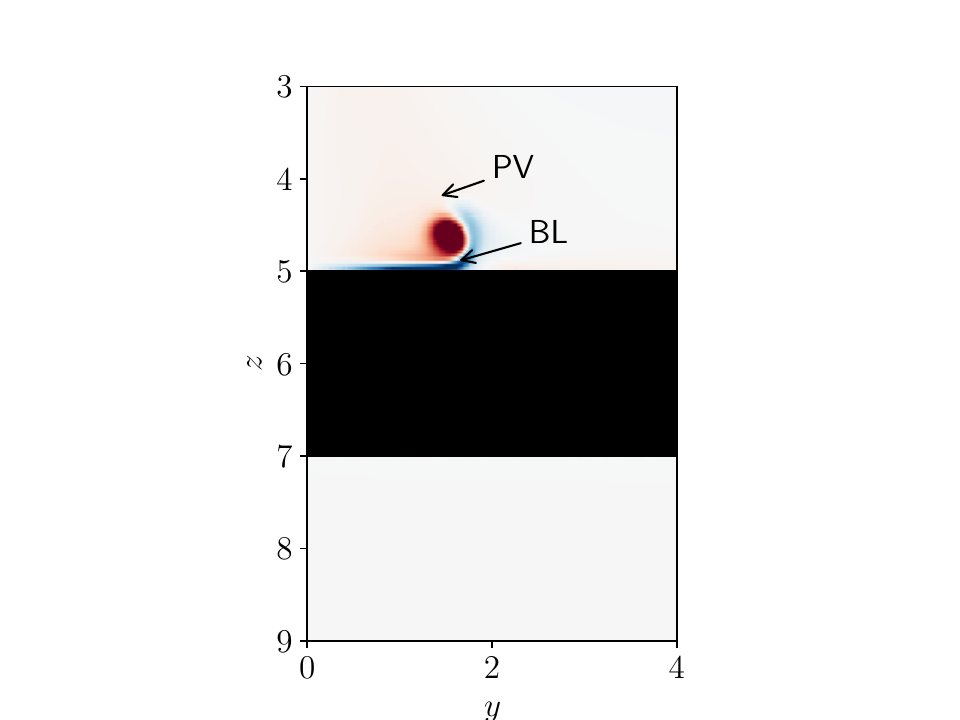}
    \includegraphics[trim={3cm 0 4cm 0},clip,height=0.25\linewidth]{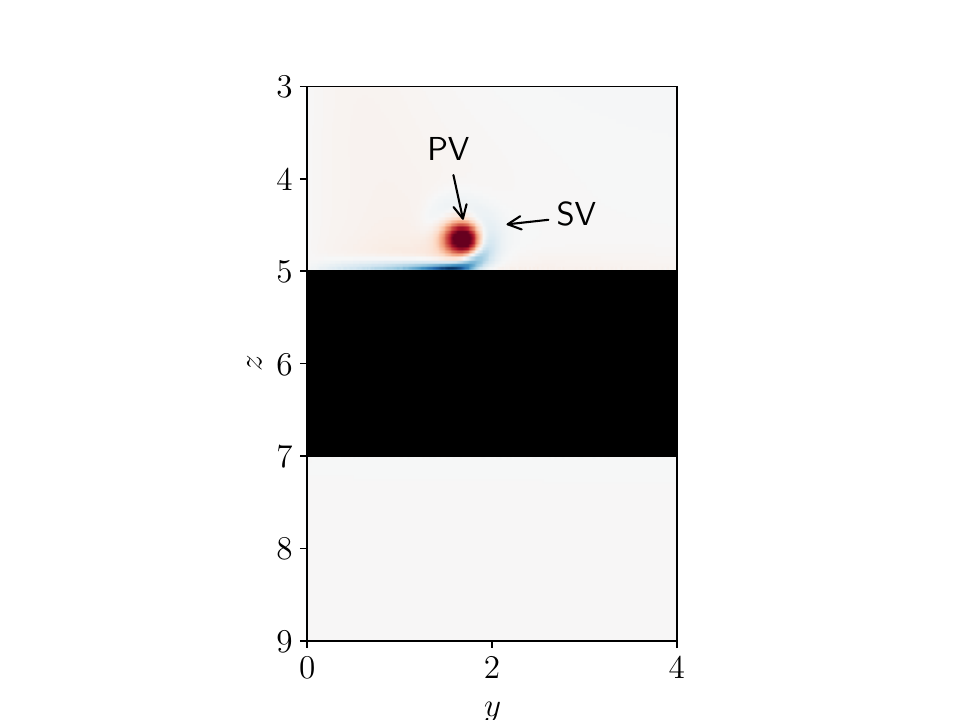}
    \includegraphics[trim={3cm 0 0cm 0},clip,height=0.25\linewidth]{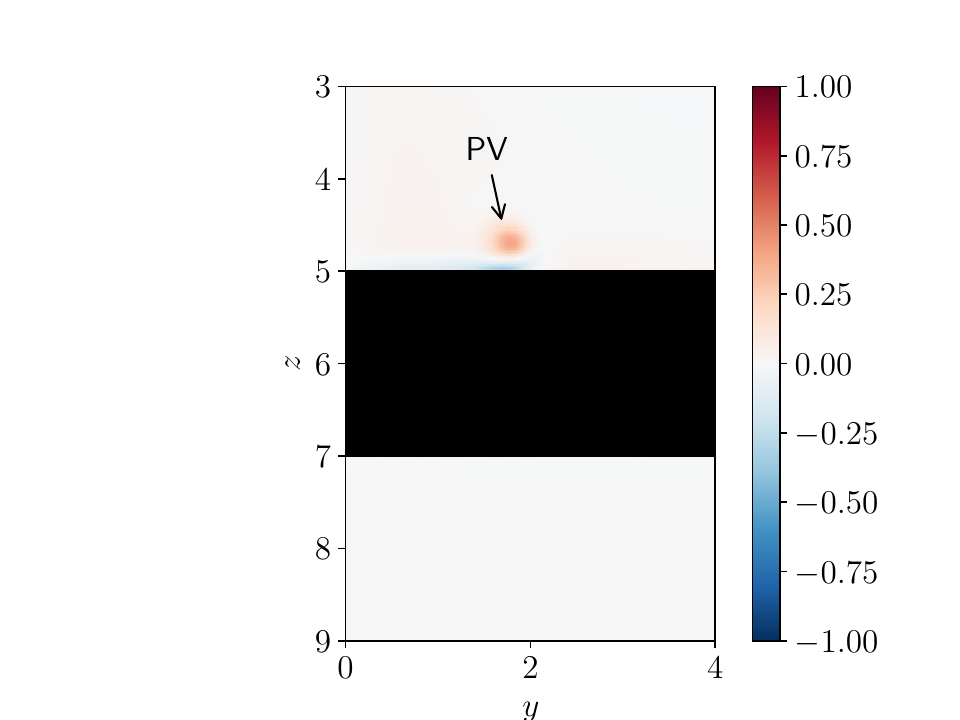}
    \caption{Temporal evolution of the azimuthal vorticity for $T_D=1$, $\Lambda=0.2$, $Re_\Gamma=1000$, through the $x$-$z$ plane (top) and $y$-$z$ plane (bottom). From left to right: $t=20$, $30$, $40$ and $60$. Legend: PV, primary vortex; BL, boundary layer; SV, secondary vorticity.}
    \label{fig:cr0p2-rcyl1-wth-re1000}
\end{figure}

Figures \ref{fig:cont-cr0p2-rcyl1-wth-re1000} and \ref{fig:cr0p2-rcyl1-wth-re1000} present the case $T_D=1$, where the signatures of the curved-wall regime are very apparent. The first stages of the interaction increasingly resembles that of a ring with a flat wall. However, the phenomenology is richer. In the $x$–$z$ plane, the detached boundary layer of secondary vorticity deflects the primary vortex even more strongly, while in the $y$–$z$ plane the primary vorticity slides along the wall, with a stronger deceleration by the oppositely signed boundary-layer vorticity that lifts it away from the surface. These two phenomena illustrate the progressive transition from the cutting to the curved-wall regime: as the object diameter increases, the boundary-layer vorticity strengthens, and in planes perpendicular to the cylinder it can significantly deflect the ring from its initial trajectory. The detached vorticity forms a loop, as observed in figure~\ref{fig:crs-rcyls-re1000}, which at lower Reynolds numbers dissipates, but at higher $Re_\Gamma$ may reconnect with itself and be ejected as a secondary ring (figure~\ref{fig:crs-rcyls-re2000}). This process was previously analyzed in detail by \cite{new2017head,new2021large} (cf.~figure~20 in \cite{new2017head} for a schematic representation). This differs from the cutting regime, where the secondary ring forms primarily through reconnection between the boundary-layer vorticity and the primary vortex ring itself.

In summary, the cutting regime is characterized by the destruction of the primary vortex ring upon interaction with the obstacle, accompanied by the formation of secondary structures originating from both ring and boundary-layer vorticity. As the obstacle diameter increases, the boundary-layer vorticity progressively dominates, and the interaction transitions smoothly into the wall regime. While the transition between regimes is primarily governed by the value of $T_D$, increasing the Reynolds number alters the overall flow phenomenology: stretching and reconnection processes introduce additional complexity, allowing weak tertiary rings to reappear or secondary rings formed solely from boundary-layer vorticity to detach from the primary structure. As the wall regime has been analyzed in detail by \cite{new2017head,new2020collision}, we now turn to a comparison between vortex rings and vortex tubes interacting with cylinders to elucidate the role of flow topology.

\section{The role of body and vortex topology}
\label{sec:topology}

Having established how the vortex–cylinder interaction depends on geometric parameters in the ring configuration, we now compare the flow phenomenology with that of normal body–vortex (tube) interaction (BVI). Mathematically, BVI can be regarded as a limiting case of the ring–cylinder interaction in which $T_D \to 0$ while $T_\sigma$ remains finite as $\Lambda \to 0$. Similarly, the interaction of a vortex ring with a wall corresponds to the case where both $T_D$ and $T_\sigma \to \infty$ while $\Lambda$ remains finite. However, taking the limits of the parameters is not enough to capture the full picture: the topology of both the vortex and the body fundamentally influences the interaction, determining which features are unique to ring geometry and which are generic to vortex–body encounters.

\begin{figure}
    \centering
    \includegraphics[width=0.9\linewidth]{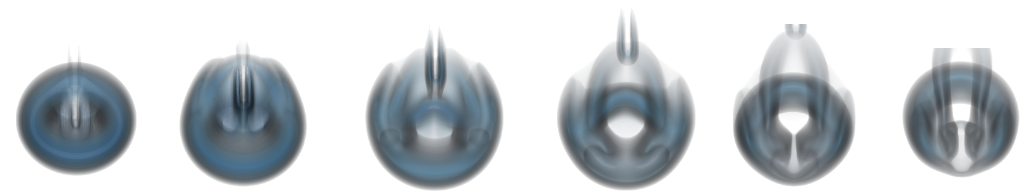}
    \caption{Volumetric visualization of the instantaneous vorticity modulus showing the impact of a vortex ring with $\Lambda=0.2$ and $Re_\Gamma=1000$ upon a half a cylinder with $T_D=0.1$. Time ranges from left to right from $t=30$ to $80$ in intervals of $10$. }
    \label{fig:ww-halfwire}
\end{figure}

To illustrate this, figure~\ref{fig:ww-halfwire} shows the evolution of vorticity during the interaction between a vortex ring with $\Lambda=0.2$ and $Re_\Gamma=1000$ and a half-cylinder of $T_D=0.1$. For comparison, figure~\ref{fig:cr0p2-rcyls-re1000} in the previous section showed the corresponding interaction with a full cylinder, resulting in a clean cutting of the ring. In the present case, the ring is first stretched at the point of contact and deformed as the remainder of the structure passes around the obstacle, eventually leading to a partial cut. However, the two sides of the ring subsequently reconnect downstream, and a deformation wave propagates along the circumference from the initial contact point until it reaches the opposite side.

\begin{figure}
    \centering
    \includegraphics[trim={2cm 0 4cm 0},clip,height=0.22\linewidth]{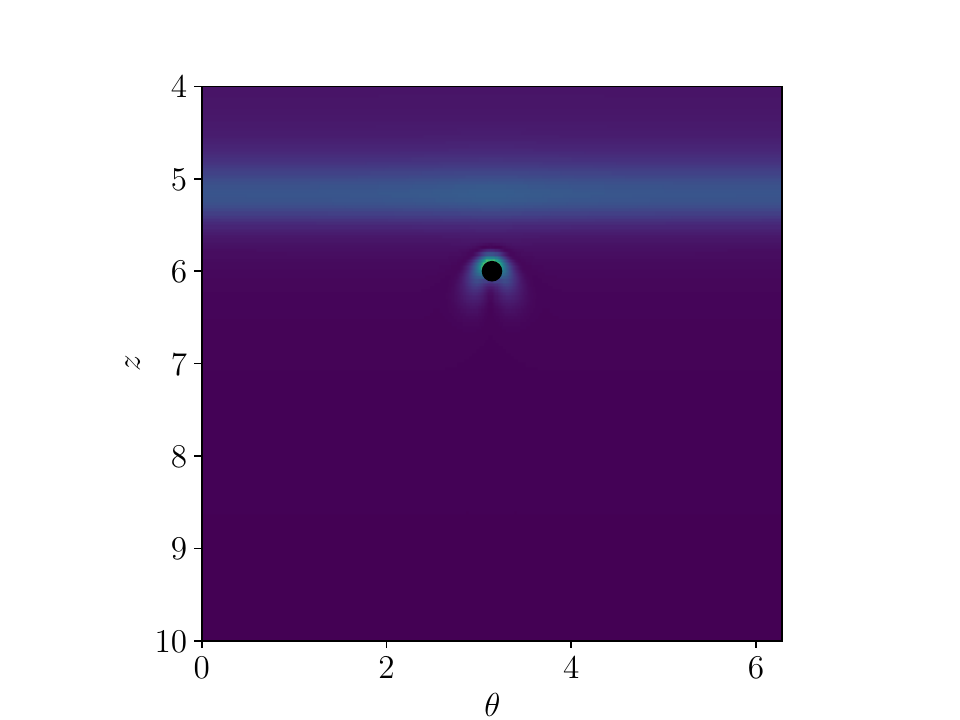}
    \includegraphics[trim={2.5cm 0 4cm 0},clip,height=0.22\linewidth]{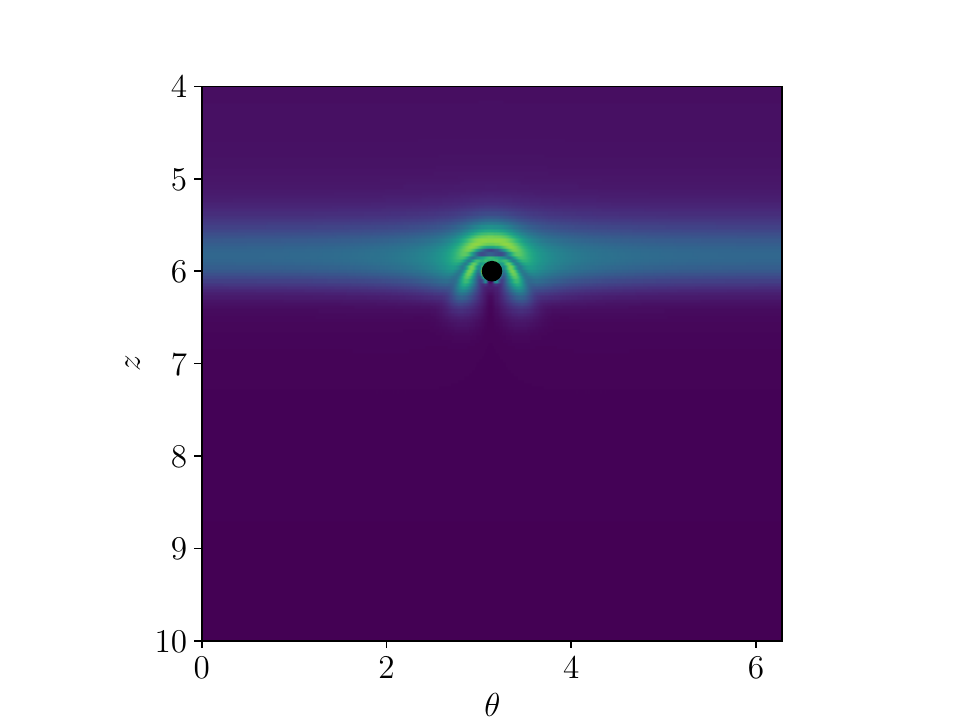}
    \includegraphics[trim={2.5cm 0 4cm 0},clip,height=0.22\linewidth]{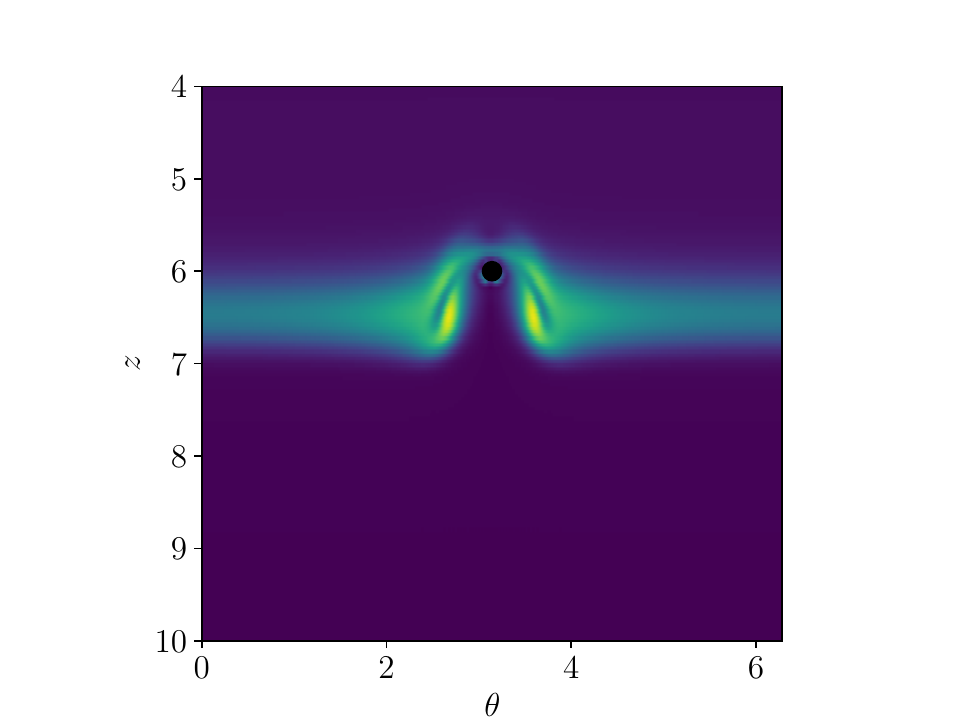}
    \includegraphics[trim={2.5cm 0 4cm 0},clip,height=0.22\linewidth]{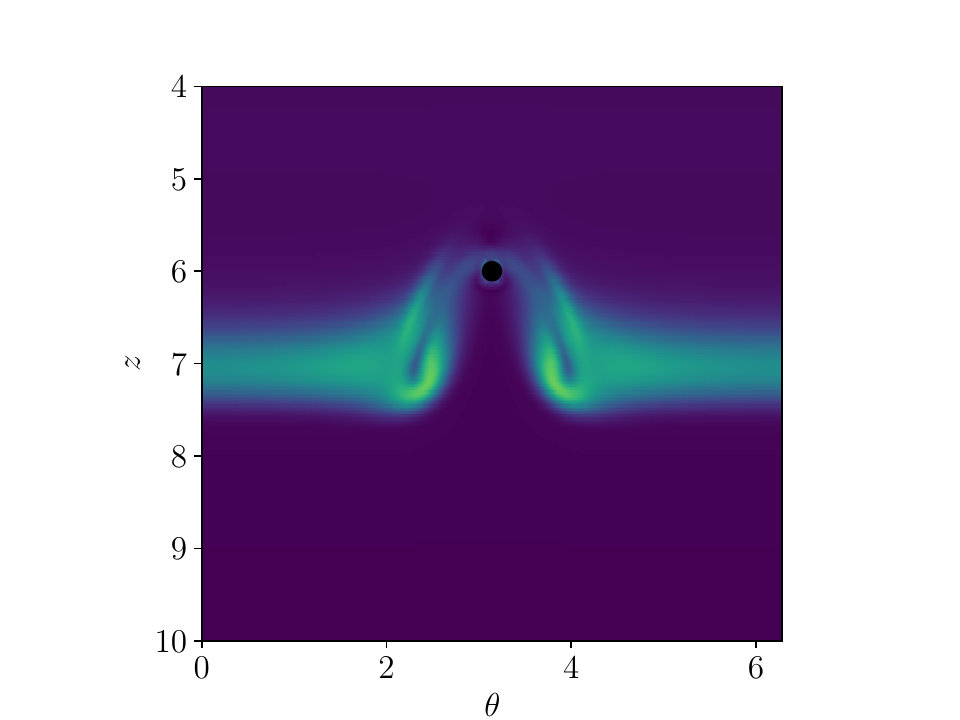}
    \includegraphics[trim={1.5cm 0 0cm 0},clip,height=0.22\linewidth]{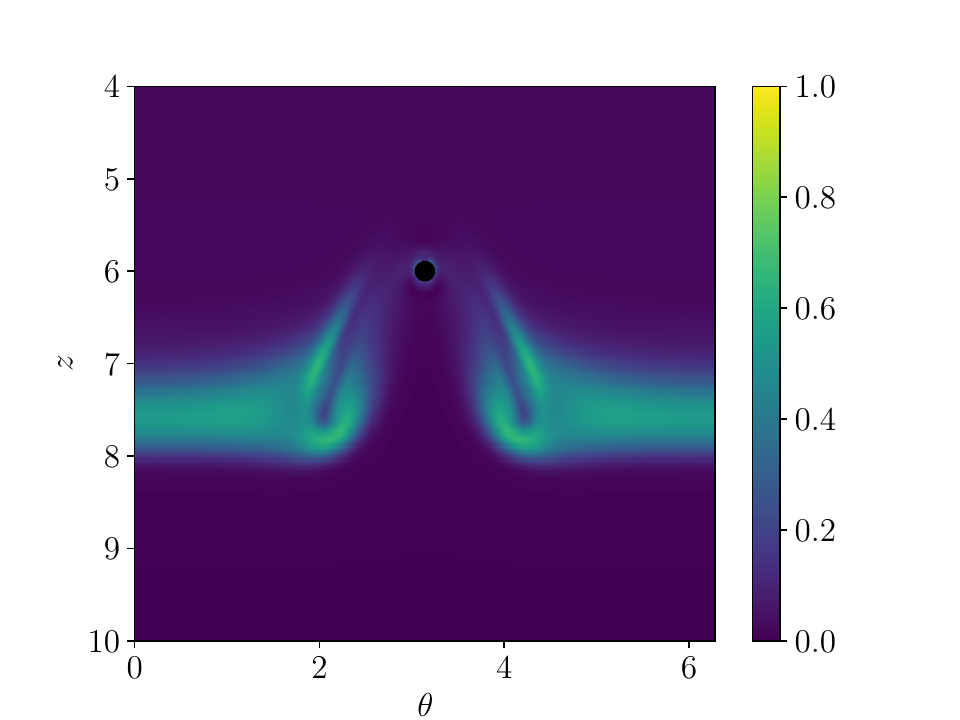}
    \caption{Evolution of the vorticity modulus at $r=1$ for the interaction with half a cylinder with $\Lambda=0.2$, $Re_\Gamma=1000$ and $T_D=0.1$. From left to right: $t=25$, $30$, $35$, $40$ and $45$.}
    \label{fig:ww-halfwire-cut}
\end{figure}

Overall, this behavior closely resembles that of normal BVI in the weak vortex regime. To further support this, figure~\ref{fig:ww-halfwire-cut} shows the temporal evolution of the vorticity modulus along a constant-radius cut through the ring center. These panels can be directly compared to figure~11 in \cite{soriano2024direct} for the case of cylinder–vortex tube interaction in the strong regime ($I_P=0.25$). The same stretching and necking dynamics are observed as the vortex approaches the object, followed by subsequent deformation and partial recovery of the core. This comparison highlights the role of topology in governing whether reconnection can occur: because the ring has two points of contact with a full cylinder, a secondary string of vorticity can form and reconnect with the ring to form a secondary ring, whereas in the case of the half-wire, the topological characteristics are different and no such reconfiguration is possible. Thus, when topology permits, rings and tubes can exhibit qualitatively similar phenomenology.

\begin{figure}
    \centering
    \includegraphics[width=0.9\linewidth]{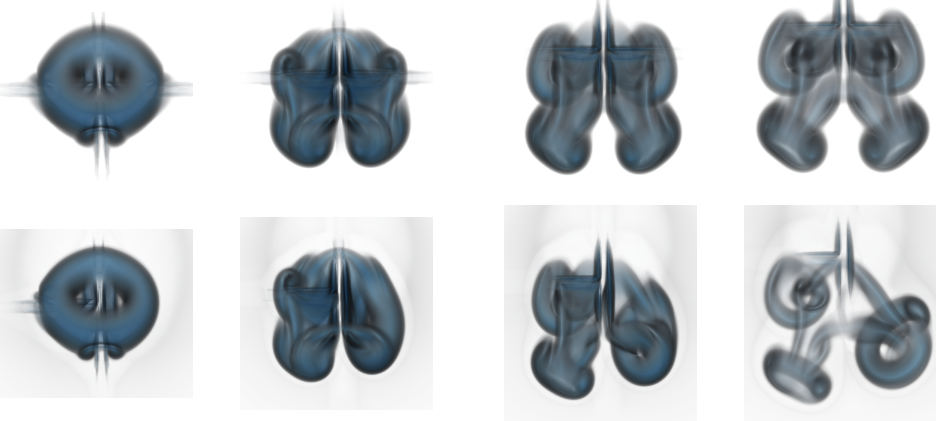}
    \caption{Volumetric visualization of the instantaneous vorticity modulus of the impact of a vortex ring with $\Lambda=0.2$ and $Re_\Gamma=1000$ upon a X (top) and T-like (bottom) structures with a cylinder with $T_D=0.1$. Time ranges from left to right from $t=30$ to $80$ in intervals of $10$. }
    \label{fig:ww-otherwire}
\end{figure}

To further explore the effect of topology, we simulate the interaction between a ring and several objects such as T-shaped structures or $N_b$-pointed stars. First, figure~\ref{fig:ww-otherwire} shows simulations in which the ring interacts with X-shaped ($N_b=4$) and T-shaped structures. In the first case, the ring is cut into four smaller rings, while in the second it splits into three rings of differing size and circulation. Using the negatively signed angular vorticity section of the ejected ring, we estimate the circulation of these structures. For the full-wire case, each secondary ring carries approximately $23\%$ of the original circulation at $t=60$. For the X-structure, each secondary ring contains roughly $15\%$, while for the T-structure at the same time instant, the two smaller rings each contain about $15\%$ and the larger one about $24\%$ of the original circulation.

If one naively totals the circulation of the four rings, it may seem that they contain more total circulation than the rings resulting from a ``cut in half''. However, there is no reason to think that the circulation would ``split'' between the rings: indeed, in the absence of viscosity each ring should retain the full circulation.  

\begin{figure}
    \centering
    \includegraphics[trim={3cm 0 4cm 0},clip,height=0.30\linewidth]{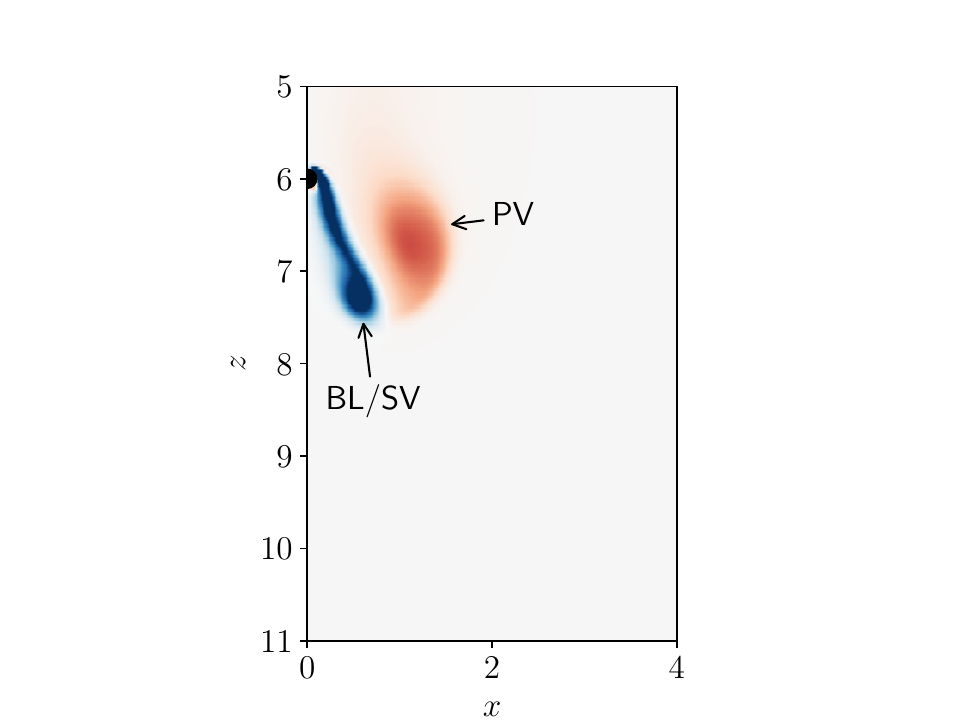}
    \includegraphics[trim={3cm 0 0cm 0},clip,height=0.30\linewidth]{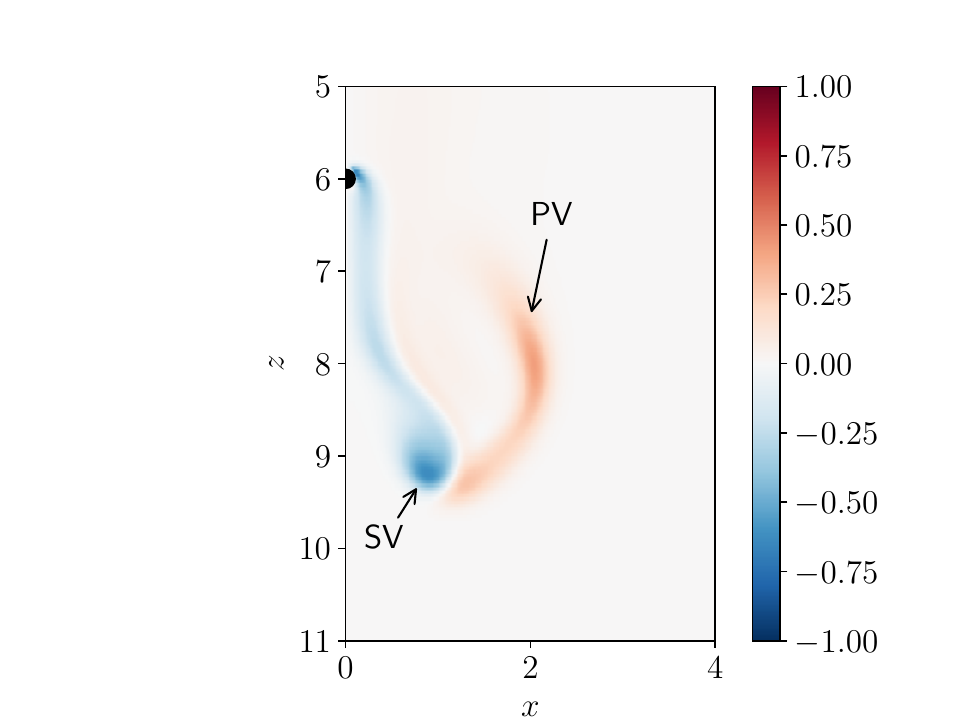}
    \includegraphics[height=0.30\linewidth]{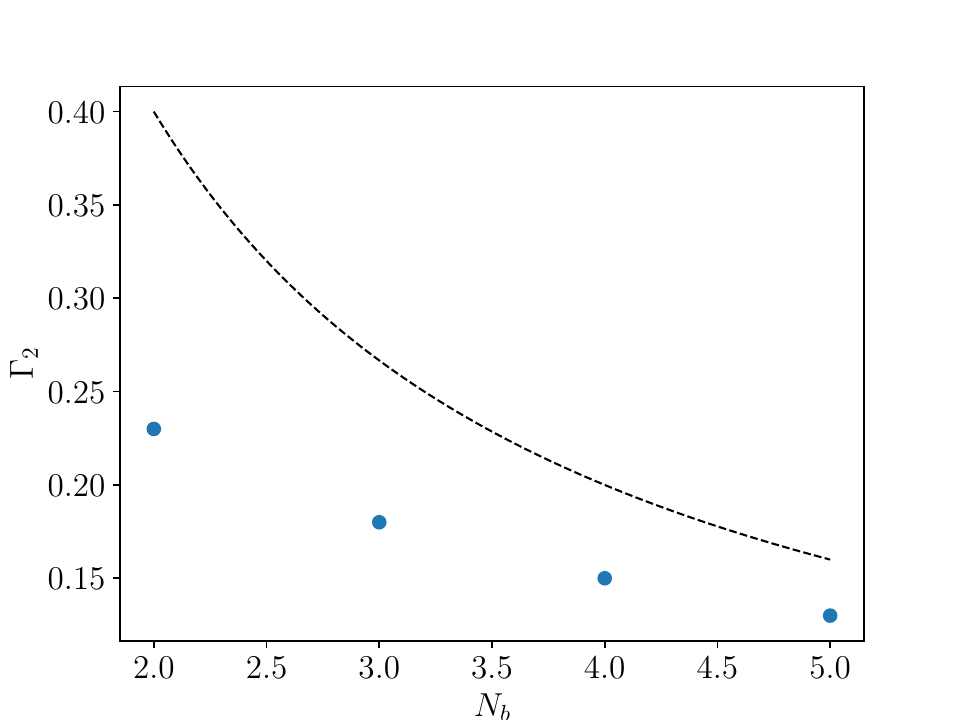}
    \caption{Left and center: instantaneous azimuthal vorticity at the mid-gap of two cylinders for $N_b=3$, with $\Lambda=0.2$, $Re_\Gamma=1000$ and $T_D=0.1$ at $t=37.5$ and $t=60$. Right: circulation of the secondary vorticity at $t=60$ for different number of $N_b$. The dotted line denotes $1/N_b$ behaviour. Labels: PV, Primary Vorticity; BL, Boundary Layer; SV, Secondary Vorticity. }
    \label{fig:nbars}
\end{figure}

To further showcase this, the left panel of figure \ref{fig:nbars} we show the instantaneous azimuthal vorticity for $N_b=3$, i.e.~a three pointed star, at the mid-point between two points. We can observe that the behaviour of the vorticity is quite similar to that seen for the interaction with a wire at similar $Re_\Gamma$, $T_D$ in Figure \ref{fig:cr0p2-rcyl0p1-wth-re1000}: the detached secondary vorticity has rolled up while the primary vorticity is more deformed. Regardless of how many rings the primary ring splits into, the vorticity distribution of both primary and secondary vorticity remains approximately equal (save viscous effects). Hence, the loss of circulation is due to increased viscous effects, not to the number of rings. 

To quantify this, in the right panel of figure \ref{fig:nbars}, we show the circulation in the ejected rings $\Gamma_2$ s a function of $N_b$, with $\Gamma_2\sim N_b^{-1}$ included as a reference. We can see that while the circulation decreases, as the number of bars/points becomes larger, the decrease in circulation of the secondary rings is less pronounced than a theoretical estimate based on circulation equipartition.

These results again underscore the central role of topology in determining the outcome of the interaction. By altering the connectivity of the obstacle, we effectively control how the primary vortex ring is partitioned into smaller components, and the circulation of each fragment depends on the angular arrangement of the cutting surfaces. In essence, the topology of the object acts as a governing parameter in the redistribution of circulation and the formation of secondary structures.

\section{Summary and Outlook}
\label{sec:cc}

The present results reveal the regime phase space of the interaction between a vortex ring and cylindrical obstacles across a wide range of geometric and dynamical parameters. At low values of $T_D$, corresponding to the ``wire'' regime, the primary vortex ring survives the encounter with only moderate deformation as reported by \cite{naitoh1995vortex,adhikari2009impact}. The interaction generates a thin boundary-layer sheet that detaches from the obstacle and travels with the primary vortex as a weak lobe of secondary vorticity. In this regime, the circulation of the primary ring remains of the same order of magnitude after the interaction.

As $T_D$ increases, the interaction transitions smoothly into the ``cutting'' regime, which has not been previously reported in the literature. Here, the primary ring can no longer reform after passing around the obstacle, and the flow reorganizes into multiple coherent structures formed through reconnection between secondary boundary-layer and primary ring vorticity. The process leads to a redistribution of circulation among secondary rings and other structures, whose evolution depends on both $T_D$ and $Re_\Gamma$. Increasing the Reynolds number enhances stretching and produces additional small-scale features while enabling weak tertiary rings to emerge from detached lobes of vorticity. 

For sufficiently large obstacles, the interaction approaches the ``wall'' regime, where the dynamics are dominated by boundary-layer vorticity and the detached shear layer strongly deflects the primary vortex. The transition between the cutting and wall regimes is again continuous and primarily governed by $T_D$. At high Reynolds numbers, the detached layers can roll up and reconnect to form secondary rings, consistent with previous experimental and numerical observations \citep{new2017head,new2021large}.

Finally, the comparison with body–vortex interactions and with modified obstacle topologies demonstrates that topology fundamentally determines the outcome of the interaction in the cutting regime. When the vortex reconnects with itself across the obstacle, as in the half-cylinder case, the encounter closely resembles canonical body–vortex interaction. Conversely, when the primary ring reconnects with secondary vorticity, the ring is partitioned into smaller structures whose circulation depends on the geometry and angular configuration of the cutting surfaces. 

Overall, the results demonstrate that the flow phenomenology of vortex–cylinder interactions can be understood as a continuum governed by three parameters: the geometric scale ratio $T_D$, the Reynolds number $Re_\Gamma$, and the topology of both the vortex and the obstacle. Together, these parameters determine whether the primary ring survives, is cut, or is fully destroyed and replaced by new coherent structures. Meanwhile, the ring slenderness $\Lambda$ plays a minor role, while the thickness ratio $T_\sigma$ was shown to a worse choice than $T_D$ to delineate regimes of the interaction between vortex rings and objects.

These findings lay the groundwork for simple models based on vortex flows to more complex and turbulent configurations such as blade-vortex interaction or gust-airfoil problems. While we have only considered cylindrical objects with a circular cross-section, we do not expect the results to significantly change if the object has a different shape. However, other effects could be added: the object could move towards the vortex such that the effective impact parameter is changed, or axial flow could be added to the ring to complicate the picture. Finally, further increasing the Reynolds number may open different behaviors not seen in the manuscript and further elucidate the similarities between vortex-ring interaction and vortex–structure interactions in natural and technological flows.

\backsection[Funding]{ROM acknowledges support from the Emergia Program of the Junta de Andalucía (Spain). We also thank the Systems Unit of the Information Systems Area of the University of Cadiz for computer resources and technical support. }

\backsection[Declaration of interests]{The authors report no conflict of interest.}

\backsection[Generative AI:]{Gemini has been used to assist in the writing of the manuscript.}

\backsection[Author ORCIDs]{R. Ostilla-M\'onico, https://orcid.org/0000-0001-7049-2432}

\appendix

\section{Mesh Independence Checks}
\label{appA}

To assess the adequateness of the grid, two additional simulations of the case with $T_D=0.4$, $\Lambda=0.2$ and $Re_\Gamma=2000$ were conducted with coarse and fine grids. The coarse grid was taken by halving the resolution of the base mesh in all directions, i.e.~$N_\theta\times N_r\times N_z=192\times192\times192$, while the fine grid was taken by increasing the number of points by a factor of $1.5\times$ in all directions, i.e.~$N_\theta\times N_r\times N_z=512\times512\times512$. The medium, or reference mesh has a resolution of $N_\theta\times N_r\times N_z=384\times384\times384$.

\begin{figure}
    \centering
    \includegraphics[trim={3cm 0 4cm 0},clip,height=0.25\linewidth]{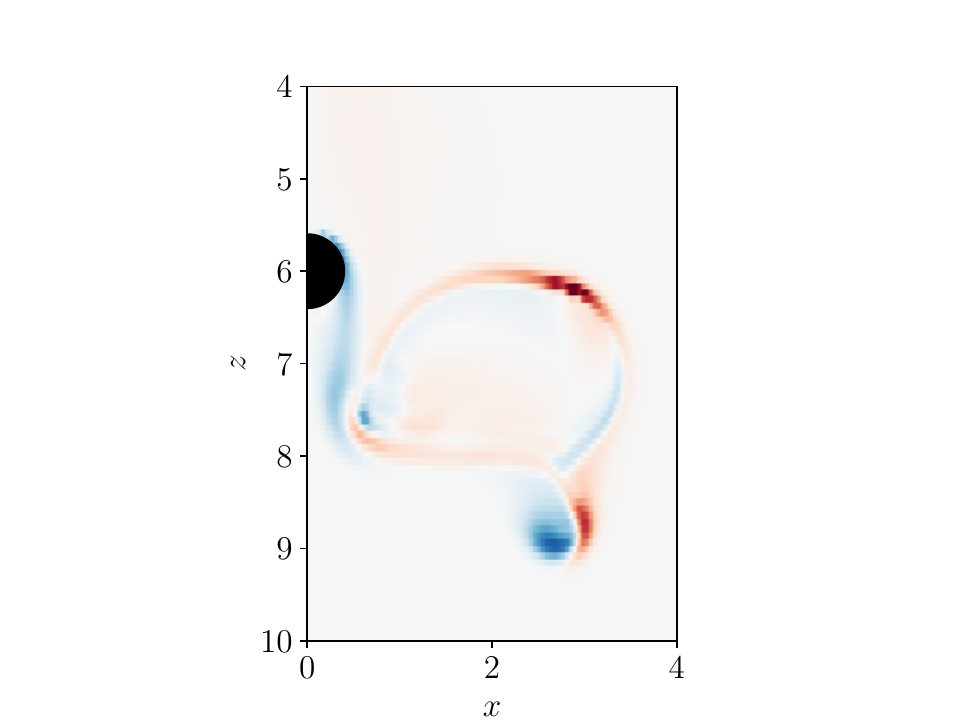}
    \includegraphics[trim={3cm 0 4cm 0},clip,height=0.25\linewidth]{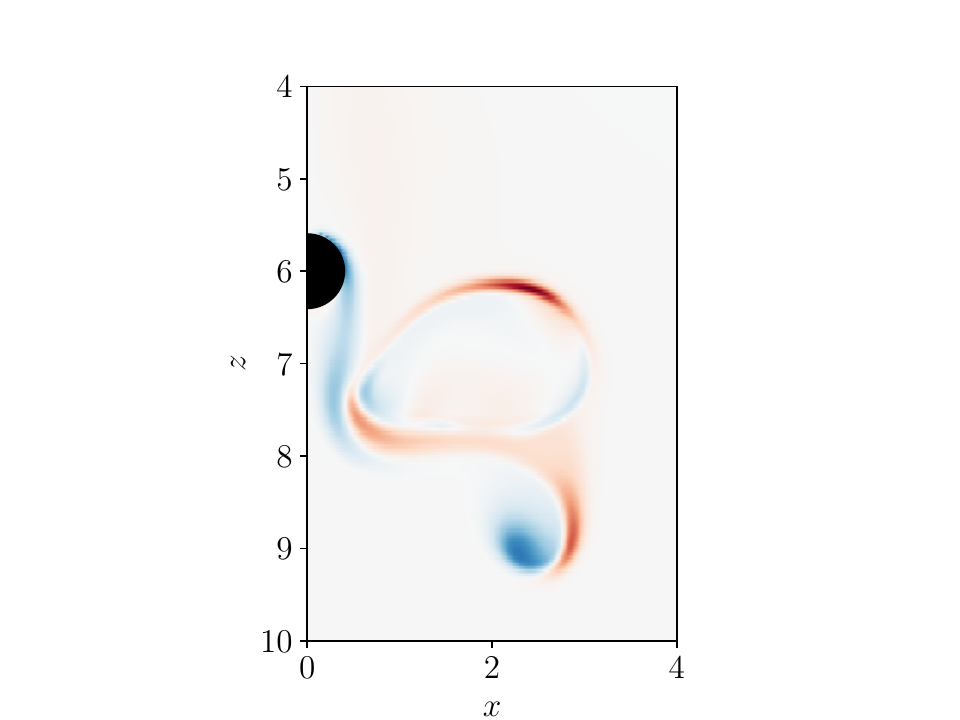}
    \includegraphics[trim={3cm 0 0cm 0},clip,height=0.25\linewidth]{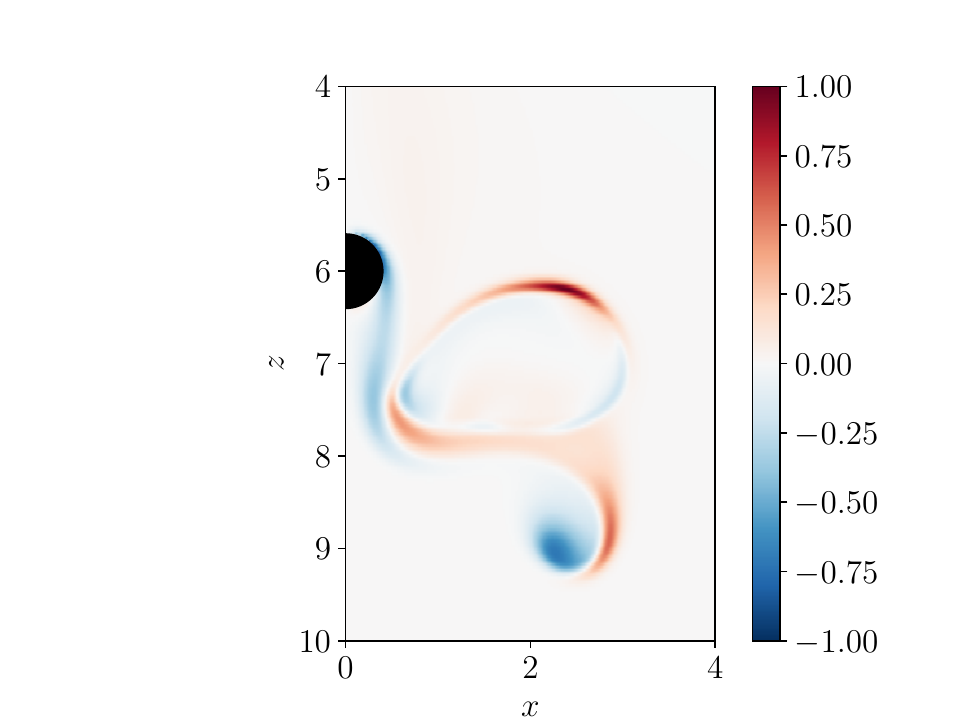}
\\
    \includegraphics[trim={3cm 0 4cm 0},clip,height=0.25\linewidth]{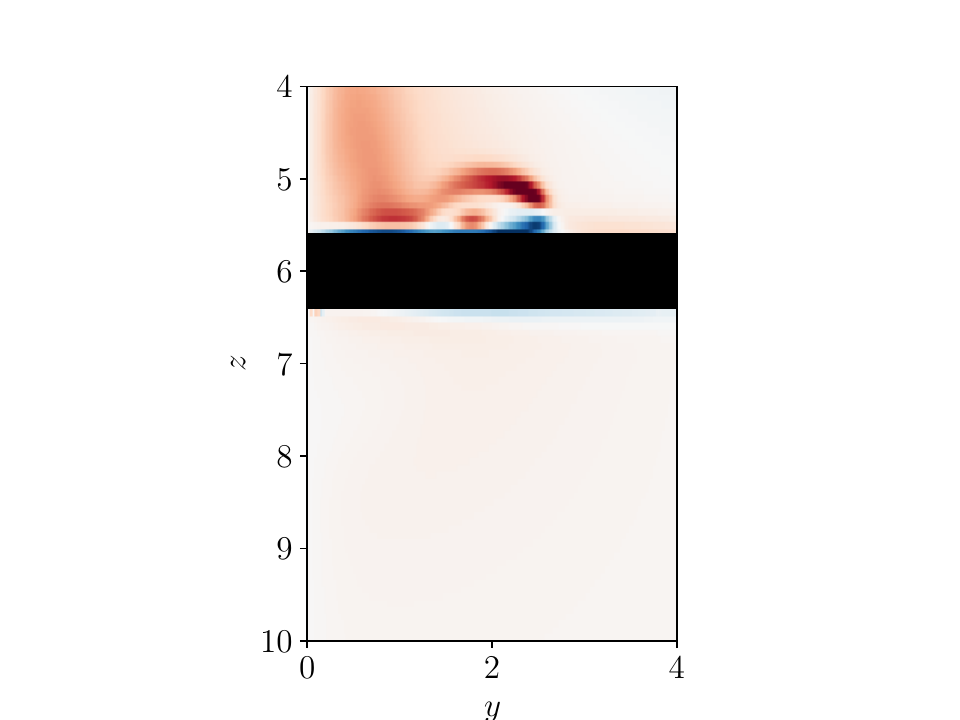}
    \includegraphics[trim={3cm 0 4cm 0},clip,height=0.25\linewidth]{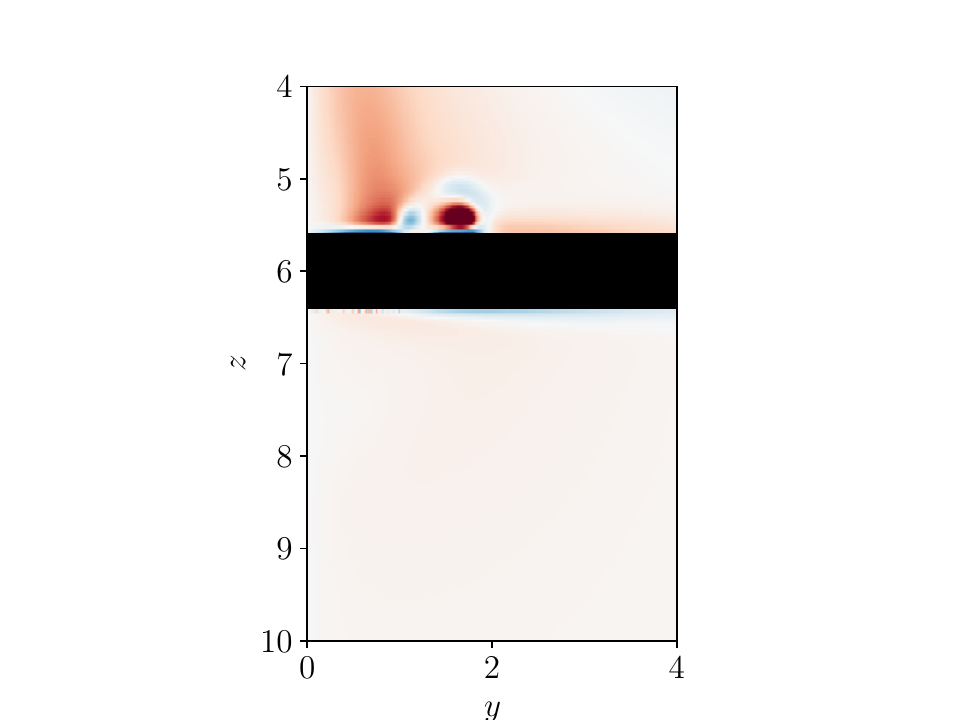}
    \includegraphics[trim={3cm 0 0cm 0},clip,height=0.25\linewidth]{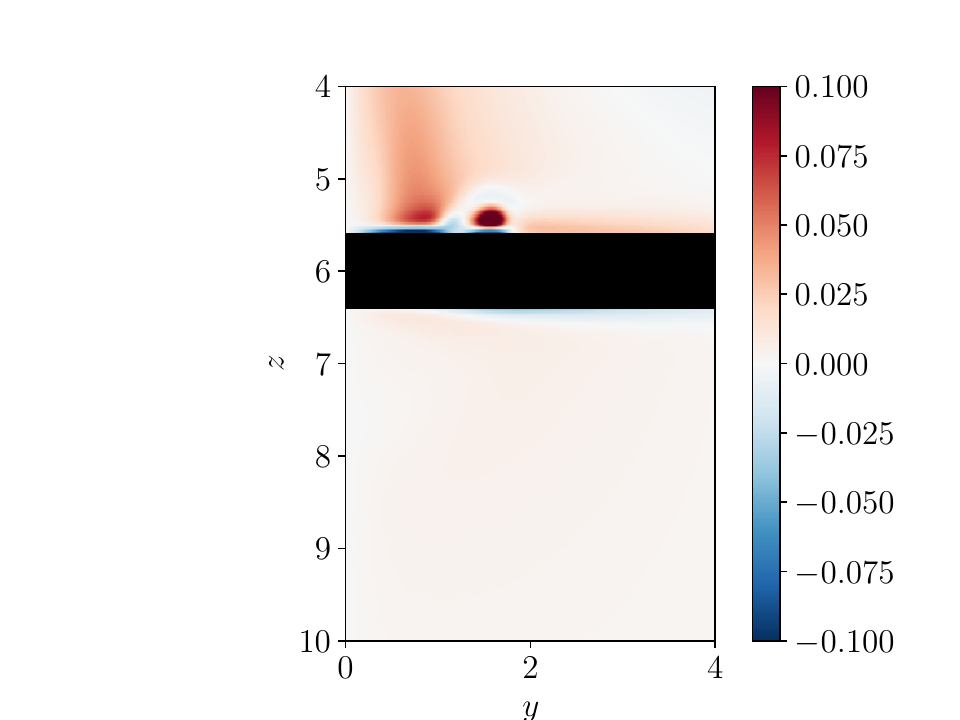}
    \caption{Instantaneous azimuthal vorticity at $t=45$ for $\Lambda=0.2$, $Re_\Gamma=2000$ and $T_D=0.4$ through the $x$-$z$ plane (top) and $y$-$z$ plane (bottom) for the three meshes simulated: coarse (left), medium/reference (center) and fine (right).  }
    \label{fig:meshchecks}
\end{figure}

Figure~\ref{fig:meshchecks} shows the instantaneous angular vorticity $\omega_\theta$ at the time of impact ($t=45$) for the three meshes across two different cuts in the $x$-$z$ plane and the $y$-$z$ plane which correspond to the cuts commonly used in the manuscript. In the $x$-$z$ plane, similar behaviour is seen for all meshes, even if the vortex sheets do not appear as sharp in the coarse mesh. However, in the $y$-$z$ plane larger differences can be appreciated between the coarse and the reference and fine meshes, showing that the resolution in the coarse mesh is insufficient to adequately capture the physics of the problem. When comparing the reference and fine meshes, we can observe that overall phenomenology is well captured, unlike what is seen for the coarser mesh, where the flow topology is changed. This gives us confidence in using the reference mesh in this manuscript.

\bibliographystyle{jfm}
\bibliography{jfm}

\end{document}